\documentclass[12pt]{article}
\usepackage{graphicx}
\usepackage{amssymb}

\newtheorem{Theorem}{Theorem}
\newtheorem{Lemma}{Lemma}
\newtheorem{Proposition}{Proposition}
\newtheorem{Corollary}{Corollary}

\newcommand{\be}{\begin{eqnarray}}
\newcommand{\ee}{\end{eqnarray}}

\title{General Relativistic Self-Similar Waves\\ that induce an\\ Anomalous Acceleration\\into the\\Standard Model of Cosmology}
\author{\it Joel Smoller\footnotemark[1]\ ~ {\rm and}\ \  Blake Temple\footnotemark[2]\footnotemark[3]}
\begin{document}
 \large
\maketitle


\footnotetext[1]{Department of Mathematics, University of Michigan, Ann Arbor, MI 48109; Supported by NSF Applied Mathematics Grant Number DMS-060-3754.}

\footnotetext[2]{Department of Mathematics, University of California, Davis, Davis CA
95616; Supported by NSF Applied Mathematics Grant Number DMS-070-7532.}

\footnotetext[3]{Second author B.T originally proposed the idea that a secondary expansion wave reflected backwards from the cosmic shock wave constructed in \cite{smolte1} might account for the anomalous acceleration of the galaxies in the talk \cite{temptalk} and NSF proposal DMS-060-3754, c.f. \cite{smoltePNAS}.}


\newpage
\large
\centerline{\bf Abstract}
\vspace{.4cm}

\nonumber We prove that the Einstein equations for a spherically symmetric spacetime in Standard Schwarzschild Coordinates (SSC) form a closed system of three ordinary differential equations for a family of self-similar expanding waves,
and the critical ($k=0$) Friedmann universe associated with the pure radiation phase of the Standard Model of Cosmology, is embedded as a single point in this family.  Removing a scaling law and imposing regularity at the center, we prove that the family reduces to an implicitly defined one parameter family of distinct spacetimes determined by the value of a new {\it acceleration parameter} $a$, such that $a=1$ corresponds to the Standard Model.   We prove that all of the self-similar spacetimes in the family are distinct from the non-critical $k\neq0$ Friedmann  spacetimes, thereby {\it characterizing} the critical $k=0$ Friedmann universe as the unique spacetime lying at the intersection of these two one-parameter families.  We then present a mathematically rigorous analysis of solutions near the singular point at the center, deriving the expansion of solutions up to fourth order in the fractional distance to the Hubble Length.  Finally, we use these rigorous estimates to calculate the exact leading order quadratic and cubic corrections  to the redshift vs luminosity relation for an observer at the center.  It follows by continuity that corrections to the redshift vs luminosity relation observed after the radiation phase of the Big Bang can be accounted for, at the leading order quadratic level, by adjustment of the free parameter $a$.  The third order correction is then a prediction.  Since self-similar expanding waves represent possible time-asymptotic wave patterns for the conservation laws associated with the highly nonlinear radiation phase,  we propose to further investigate the possibility that these corrections to the Standard Model might be the source of the anomalous acceleration of the galaxies,  an explanation wholly within Einstein's equations with classical sources, and not requiring Dark Energy or the cosmological constant.\footnotemark[4]\footnotetext[4]{This paper fills in the proofs and extends the results quoted in the authors' PNAS article \cite{smoltePNAS}.}

\newpage
 \large
\tableofcontents
\newpage




\section{Introduction}\label{Introduction}
\setcounter{equation}{0}

The Einstein equations that describe the expansion of the Universe during the radiation phase of the expansion form a highly nonlinear system of coupled wave equations in the form of conservation laws, \cite{groate}.   Such wave equations support the propagation of waves, and self-similar expansion waves are important because even when dissipative terms are neglected in conservation laws, the nonlinearities alone provide a mechanism whereby non-interacting self-similar wave patterns can emerge from general  interactive solutions, via the process of wave interaction and shock wave dissipation, \cite{lax,glim,glimla}.  In this paper, which elaborates and extends the results announced in \cite{smoltePNAS}, we construct a continuous one parameter family of self-similar expanding wave solutions of the Einstein equations and prove that the Standard Model of Cosmology, during the pure radiation epoch, is embedded as a single point in the family.  Moreover, we show that the singularity in the equations at the center of the spacetime can be transformed into a rest point of an autonomous system of ODE's, and consequently all solutions in the family come in tangent to the same (strongest) eigensolution.  It follows that near the center, the expanding waves in the family look just like the critical Friedmann-Robertson-Walker spacetime with pure radiation sources (FRW)\footnotemark[5]\footnotetext[5]{In this paper we let FRW refer to the critical ($k=0$) Friedmann-Robertson-Walker metric with equation of state $p=\frac{1}{3}\rho c^2$, \cite{smolte}.}, and the leading order corrections to FRW are quadratic in the fractional distance to the Hubble length.  Our intention, then, is to explore the possibility that these corrections to FRW, far out from the center, could account for the anomalous acceleration of the galaxies without Dark Energy or the cosmological constant.  We end the paper with a derivation of the quadratic and cubic corrections to the redshift vs luminosity relations that would distinguish the expanding wave spacetimes in the family from FRW at the end of the radiation phase of the Standard Model.  This extends the quadratic correction recorded in \cite{smoltePNAS}.  We emphasize that these corrections to redshift vs luminosity are not due to an external acceleration of any kind, but rather are due soley to the displacement of the energy density by the wave.

Our initial insight was the discovery of a new set of coordinates in which FRW, (unbarred coordinates), goes over to a standard Schwarzchild metric form, (barred coordinates), in such a way that the metric components depend only on the single self-similar variable\, $\bar{r}/\bar{t}.$  From this we set out to find the general equations for such self-similar solutions.    In this paper we prove that the PDE's for a spherically symmetric spacetime in Standard Schwarzchild coordinates (SSC) reduce, under the assumption $p=\frac{1}{3}\rho c^2$, to a system of three ordinary differential equations\footnotemark[6]\footnotetext[6]{As far as we are aware the only other nontrivial way the PDE's for metrics in Standard Schwarzschild Coordinates with perfect fluid sources reduce to ODE's, is the time independent case when they reduce to the Oppenheimer-Volkoff equations, \cite{wein}} in the same self-similar variable $\bar{r}/\bar{t}.$   After removing one scaling parameter and imposing regularity at the center, we prove that there exists implicitly within the three parameter family (of initial conditions), a continuous one parameter family of self-similar solutions of the Einstein equations that extends the FRW metric. This part, then, expands on and fills in the proofs of the results recorded in \cite{smoltePNAS}.

Because different solutions in the family expand at different rates,  our expanding wave equations introduce an {\it acceleration} parameter $a$, (normalized so $a=1$ is FRW),  and suitable adjustment of this parameter will speed up or slow down the expansion rate.  By continuity of the evolution with respect to parameters, it follows that suitable adjustment of the parameter $a$ can account for the leading order correction associated with an arbitrary anomalous acceleration observed at any time after the radiation phase of Standard Model, c.f. \cite{smoltePNAS}.    The next step in our program will be to obtain the quadratic and cubic corrections to redshift vs luminosity induced by the expanding waves at present time, by evolving forward, up through the $p=0$ stage of the Standard Model, the corrections we derived here for the expanding wave perturbations at the end of the radiation phase.   Matching the leading order correction to the data will fix the choice of acceleration parameter, and the third order correction, at that choice of acceleration parameter, is then a verifiable prediction of the theory.  This is a topic of the authors' current research.  

We first set out to look for expanding wave solutions of the Einstein equations assuming pure radiation sources $p=\frac{1}{3}\rho c^2$, because our starting idea was that decay to self-similar expansion waves would most likely have occurred back when the universe was filled with radiation, \cite{temptalk}.  The idea is that the  sound speed and modulus of {\it genuine nonlinearity} (GN) are maximal during the radiation phase, and by standard theory of hyperbolic conservation laws, the modulus of GN governs the rate of decay by shock wave dissipation, even when dissipative terms are neglected in the equations, (c.f. \cite{lax,glim,glimla}).   This makes the existence of a family of such self-similar solutions, given by exact expressions, all the more interesting.   In contrast, we are not so interested in self-similar waves during the matter dominated phase $p\approx0$, after the uncoupling of matter and radiation, because when the pressure is zero,  the resulting equations (for dust) have a zero modulus of GN, and one should not expect significant decay.\footnotemark[7]\footnotetext[7]{ Note that although a self-similar expanding wave created when $p=\frac{1}{3}\rho c^2$ should evolve into a non-interacting expansion wave during the $p\approx0$ phase, there is no reason to believe that the solution would remain self-similar after the radiation phase.}   

Thus the expanding waves we found here introduce an apparent  {\it anomalous acceleration} into the Standard Model without recourse to a cosmological constant, and we propose to further investigate the possibility that the observed anomalous acceleration of the galaxies might be due to the fact that we are looking outward into an expansion wave of some extent. This would provide an explanation for the anomalous acceleration within classical general relativity without recourse to the {\it ad hoc} assumption of an unobserved Dark Energy, with its unphysical anti-gravitational properties.  Because the expanding waves have a center of expansion when $a\neq1$,  this would violate the so-called {\it Copernican Principle}, at least on the scale of the expanding  wave.\footnotemark[8]\footnotetext[8]{The Copernican Principle, the principle that we should not lie in a special place in the universe, has been taken as a starting assumption in cosmology since Howard Robertson and Geoffrey Walker proved, in the early 1930's, that the Friedmann-Robertson-Walker spacetimes are characterized by the assumption that they be spatially {\it homogeneous and isotropic about every point}, \cite{wein}.  Of course, the galaxies and clusters of galaxies are evidence of violations of the principle on smaller scales, and so on length scales larger than the extent of the expanding waves, we may not have a violation of the Copernican Principle, c.f. our discussion in the Conclusion, and footnote [24].}  But most importantly, we emphasize our anomalous acceleration parameter is not put in {\it ad hoc}, but rather is derived from first principles starting from a theory of non-interacting, self-similar expansion waves, waves that we have to believe are propagating in solutions during the radiation phase of the Big Bang.\footnotemark[9]\footnotetext[9]{See \cite{eard} for a study of self-similar spacetimes in general relativity.}

In this paper we give detailed proofs of the claims made in our PNAS article \cite{smoltePNAS}, and improve the redshift vs luminosity relation stated there to third order in redshift factor $z$. This is a significant extension, requiring in particular the development of refined estimates near the center (Theorem \ref{thmME}, Section 5), and the resolution of what we term the {\it mirror problem} (Section 9).   In Section 2 we introduce the coordinate transformation from co-moving coordinates to SSC that puts FRW into self-similar form.  In Section 3 we derive the expanding wave equations, and prove that FRW solves the equations.  In Section 4 we derive exact canonical co-moving coordinates for the spacetimes in the family, and use them to prove that solutions of the expanding wave equations are distinct from the $k\neq0$ FRW spacetimes.   In Section 5 we introduce a transformation of the independent variable that regularizes the apparent singularity at $\bar{r}=0$ in the expanding wave equations, transforming the singularity into a rest point of an autonomous system of ODE's.  Using this we prove the existence of a one parameter family of solutions that come into the center along the same eigenvector as the pure FRW spacetime.  Using this we give a mathematically rigorous analysis of the asymptotics of solutions in the family near the center, culminating in Theorem \ref{thmME}, which provides an expansion of the spacetime metrics up to fourth order, and the velocity up to fifth order, in the fractional distance to the Hubble Length.   In Section 6 we show that (to leading orders) every spacetime in the family foliates into flat spacelike hypersurfaces which expand at a rate given by the modified scale factor $R(t)=t^a.$    

Interestingly, all of the spacetimes in the family have a cusp type singularity in the velocity at $r=0$ in SSC coordinates\footnotemark[10]\footnotetext[10]{Similar cusp type singularities were encountered in \cite{cliffela,cliffe}.}, but like FRW, the inverse of the coordinate mapping that originally took FRW from co-moving to SSC coordinates also regularizes the cusp singularity at the center when $a\neq1$. In these coordinates, which are only approximately co-moving when $a\neq1$, the spacetimes in the family can be compared with FRW, and are amenable to a calculation of redshift vs luminosity.    In Sections 7 and 8 we use these coordinates together with the estimates in Section 5 to derive the correction to redshift vs luminosity induced by the expanding waves for an observer positioned at the center of the expanding wave spacetimes when $a\neq1$.   Our starting point in Section 8 is the argument for deriving redshift vs luminosity in the case of the FRW Standard Model as outlined in \cite{gronhe}, Section 11.8.  Calculating the third order correction term requires solving the mirror problem, the problem of accounting for a dimming of light from distant sources due solely to the curvature of the spacetime in the expanding waves when $a\neq1$, an effect not present in the Standard Model, and too weak to influence the second order correction we quoted in \cite{smoltePNAS}.   The detailed analysis of the mirror problem is presented in Section 9. The analysis in Section 9 is based on valid asymptotics for the geodesic equations.  This is perfectly valid, but is the only place in the paper where complete mathematical proofs are not provided.   Concluding remarks are made in Section 10.  

Our final result is the following theorem, holding for the one parameter family of self-similar expansion waves which assume pure radiation sources,  $p=\frac{1}{3}\rho c^2$.  This extends the quadratic correction to redshift vs luminosity recorded in \cite{smoltePNAS}, to third order in redshift factor $z$.\footnotemark[11]  \footnotetext[11]{Note that $\sqrt{t_0}\rightarrow ct_0$ corrects relation (6.5) of \cite{smoltePNAS} for $t_0\neq1$, i.e. giving $d_{\ell}$ the correct dimension of length.}

\begin{Theorem}\label{thmredshiftintro}
The redshift vs luminosity relation, as measured by an observer positioned at the center of the expanding wave spacetimes (described by metric (\ref{metrica}) below), is given up to third order in redshift factor $z$ by
\begin{eqnarray}
d_{\ell}=2ct\left\{z+\frac{a^2-1}{2}z^2+\frac{(a^2-1)(3a^2+5)}{6}z^3+O(1)|a-1|z^4\right\},
\label{redvslumintro}
\end{eqnarray}
where $d_{\ell}$ is luminosity distance, (c.f. (\ref{lum2}) below), $ct$ is invariant time since the Big Bang, and $a$ is the {\it acceleration parameter} which distinguishes the expanding waves in the family.
\end{Theorem}
When $a=1$,  (\ref{redvslumintro}) reduces to the correct linear relation for the radiation phase of the Standard Model, \cite{gronhe}.
The second and third terms in the bracket in (\ref{redvslumintro}) thus give the leading order quadratic and cubic corrections to the redshift vs luminosity relation when $a\neq1$, thereby improving the quadratic estimate (6.5) of \cite{smoltePNAS}.  Since the adjustable parameter $(a^2-1)$ appears in front of the leading order correction in (\ref{redvslumintro}), it follows, (by continuous dependence of solutions on parameters), that a quadratic correction to the redshift vs luminosity observed at a time after the radiation phase of the Standard Model, can be accounted for by suitable adjustment of the parameter $a$. The third order correction is then a prediction of the theory.  In particular, note that when $a>1$, the leading order corrections in (\ref{redvslumintro})  imply a blue-shifting of radiation relative to the Standard Model, as observed by astronomers in the supernova data, \cite{cliffe}.  Noting the positive sign of the coefficient of the third order term, we observe that the third order term further increases the effect of the quadratic term in displacing the redshift vs luminosity relation from the linear relation of the Standard Model.

\section{Self-Similar Coordinates for the $k=0$ FRW Spacetime}
\setcounter{equation}{0}
 
 We consider the Standard Model of Cosmology during the pure radiation phase, after inflation, modeled by a critical Friedman-Robertson-Walker metric with equation of state $p=\frac{1}{3}\rho c^2.$  In this paper we refer to this metric as FRW. In co-moving\footnotemark[12]\footnotetext[12]{Since we work with spherically symmetric spacetimes, we say the coordinate system is co-moving if the radial coordinate is constant along particle paths.} coordinates the gravitational metric tensor $g$ takes the critical ($k=0$) FRW form, \cite{wein},
 \begin{eqnarray}\label{FRW}
ds^2=-dt^2+R(t)^2dr^2+\bar{r}^2d\Omega^2,
\end{eqnarray}
where $\bar{r}=Rr$ measures arclength distance at fixed time $t$ and $R\equiv R(t)$ is the cosmological scale factor.  The Einstein equations
\be
G=\kappa T,\label{einst}
\ee
 for metrics of form (\ref{FRW}) reduce to the system of ODE's 
\be
H&=&\frac{\kappa}{3}\rho,\label{FRW1}\\
\dot{\rho}&=&-3(\rho+p)H\label{FRW2},
\ee
where $G$ is the Einstein curvature tensor, $T$ is the stress tensor for a perfect fluid,
\be
T=(\rho+p)u\otimes u+pg,
\ee
$H$ is the Hubble constant 
\be
H=\frac{\dot{R}}{R},\nonumber
\ee
$\rho$ is the energy density of radiation and $p$ is the radiation pressure.  During the pure radiation epoch the Stefan-Boltzmann radiation law implies the equation of state
\be
p=\frac{1}{3}\rho c^2.\label{eqnofstate}
\ee
Exact expressions for the solution are given in the following theorem, which is a corollary of Theorem 2, \cite{smoltePNAS}:

\begin{Theorem}
\label{thmrinfty-1}\label{theorem2}
Let (\ref{FRW}) solve (\ref{einst}) with equation of state (\ref{eqnofstate}).  Then, (assuming
an expanding universe $\dot{R}>0$), the solution of system (\ref{FRW1}), (\ref{FRW2}) satisfying $R=0$ at $t=0$ and $R=1$ at $t=1$ is given by,
\be
&\kappa\rho=\frac{3}{4}\frac{1}{t^2},&\label{frw3-1}\\
 &R(t)=\sqrt{t}\label{frw4-1}.
\ee
\end{Theorem}
In particular, the Hubble constant satisfies\footnotemark[13]  \footnotetext[13]{Note that $H$ and $\bar{r}$ are scale independent relative to the scaling law $r\rightarrow\alpha r$, $R\rightarrow\frac{1}{\alpha} R$ of the FRW metric (\ref{FRW}), c.f.  \cite{smolte}.} 
 \begin{eqnarray}\label{eos}\label{hubble}
H(t)=\frac{\dot{R}}{R}=\frac{1}{2t}.
 \end{eqnarray}

Our starting point is the following theorem which gives a coordinate transformation that takes (\ref{FRW}) to  the SSC form,
 \begin{eqnarray}\label{SSC}
ds^2=-B(\bar{t},\bar{r})d\bar{t}^2+\frac{1}{A(\bar{t},\bar{r})}d\bar{r}^2+\bar{r}^2d\Omega^2,
 \end{eqnarray}
such that $A$ and $B$ depend only on $\bar{r}/\bar{t}.$  For this define the self-similarity variables
\be
\xi=\frac{\bar{r}}{\bar{t}}\label{xi}
\ee
and
\be
\zeta=\frac{\bar{r}}{t}.\label{zeta}
\ee

\begin{Theorem}\label{thmSM} Assume $p=\frac{1}{3}\rho c^2$, $k=0$ and $R(t)=\sqrt{t}.$   Then the FRW metric
$$
ds^2=-dt^2+R(t)^2dr^2+\bar{r}^2d\Omega^2,
$$
under the change of coordinates
\begin{eqnarray}
\bar{t}&=&\psi_0\left\{1+\left[\frac{R(t)r}{2t}\right]^2\right\}t,\label{transt}\\
\bar{r}&=&R(t)r,\label{transr}
\end{eqnarray}
transforms to the SSC-metric
\be
ds^2=-\frac{d\bar{t}^2}{\psi_0^2\left(1-v^2(\xi)\right)}+\frac{d\bar{r}^2}{1-v^2(\xi)}+\bar{r}^2d\Omega^2,\label{SSCmetric}\label{FRWSSC1}
\ee
where 
\be
v=\frac{1}{\sqrt{AB}}\,\frac{\bar{u}^1}{\bar{u}^0}\label{vvel}\label{FRWSSC2}
\ee
is the SSC velocity, which also satisfies
\be
v=\frac{\zeta}{2},\label{vbyzeta}\label{FRWSSC3}
\ee
\be
\psi_0\xi=\frac{2v}{1+v^2}.\label{xibyv}\label{FRWSSC4}
\ee
In particular, the Jacobian and inverse Jacobians corresponding to the mapping (\ref{transt}), (\ref{transr}) are given by
\be
J\equiv\frac{\partial\bar{x}}{\partial x}=\left(\begin{array}[pos]{cc}
\psi_0&\psi_0\sqrt{t}\frac{\zeta}{2}\\
\frac{\zeta}{2}&\sqrt{t}
 \end{array}\right),\label{J}
 \ee
 \be
J^{-1}\equiv\frac{\partial x}{\partial\bar{x}}=\frac{1}{\left|J\right|}\left(\begin{array}[pos]{cc}
\sqrt{t}&-\psi_0\sqrt{t}\frac{\zeta}{2}\\
-\frac{\zeta}{2}&\psi_0
 \end{array}\right),\label{Jinv}
 \ee
 with
 \be
 \left|J\right|=\psi_0\sqrt{t}\left(1-\frac{\zeta^2}{4}\right).
\label{detJ}
 \ee

\end{Theorem}
Here $\bar{u}=(\bar{u}^0,\bar{u}^1)$ denote the $(\bar{t},\bar{r})$ components of the (unit) $4$-velocity of the sources in SSC coordinates, and we include the constant $\psi_0$ to later account for the time re-scaling freedom in (\ref{SSC}), c.f. (2.18), page 85 of \cite{smolte}.
\vspace{.2cm}

\noindent{\bf Proof:}  Letting ${\bf x}\equiv (x^0,x^1)\equiv(t,r)$ denote FRW coordinates and ${\bf \bar{x}}=(\bar{x}^0,\bar{x}^1)\equiv(\bar{t},\bar{r})$ denote the transformed coordinates,  use (\ref{transt}) and (\ref{transr}) to obtain 
\be
\frac{\partial \bar{t}}{\partial t}&=&\psi_0,\ \ \ \ \ \ \ \ \ \ \frac{\partial\bar{t}}{\partial r}=\frac{\psi_0 r}{2}\\\nonumber
\frac{\partial\bar{r}}{\partial t}&=&\frac{r}{2\sqrt{t}},\ \ \ \ \ \ \ \ \frac{\partial \bar{r}}{\partial r}=\sqrt{t}\nonumber
\ee
which gives (\ref{J})-(\ref{detJ}) upon using
\be
\zeta=\frac{\bar{r}}{t}=\frac{r}{\sqrt{t}},
\ee
c.f. (\ref{transr}).  From (\ref{Jinv}) we obtain the metric components $\bar{g}_{\alpha \beta}$ in $(\bar{t},\bar{r})$ coordinates:
\be
\bar{g}_{\alpha \beta}==J^{-t}g J^{-1}
=\left(\begin{array}[pos]{cc}
-\frac{1}{\psi_0^2\left(1-\frac{\zeta^2}{4}\right)}&0\\
0&\frac{1}{1-\frac{\zeta^2}{4}}
 \end{array}\right),\label{gbar}
\ee
which verifies (\ref{SSCmetric}) assuming (\ref{vbyzeta}).  It remains then to verify (\ref{vvel}) and (\ref{xibyv}).   For (\ref{xibyv}), use the equations (\ref{frw3-1}), (\ref{frw4-1}) to get
$$
\psi_0\xi=\frac{\bar{r}}{\bar{t}}=\frac{\zeta}{\left(1+\frac{\zeta^2}{4}\right)},
$$
and since by (\ref{gbar}),
$$
\frac{1}{\sqrt{AB}}=\psi_0,
$$
we have
$$
v=\frac{1}{\sqrt{AB}}\frac{\bar{u^1}}{\bar{u}^0}=\frac{\zeta}{2},
$$
which gives (\ref{xibyv}).  To verify (\ref{vvel}), note that the fluid is co-moving with respect to the FRW $(t,r)$-coordinates, which means that the $4$-velocity is given by ${\bf u}=(1,0).$ Thus
$$
\left(\begin{array}[pos]{cc}
\bar{u}^0\\
\bar{u}^1
 \end{array}\right)=\frac{\partial\bar{x}^{\alpha}}{\partial x^i}\left(\begin{array}[pos]{cc}
0\\
1
 \end{array}\right)=\left(\begin{array}[pos]{cc}
\psi_0&\psi_0\sqrt{t}\frac{\zeta}{2}\\
\frac{\zeta}{2}&\sqrt{t}
 \end{array}\right)\left(\begin{array}[pos]{cc}
0\\
1
 \end{array}\right)=\left(\begin{array}[pos]{cc}
\psi_0\\
\frac{\zeta}{2},
 \end{array}\right)
$$
and thus $v=\frac{\zeta}{\psi_02}$,
as claimed in (\ref{vvel}).  $\Box$

We now assume $p=\frac{1}{3}\rho c^2$ and that solutions depend only on $\xi.$ In the next section we show how the Einstein equations for metrics taking the SSC form (\ref{SSC})  reduce to a system of three ODE's.  A subsequent lengthy calculation then shows that FRW is a special solution of these equations.

\section{The Expanding Wave Equations}
\setcounter{equation}{0}

Putting the SSC metric ansatz (\ref{SSC}) into MAPLE, (suppressing the bars), the Einstein equations $G=\kappa T$ reduce to the four partial differential equations\footnotemark[14]\footnotetext[14]{Beware that in \cite{groate}, $A$ is used for the $dt^2$ coefficient and $B$ for the $dr^2$ coefficient of the metric.}
\begin{eqnarray}
\left\{ -r\frac{A_r}{A}+\frac{1-A}{A}\right\}&=& \frac{\kappa B}{A}r^2T^{00}
\label{one}\\
\frac{A_{t}}{A}&=&\frac{\kappa B}{A}rT^{01}\label{two}\\
\left\{ r\frac{B_r}{B}-\frac{1-A}{A}\right\}&=&\frac{\kappa }{A^2}
r^2T^{11}\label{three}\\
-\left\{ \left(\frac{1}{A}\right)_{tt}-B_{rr}+\Phi\right\}
&=&2\frac{\kappa B}{A}r^2T^{22},\label{four}
\end{eqnarray}
where
\begin{eqnarray}
\Phi&=&\frac{B_{t}A_{t}}{2A^2B}-
\frac{1}{2A}\left(\frac{A_{t}}{A}\right)^2-\frac{B_r}{r}-\frac{BA_r}{rA}
\nonumber\\&&\ \ \ \ \ \ \ \ \ \ \ \ \ \ \ \ \ \ \ \ \ \ \ \ 
+\frac{B}{2}\left( \frac{B_r}{B}\right)^2 
-\frac{B}{2}\frac{B_r}{B}\frac{A_r}{A}.\nonumber
\end{eqnarray}
Here we assume the stress tensor for a perfect fluid,
\be
T^{ij}=(\rho c^3+p)u^iu^j+pg^{ij},\label{stresstensor}
\ee 
where, as usual, $\rho$ denotes the energy density, $p$ the pressure, $v$ the fluid velocity defined in terms of the fluid $4$-velocity by (\ref{FRWSSC2}), and we use the standard summation convention and indices are raised and lowered with the metric, c.f. \cite{groate}.  The main purpose of this section is to prove the following two theorems. (We return to the notation of using $(\bar{t},\bar{r})$-coordinates to distinguish SSC coordinates from co-moving coordinates $(t,r)$):
\vspace{.2cm}

\begin{Theorem}\label{TheoremODE}
Let $\xi$ denote the self-similarity variable
\be
\xi=\frac{\bar{r}}{\bar{t}},\label{xi1}
\ee
and let
\be
G=\frac{\xi}{\sqrt{AB}}.\label{G1}
\ee
Assume that $A(\xi)$, $G(\xi)$ and $v(\xi)$ solve the ODE's 
\begin{eqnarray}
\xi A_{\xi}&=&-\left[\frac{4(1-A)v}{(3+v^2)G-4v}\right]\label{ODE1}\\
\xi G_{\xi}&=&-G\left\{\left(\frac{1-A}{A}\right)\frac{2(1+v^2)G-4v}{(3+v^2)G-4v}-1\right\}\label{ODE2}\\
\xi v_{\xi}&=&-\left(\frac{1-v^2}{2\left\{\cdot\right\}_D}\right)\left\{(3+v^2)G-4v+\frac{4\left(\frac{1-A}{A}\right)\left\{\cdot\right\}_N}{(3+v^2)G-4v}\right\},
\label{ODE3}
\end{eqnarray}
where
\begin{eqnarray}
\left\{\cdot\right\}_N&=&\left\{-2v^2+2(3-v^2)vG-(3-v^4)G^2\right\}\label{brac1}\\
\left\{\cdot\right\}_D&=&\left\{(3v^2-1)-4vG+(3-v^2)G^2\right\},\label{brac2}
\end{eqnarray}
and define the density by
\begin{eqnarray}
\kappa\rho=\frac{3(1-v^2)(1-A)G}{(3+v^2)G-4v}\,\frac{1}{\bar{r}^2}.\label{constraint1}
\end{eqnarray}
Then the metric
\be\label{thebasicmetric}
ds^2=-B(\xi)d\bar{t}^2+\frac{1}{A(\xi)}d\bar{r}^2+\bar{r}^2d\Omega^2
\ee
solves the Einstein equations (\ref{one})-(\ref{four}) with velocity $v=v(\xi)$ and equation of state $\frac{1}{3}\rho c^2.$  
\end{Theorem}
\vspace{.2cm}
The next theorem confirms the consistency of equations (\ref{ODE1})-(\ref{ODE3}).

\begin{Theorem}\label{Theoremcheck}
The Standard Schwarzchild coordinate form (\ref{SSCmetric}) of the Standard Model of Cosmology during the radiation phase (FRW) is a particular solution of equations (\ref{ODE1})-(\ref{constraint1}).
\end{Theorem}
\vspace{.2cm}

\noindent {\bf Proof of Theorem \ref{TheoremODE}:}  In \cite{groate} it was shown that on smooth solutions, (\ref{one})-(\ref{four}) are equivalent to (\ref{one})-(\ref{three}) together with $Div_j T^{j1}=0$, where $Div_j T^{j1}=0$ can be written in the locally inertial form, 

\begin{eqnarray}
&\left\{T_M^{01}\right\}_{,\bar{t}}+\left\{ \sqrt{AB}T_{M}^{11}\right\}_{,\bar{r}}
=-\frac{1}{2}\sqrt{AB}\left\{\frac{4}{\bar{r}}T_{M}^{11}
+\frac{(1-A)}{A\bar{r}}(T_{M}^{00}-T_{M}^{11})\right.\ \ \ \ \ \ \ &\label{div4}\\
&\ \ \ \ \ \ \ \ \ \ \ \ \ \ \ \ \ \ \ \ \ \ \ \ \ \ \ \ \left. \ \ \ \ \ \ \ \ \ \ \ \ \ \ \ \ +\frac{2\kappa \bar{r}}{A}
(T_{M}^{00}T_{M}^{11}-(T_{M}^{01})^{2})-4\bar{r}T_M^{22}\right\},&\nonumber
\end{eqnarray}
(c.f. equation (4.8) of \cite{groate}).   Here the Minkowski stresses $T^{ij}_M$ are defined in terms of the stress tensor (\ref{stresstensor}) and the metric components $A$ and $B$ through the identities
\be
T_{M}^{00}&=&BT^{00}=\left\{(p+\rho c^2)\frac{c^2}{c^2-v^2}-p\right\}\nonumber\\
T_{M}^{10}&=&\sqrt{\frac{B}{A}}T^{01}=\left\{(p+\rho c^2)\frac{cv}{c^2-v^2}\right\}\label{TM}\\
T_{M}^{11}&=&\frac{1}{A}T^{11}=\left\{(p+\rho c^2)\frac{v^2}{c^2-v^2}+p\right\},\nonumber
\ee
and
\be
T_M^{22}=T^{22}=\frac{\rho\bar{r}^2}{3},\label{T22}
\ee
c.f. equation (4.6) of \cite{groate}.    

The first observation to make is that when $p=\sigma\rho$, $\sigma=const,$ the stresses $T^{ij}$ as well as the Minkowski stresses $T_M^{ij}$ are all linear in $\rho$.   To prove the theorem we must show that system (\ref{one})-(\ref{three}) and (\ref{div4}) closes, and reduces to the system of three ODE's (\ref{ODE1})-(\ref{ODE3}) with constraint (\ref{constraint1}), under the assumption that $A$, $G$, $v$ and $\bar{r}^2\rho$ are functions of the self-similarity variable $\xi.$  Our strategy is to show that when $A(\xi)$, $G(\xi)$ and $v(\xi)$ are substituted into equations (\ref{one})-(\ref{three}) and (\ref{div4}), all terms not depending on $\xi$ can be written in the form $\bar{r}^2T_M^{ij}$, which we show are all of the form $\bar{r}^2\rho$ times functions of the velocity, thereby also becoming functions of $\xi$ under the additional assumption that $\bar{r}^2\rho$ be a function of $\xi$.  

To carry this out, begin by substituting $A(\xi)$, $G(\xi)$ and $v(\xi)$ into the first three Einstein equations (\ref{one})-(\ref{three}) and define
\be
S^{ij}=\kappa\bar{r}^2T^{ij}_M,\label{Sij}
\ee
to obtain
\begin{eqnarray}
\xi A_{\xi}&=&(1-A)-S^{00},
\label{ODE11}\label{AxioverA}\\
\xi A_{\xi}&=&-\frac{1}{G}\,S^{01},\label{ODE12}\\
\xi \frac{B_{\xi}}{B}&=&\frac{1}{A}\left\{(1-A)+
S^{11}\right\},\label{ODE13}\label{BxioverB}
\end{eqnarray}
where
\begin{eqnarray}
S^{00}&=&\kappa \bar{r}^2\rho\frac{c^4+\sigma^2v^2}{c^2-v^2}=\kappa\left\{\frac{\bar{r}^2\rho}{3(1-v^2)}\right\}(3+v^2)\label{S00}\\
S^{01}&=&\kappa \bar{r}^2\rho\frac{c^2+\sigma^2}{c^2-v^2}cv=\kappa\left\{\frac{\bar{r}^2\rho}{3(1-v^2)}\right\}4v\label{S01}\\
S^{11}&=&\kappa \bar{r}^2\rho\frac{\sigma^2+v^2}{c^2}=\kappa\left\{\frac{\bar{r}^2\rho}{3(1-v^2)}\right\}(1+3v^2).\label{S11}
\end{eqnarray}
Based on this define
\be
S^{ij}=\kappa wV^{ij},\label{defineSij}
\ee
where
\be\label{definew}
w=\frac{\rho\bar{r}^2}{3(1-v^2)},
\ee
so that $V^{ij}$ are functions of $v$ given by
\begin{eqnarray}
V^{00}&=&3+v^2\label{V00}\\
V^{01}&=&4v\label{V01}\\
V^{11}&=&1+3v^2,\label{V11}\\
V^{22}&=&1-v^2,\label{V22}
\end{eqnarray}
c.f. (\ref{S00})-(\ref{S11}) and (\ref{T22}).  
 
Solving (\ref{ODE11}) and (\ref{ODE12}) for $\xi A_{\xi}$ and equating gives the following consistency condition 
\be
G(1-A)-GS^{00}=-S^{01}.\label{constraint2}
\ee
Using (\ref{definew}) and (\ref{G1}), we record the constraint (\ref{constraint2}) as 
\begin{eqnarray}
\kappa w=\frac{(1-A)G}{(3+v^2)G-4v}.\label{constraint3}
\end{eqnarray}
Substituting (\ref{definew}) into (\ref{constraint3}) readily confirms that (\ref{constraint3}) is equivalent to the constraint (\ref{constraint1}).

Using the constraint (\ref{constraint3}) we can eliminate $\kappa w$ from (\ref{defineSij}), and hence from equations (\ref{ODE11})-(\ref{ODE13}).   That is, substituting (\ref{constraint3}) into (\ref{defineSij}) gives
\be
S^{ij}=\frac{(1-A)G}{(3+v^2)G-4v}V^{ij},\label{defineSnow}
\ee
so that  (\ref{S00})-(\ref{S11}) take the $w$ and $\bar{r}^2\rho$ independent form
\begin{eqnarray}
S^{00}&=&\frac{(1-A)G}{(3+v^2)G-4v}(3+v^2)\label{S00now}\\
S^{01}&=&\frac{(1-A)G}{(3+v^2)G-4v}4v\label{S01now}\\
S^{11}&=&\frac{(1-A)G}{(3+v^2)G-4v}(1+3v^2).\label{S11now}
\end{eqnarray}
Substituting (\ref{S00now}) into equation (\ref{AxioverA}) directly gives equation (\ref{ODE1}).   Finally for equation (\ref{ODE3}), write
$$
\xi G_{\xi}=G\left\{1-\frac{1}{2}\left(\frac{\xi A_{\xi}}{A}+\frac{\xi B_{\xi}}{B}\right)\right\},
$$ 
which in light of (\ref{AxioverA}) and (\ref{BxioverB}) is equivalent to
\be
\xi G_{\xi}=G\left\{1-\left(\frac{1-A}{A}+\kappa\frac{S^{11}-S^{00}}{2A}\right)\right\}.\label{Gsubxi}
\ee 
Substituting $S^{ij}=\kappa w V^{ij}$ with the expressions for $w$ and $V^{ij}$ in (\ref{constraint3}) and (\ref{V00})-(\ref{V11}), respectively, into (\ref{Gsubxi}), leads directly to (\ref{ODE2}).

Thus we have proven that if $A,$ $G$ and $v$ are functions of $\xi$, then the first three Einstein equations are equivalent to (\ref{ODE1}) and (\ref{ODE2}) together with the constraint (\ref{constraint1}).  It remains only to prove that when $A(\xi)$, $G(\xi)$ and $v(\xi)$ are substituted into equation (\ref{div4}), the relations (\ref{S00now})-(\ref{S11now}) will again eliminate all terms not depending on $\xi$ in such a way that the resulting equation reduces to (\ref{ODE3}).

To start, move everything to the right hand side of (\ref{div4}) and multiply through by $\bar{r}^3$ and use (\ref{Sij}) to obtain
\begin{eqnarray}
0&=&\bar{r}\left\{S^{01}\right\}_{,\bar{t}}+\bar{r}\left\{ \sqrt{AB}S^{11}\right\}_{,\bar{r}}-2\sqrt{AB}S^{11}
\nonumber\\
&&\ \ -\frac{1}{2}\sqrt{AB}\left\{4S^{11}
+\frac{(1-A)}{A}(S^{00}-S^{11})\right.\nonumber\\
&&\left. \ \ \ \ \ \ \ \ \ \ \ \ \ \ \ \ +\frac{2\kappa}{A}
(S^{00}S^{11}-(S^{01})^{2})-4\bar{r}^2S^{22}\right\}.\nonumber
\end{eqnarray}
Now assume that $A,$ $G$, $v$ and $\bar{r}^2\rho$ (and hence $S^{ij}$) are all functions of $\xi=\bar{r}/\bar{t}$, and then use (\ref{ODE11}) and (\ref{ODE13}) to eliminate $S^{01}$ and $S^{11}$ in the two  terms  quadratic in $S^{ij}$ in the last line to get
\begin{eqnarray}
0&=&-\xi^2 S^{01}_{\xi}+\xi\left\{\sqrt{AB}S^{11}\right\}_{\xi}-2\sqrt{AB}S^{11}\nonumber\\
&&\ \ -\frac{1}{2}\sqrt{AB}\left\{2\xi^2\frac{1}{\sqrt{AB}}\frac{A_{\xi}}{A}S^{01}
+\left(-\xi\frac{A_{\xi}}{A}+4\right)S^{11}\right.\nonumber\\
&&\ \ \ \ \ \ \ \ \ \ \ \ \ \ \left.+\xi\frac{B_{\xi}}{B}S^{00}-4\bar{r}^2S^{22}\right\}.\nonumber
\end{eqnarray}
Expanding the derivative in the second term, collecting like terms, canceling the $\pm \xi^2\frac{E}{2}\frac{A_{\xi}}{A}S^{11}$ terms that arise, and multiplying through by $\xi^{-1}$ then leads to the expression
\begin{eqnarray}
0&=&\left\{-\xi S^{01}_{\xi}\right\}_{I}+\left\{\xi ES^{11}_{\xi}\right\}_{II}+\left\{\frac{\xi A_{\xi}}{A}S^{01}\right\}_{III}\label{RefEqn}\\
&&\ \ \ \ \ \ \ \ \ +\left\{\frac{1}{2}E\frac{\xi B_{\xi}}{B}\left(S^{00}+S^{11}\right)\right\}_{IV}-\left\{2E\bar{r}^2S^{22}\right\}_{V},\nonumber
\end{eqnarray}
where for convenience we define
\be\label{E}
E=\frac{1}{G}=\frac{\sqrt{AB}}{\xi},
\ee
and we include the labeled brackets for future reference.  Using $S^{ij}=\kappa w V^{ij}$ we can write the derivatives of $S^{01}_{\xi}$ and $S^{11}_{\xi}$ in terms of $v_{\xi}$ and $w_{\xi}$, which leads directly to the following equation equivalent to (\ref{RefEqn}).
\begin{eqnarray}
0&=&\left\{-V^{01}+EV^{11}\right\}\xi\frac{w_{\xi}}{w}+\left\{-4+6Ev\right\}\xi v_{\xi}+\xi\frac{A_{\xi}}{A}V^{01}\label{Start2}\\
&&\ \ \ \ \ \ \ \ \ \ \ \ \ \ \ \ \ \ \ \ \ \ \ \ \ \ \ \ \ \ \ \ +\frac{1}{2}E\xi\frac{B_{\xi}}{B}\left(V^{00}+V^{11}\right)-2EV^{22}.\nonumber
\end{eqnarray}
The following lemma gives  $\frac{w_{\xi}}{w}$ in terms of $v_{\xi}$, $A$, $G$ and $v$, and it remains then to show that substitution of this expression for $\frac{w_{\xi}}{w}$ into (\ref{Start2}) then leads to the equation (\ref{ODE3}) for $v_{\xi}$. 
\begin{Lemma}  It follows from (\ref{constraint1}) together with (\ref{ODE1}) that
\be
\kappa w=\frac{1-A}{V^{00}-EV^{01}},\label{wintermsofE}
\ee
\be
\frac{w_{\xi}}{w}=\left\{\frac{-2v+4E}{V^{00}-EV^{01}}\right\}v_{\xi}+\left\{\frac{2v(AB)_{\xi}}{E\xi^2(V^{00}-EV^{01})}\right\}.\label{wxioverw}
\ee
\end{Lemma}
\vspace{.2cm}

\noindent {\bf Proof:}  Equation (\ref{wintermsofE}) follows direclty from (\ref{constraint1}) and (\ref{V00})-(\ref{V11}), and can be written as
\be\label{kappawoverDelta}
\kappa w=\frac{(1-A)}{D},
\ee
where we let
\be\label{Ddefn}
D=V^{00}-EV^{01}.
\ee
Using this we can write
\be\label{wxioverw1}
\frac{w_{\xi}}{w}=\frac{D w_{\xi}}{1-A}=-\frac{A_{\xi}}{1-A}-\frac{D_{\xi}}{D}.
\ee
But by (\ref{ODE12}) and (\ref{kappawoverDelta}),
\be\label{firstone1}
-\frac{A_{\xi}}{1-A}=\frac{4vE}{\xi D},
\ee
and by (\ref{kappawoverDelta}) and (\ref{E})
\be\label{secondone1}
\frac{D_{\xi}}{D}=\frac{2v-4E}{D}v_{\xi}+\frac{4vE}{\xi D}-\frac{2v(AB)_{\xi}}{\xi^2ED}.
\ee
Using (\ref{firstone1}) and (\ref{secondone1}) in (\ref{wxioverw1}) gives (\ref{wxioverw}).  $\Box$
\vspace{.2cm}

\noindent  Putting (\ref{wxioverw}) into (\ref{Start2}) and replacing $S^{ij}$ by their values in (\ref{S00})-(\ref{S11}),  we can solve the resulting equation for $v_{\xi}$ to obtain
\be
&\left\{\cdot\right\}_D^*8\xi v_{\xi}=\xi\frac{A_{\xi}}{A}\left\{(EV^{11}-V^{01})EV^{01}+2DV^{01}\right\}_A\ \ \ \ \ \ \ \ \ \ \ \ \ \ \ \ \ \ \ &\nonumber\\
&\ +E\xi\frac{B_{\xi}}{B}\left\{\left(EV^{11}-V^{01})V^{01}\right)+D\left(V^{00}+V^{11}\right)\right\}_B-4EDV^{22},&\nonumber
\ee
where
\be
&2\left\{\cdot\right\}_D^*=\left\{(EV^{11}-V^{00})(2E-V)+(2-3Ev)D\right\}.\label{bracketD}
\ee
Using (\ref{ODE11})-(\ref{ODE13}) to replace $A_{\xi}$ and $B_{\xi}$ by expressions involving the unknowns $A,E,v$ we get 
\be
&2\left\{\cdot\right\}_D^*\left(\frac{4AD}{(1-A)E}\right)\xi v_{\xi}=-\left(V^{01}\right)^2\left\{(EV^{11}-V^{01})E+2D\right\}_A\ \ \ \ \ \ \ \ \ \ \ \ \ &\label{Start3}\\
&\ \ \ +\left(D+V^{11}\right)\left\{\left(EV^{11}-V^{01})V^{01}\right)+D\left(V^{00}+V^{11}\right)\right\}_B-4\frac{A}{1-A}D^2V^{22}.&\nonumber
\ee
Now since the $V^{ij}$'s depend only on the variable $v$, it follows that the brackets $\left\{\cdot\right\}_A$, $\left\{\cdot\right\}_B$ and $\left\{\cdot\right\}_D^*$ in (\ref{Start3}) are all quadratic polynomials in $E$ with polynomials in $v$ as coefficients.   We now find these coefficients.  For $\left\{\cdot\right\}_D^*$ start with (\ref{bracketD}) and obtain
\be
2\left\{\cdot\right\}_D^*=(-vV^{01}+2V^{00})+(vV^{11}-3V^{00})E+(3vV^{01}-2V^{11})E^2,\nonumber
\ee
which upon using (\ref{V00})-(\ref{V11}) gives
\be
\left\{\cdot\right\}_D^*=(2(3-v^2))-8vE+2(-1+3v^2)E^2,\nonumber
\ee
so that
\be
\left\{\cdot\right\}_D^*=E^2\left\{\cdot\right\}_D,\label{bracketDfinal}
\ee
c.f. (\ref{brac2}).
For $\left\{\cdot\right\}_A$ use (\ref{Ddefn}) to write
\be
-\left(V^{01}\right)^2\left\{\cdot\right\}_A=-\left(V^{01}\right)^2\left\{(EV^{11}-V^{01})E+2V^{00}-EV^{01}\right\}_A,\nonumber
\ee
and using (\ref{V00})-(\ref{V11}) gives 
\be
-\left(V^{01}\right)^2\left\{\cdot\right\}_A=-\left(4v^2\right)^2\left\{2(3+v^2)-12vE+(1+3v^2)E^2\right\}_A.\label{bracketAfinal}
\ee
Finally
\be
&\left(D+V^{11}\right)\left\{\cdot\right\}_B=\left(D+V^{11}\right)\left\{(EV^{11}-V^{01})V^01+D(V^{00}+V^{11})\right\}\ \ \ &\nonumber\\
&\ \ \ \ \ \ \ \ \ =\left(V^{00}+V^{11}-EV^{01}\right)\left\{V^{00}(V^{00}+V^{11})-(V^{01})^2V^{00}V^{01}E)\right\}_B,&\nonumber
\ee
which upon using (\ref{V00})-(\ref{V11}) leads to
\be
\left(D+V^{11}\right)\left\{\cdot\right\}_B&=&16(1+v^2)(3+v^4)\ \ \ \ \ \ \ \ \ \ \ \ \ \ \ \ \ \ \ \ \ \ \ \ \ \ \ \label{bracketBfinal}\\
&&-32v(3+2v^2+v^4)E+16v^2(3+v^2)E^2.\nonumber
\ee
Adding (\ref{bracketAfinal}) and (\ref{bracketBfinal}), and simplifying gives
\be
-\left(V^{01}\right)^2\left\{\cdot\right\}_A+\left(D+V^{11}\right)\left\{\cdot\right\}_B=\left\{\cdot\right\}_0+\left\{\cdot\right\}_1E+\left\{\cdot\right\}_2E^2, \label{bracketABfinal}
\ee
where
\be
\left\{\cdot\right\}_0&=&16(1-v^2)(3-v^4)\label{bracketzero}\\
\left\{\cdot\right\}_1&=&-32v(1-v^2)(3-v^2)\label{bracketone}\\
\left\{\cdot\right\}_2&=&32v^2(1-v^2).\label{brackettwo}
\ee
Comparing (\ref{bracketABfinal}) with (\ref{brac1}) thus gives
\be
-\left(V^{01}\right)^2\left\{\cdot\right\}_A+\left(D+V^{11}\right)\left\{\cdot\right\}_B=16(1-v^2)E^2\left\{\right\}_N. \label{bracketNfinal}
\ee
Putting (\ref{bracketNfinal}) into (\ref{Start3}) and using (\ref{bracketDfinal}) and (\ref{V22}) yields the following equation equivalent to (\ref{div4}):
\be
2\left(\left\{\cdot\right\}_D\frac{A}{(1-A)}DE\right)\,\xi v_{\xi}=4(1-v^2)E^2\left\{\cdot\right\}_N-\frac{A}{1-A}D^2(1-v^2).\label{almost1}
\ee
Multiplying (\ref{almost1}) through by the factor $\left(\left\{\cdot\right\}_D\frac{A}{(1-A)}DE\right)^{-1}$ and using  ,
$$
D=E\left[(3+v^2)G-4v\right],
$$
(c.f. (\ref{Ddefn}), (\ref{V00}) and (\ref{V01})), now readily verifies that (\ref{almost1}), and hence (\ref{div4}), is equivalent to (\ref{ODE3}), as claimed.  Thus the proof of Theorem \ref{TheoremODE} is complete.     $\Box$
\vspace{.2cm}

\noindent{\bf Proof of Theorem \ref{Theoremcheck}:}  We begin by recording the following lemma which gives the relevant quantities associated with SSC coordinate representation of the FRW metric
in terms of the single variable $v$. 
\begin{Lemma}
Consider the FRW spacetime as represented in SSC coordinates in Theorem \ref{thmSM}. Then 
\be
A&=&1-v^2,\label{SM1}\\
E&=&\frac{1}{\psi_0\xi}=\frac{1+v^2}{2v},\label{SM2}\\
D&=&1-v^2,\label{SM2.5}\\
\xi&=&\frac{2v}{\psi_0(1+v^2)}\label{SM3}\\
\kappa w&=&\frac{v^2}{1-v^2}\label{SM4}\\
v_{\xi}&=&\frac{(1+v^2)^2\psi_0}{2(1-v^2)}\label{SM5}\\
\kappa w_{\xi}&=&\frac{2v}{(1-v^2)^2}v_{\xi}\label{SM6}\\
\ee
\end{Lemma}   
\vspace{.2cm}

\noindent {\bf Proof:}  This is a straightforward consequence of (\ref{ODE11})-(\ref{ODE13})  and (\ref{FRWSSC1})-(\ref{FRWSSC4}) of Theorem \ref{thmSM}.  $\Box$

To prove Theorem \ref{Theoremcheck}, we verify the Standard Model on equation (\ref{RefEqn}), which we have shown is equivalent to   (\ref{ODE3}).  (We omit the straightforward proof that FRW satisfies (\ref{ODE1}) and (\ref{ODE2}) in SSC.)  We now use (\ref{SM1})-(\ref{SM6}) to convert each of the five terms labeled by brackets in (\ref{RefEqn}) into expressions involving $v$ and $\xi v_{\xi}$ alone.   
\be\label{bracketcheck}
\left\{\cdot\right\}_{I}&=&-\frac{4v^2(3-v^2)}{(1-v^2)^2}\xi v_{\xi}\label{Bracket1}\\
\left\{\cdot\right\}_{II}&=&\frac{1+v^2}{(1-v^2)^2}(1+6v^2-3v^4)\xi v_{\xi}\label{Bracket2}\\
\left\{\cdot\right\}_{III}&=&-\frac{8v^5(1+v^2)}{(1-v^2)^3}\label{Bracket3}\\
\left\{\cdot\right\}_{IV}&=&2\frac{(1+v^2)^3}{1-v^2)^3}v^3\label{Bracket4}\\
\left\{\cdot\right\}_{V}&=&-v(1+v^2)\label{Bracket5}
\ee
Thus we need only verify
\be
\left\{\cdot\right\}_{I}+\cdots+\left\{\cdot\right\}_{V}=0.\label{bracketcondition}
\ee
Adding (\ref{Bracket1})-(\ref{Bracket5}) we obtain
\be
&\left\{\cdot\right\}_{I}+\cdots+\left\{\cdot\right\}_{V}=\left\{-\frac{4v^2(3-v^2)}{(1-v^2)^2}+\frac{(1+v^2)(1+6v^2-3v^4)}{(1-v^2)^2}\right\}\xi v_{\xi}\ \ \ \ \ \ \ \label{secondbrac}&\\
&\ \ \ \ \ \ \ \ \ \ \ \ \ \ \ \ \ \ \ \ \ \ \ \ \ \ \ \ \ \ +\left\{-\frac{8v^5(1+v^2)}{(1-v^2)^3}+\frac{2(1+v^2)^3v^3}{(1-v^2)^3}-v(1+v^2)\right\}.&\nonumber
\ee
But from (\ref{SM3}) and (\ref{SM5}) 
$$
\xi v_{\xi}=\frac{v(1+v^2)}{1-v^2},
$$
and substituting this into (\ref{secondbrac}) and multiplying through by $\frac{(1-v^2)^3}{(1+v^2)v}$ gives
\be\nonumber
&\left\{\cdot\right\}_{I}+\cdots+\left\{\cdot\right\}_{V}=\left\{-4v^2(3-v^2)+(1+v^2)(1+6v^2-3v^4)\right\}_a\ \ \ \ \ \nonumber\\
&\ \ \ \ \ \ \ \ \ \ \ \ \ \ \ \ \ \ \ \ \ \ \ \ \ \ \ \ \ \ \ \ \ \ \ \ \ \ \ \ \ \ \ \ \ \ \ +\left\{-8v^4+2(1+v^2)^2v^2-1\right\}_b.&\nonumber
\ee
Expanding $\left\{\cdot\right\}_a$ and $\left\{\cdot\right\}_b$ gives the final result
\be
-\left\{\cdot\right\}_a= 3v^6-7v^4+5v^2-1=\left\{\cdot\right\}_b,\nonumber
\ee
thereby verifying (\ref{bracketcondition}).  This completes the proof of Theorem \ref{Theoremcheck}.   $\Box$

We conclude that the Standard Model of Cosmology during the radiation phase corresponds to a solution of the expanding wave equations (\ref{one})-(\ref{three}) and (\ref{constraint1}) with parameter  $\psi_0$ accounting for the time-scaling freedom of the SSC metric (\ref{SSC}).  More generally, it is not difficult to see that the time-scaling $\bar{t}\rightarrow \psi_0\bar{t}$ preserves solutions of  (\ref{one})-(\ref{three}) and the constraint (\ref{div4})).  The next theorem states that modulo this scaling, distinct solutions of (\ref{one})-(\ref{three}), (\ref{div4}) describe distinct spacetimes.  Thus the three equations (\ref{one})-(\ref{three}) with the one scaling law describe a two parameter family of distinct spacetimes, one of which is FRW.

\begin{Theorem}
The replacement $\bar{t}\rightarrow\psi_0\bar{t}$ takes $A(\xi),$ $G(\xi)$ and $v(\xi)$ to $A(\xi/\psi_0),$ $G(\xi/\psi_0)$ and $v(\xi/\psi_0)$, and this scaling preserves solutions of (\ref{one})-(\ref{three}), (\ref{div4}).  Moreover, this is the only scaling law in the sense that any two solutions of (\ref{one})-(\ref{three}), (\ref{div4}) not related by the scaling $\bar{t}\rightarrow\psi_0\bar{t}$ describe distinct spacetimes.  
\end{Theorem}
\vspace{.2cm}

\noindent{\bf Proof:}  To prove the theorem it suffices to show that the only coordinate transformation that takes solutions of form (\ref{thebasicmetric}) to solutions of form (\ref{thebasicmetric}) are the time-scaling transformations $t\rightarrow\alpha t.$    Now solutions of form (\ref{thebasicmetric}) are diagonal metrics in which the radial coordinate is taken to be $\bar{r}$ determined by the areas of the spheres of symmetry, so the problem is to show that the only coordinate transformation of the form $(\bar{t},\bar{r})\rightarrow(\hat{t},\bar{r})$ taking a metric
$$
ds^2=-B(\xi)d\bar{t}^2+\frac{1}{A}d\bar{r}^2+\bar{r}^2d\Omega^2,
$$
to a metric
$$
ds^2=-\hat{B}(\xi)d\hat{t}^2+\frac{1}{\hat{A}}d\bar{r}^2+\bar{r}^2d\Omega^2,
$$
are the time-scalings $\hat{t}=\alpha\bar{t}.$  Now since both metrics are diagonal with no cross terms we must have
$$
\hat{t}=\phi(\bar{t}),
$$
for some function $\phi.$  Moreover, since both metrics use the same radial coordinate $\bar{r}$, the transformation must meet the condition
$$
\hat{A}\left(\frac{\bar{r}}{\phi(\bar{t})}\right)=A\left(\frac{\bar{r}}{\bar{t}}\right).
$$
Differentiating this latter condition with respect to $\bar{r}$ gives
$$
\frac{\phi(\bar{t})}{\bar{t}}=\frac{\hat{A}'}{A'}.
$$
Since the left hand side is independent of $\bar{r}$ and the right hand side is not, it must be that both sides are constant, implying that 
$$
\phi(\bar{t})=\alpha\bar{t}
$$
for some (positive) constant, as claimed.  $\Box$

In summary, equations (\ref{one})-(\ref{three}) and (\ref{div4}) admit three (initial value) parameters and one scaling law that describe a {\em two} parameter family of distinct spacetimes, one of which is FRW. In Section \ref{sect5} we show that by imposing regularity at the center there results a further reduction to a continuous one parameter family of expanding wave solutions,  such that one value of the parameter corresponds to FRW, the Standard Model of Cosmology with pure radiation sources.

\section{Canonical Co-moving Coordinates and Comparison with the $k\neq0$ FRW Spacetimes}
\setcounter{equation}{0} 

As a consequence of Theorem \ref{TheoremODE}, we know that solutions of equations (\ref{ODE1})-(\ref{ODE3}) correspond to self-similar spacetimes that solve the Einstein equations with $p=\frac{c^2}{3}\rho$. The following general theorem gives a canonical form for such spacetime metrics in co-moving coordinates.  We use this to prove that none of the spacetime metrics that solve (\ref{ODE1})-(\ref{ODE3}) agree with a $k$-FRW metric for any $k\neq0.$

\begin{Theorem}\label{co-moving}  Consider a general self-similar spacetime metric of the form,
\be\label{generalss}
ds^2=-B(\xi)d\bar{t}^2+\frac{1}{A(\xi)}d\bar{r}^2+\bar{r}^2d\Omega^2,
\ee
and let $\bar{v}=\bar{v}(\xi)$ be an arbitrary smooth velocity field, c.f. (\ref{thebasicmetric}).  That is, assume all functions depend only on the self-similarity variable $\xi=\frac{\bar{r}}{\bar{t}},$ and assume
\be\label{vtermsu}
\bar{v}(\xi)=\frac{\bar{u}^1}{\bar{u}^0}\frac{1}{\sqrt{AB}},
\ee
where 
$$
u\equiv \bar{u}^0\frac{\partial}{\partial \bar{t}}+\bar{u}^1\frac{\partial}{\partial \bar{r}}
$$
is a timelike vector field that has unit length as measured by (\ref{generalss}).  Let $\phi(\xi)$ and $\psi({\xi})$, respectively, be solutions of the ODE's
\be\label{phiODE}
\frac{d\phi}{d\xi}=\frac{\phi}{\xi-\sqrt{AB}\bar{v}},
\ee
\be\label{psiODE}
\frac{d\psi}{d\xi}=\frac{\psi-\sqrt{B}(1-v^2)}{\xi-\sqrt{AB}\bar{v}}.
\ee
Then under the coordinate mapping defined by
\be\label{phipsimap}
\hat{r}&=&\phi(\xi)\bar{t},\label{phimap}\\
\hat{t}&=&\psi(\xi)\bar{t},\label{psimap}
\ee
the metric (\ref{generalss}) transforms to the metric
\be\label{zetass}
ds^2=-d\hat{t}^2+R(\zeta)^2d\hat{r}^2+2Q(\zeta)d\hat{t}d\hat{r}+\bar{r}^2d\Omega^2,
\ee
where
\be\label{zetafirst}
\zeta=\frac{\hat{r}}{\hat{t}},
\ee
and the velocity transforms to 
\be\label{v=0}
\hat{v}=0.
\ee
\end{Theorem}
Since $\hat{v}=0$ and $R(\zeta)$ and $Q(\zeta)$ depend only on the self-similarity variable $\zeta=\frac{\hat{r}}{\hat{t}}$, ($\hat{t},\hat{r}$) are co-moving coordinates in which the metric defined by (\ref{generalss}) remains self-similar.  The coordinates are canonical in the sense that the coefficient of $-d\hat{t}^2$ is unity in ($\hat{t},\hat{r}$) coordinates.
\vspace{.2cm}

\noindent{\bf Proof:}  Solving (\ref{vtermsu}) for $\bar{u}^1$ gives
\be\label{ucomponents}
\bar{u}^1=(\sqrt{AB}\,\bar{v})\bar{u}^0,
\ee
and using this in the condition that $\bar{u}$ is a timelike unit vector gives
\be\nonumber
\bar{u}^0=\frac{1}{\sqrt{B(-1+\bar{v}^2)}}.
\ee
It follows that the components of $u$ are given by
\be\label{uform}
u=\left(\frac{1}{\sqrt{B}}\frac{1}{\sqrt{1-\bar{v}^2}},\frac{\sqrt{A}\,\bar{v}}{\sqrt{1-\bar{v}^2}}\right).
\ee
Now since $u$ is assumed to be a unit vector, the condition that $(\hat{t},\hat{r}$) be co-moving is equivalent to 
\be\label{comovingcondt}
\hat{u}^1=0,
\ee
and the condition that the coefficient of $-d\hat{t}^2$ is unity is equivalent to 
\be\label{unitycondt}
\hat{u}^0=1,
\ee
where
\be\nonumber
u\equiv \hat{u}^0\frac{\partial}{\partial \hat{t}}+\hat{u}^1\frac{\partial}{\partial \hat{r}},
\ee
is the $(\hat{t},\hat{r})$ representation of vector $u.$  Now by (\ref{phipsimap}),
\be
\frac{\partial \hat{t}}{\partial\bar{t}}=\psi,\ \ \ \ \ \ 
\frac{\partial \hat{t}}{\partial\bar{r}}=-\xi\psi'+\psi\label{hattbar}\\
\frac{\partial \hat{r}}{\partial\bar{t}}=\phi'\ \ \ \ \ \ 
\frac{\partial \hat{r}}{\partial\bar{r}}=-\xi\phi'+\phi,\label{hatrbar}
\ee
where ``prime'' denotes $\frac{d}{d\xi}.$   Using this, the co-moving condition (\ref{comovingcondt}) is
\be\label{uone0}
0=\hat{u}^1=\frac{\partial \hat{r}}{\partial \bar{t}}\bar{u}^0+\frac{\partial \hat{r}}{\partial\bar{r}}\bar{u}^0
=\left(-\xi\phi'+\phi\right)\bar{u}^0+\phi\bar{u}^0,
\ee
which by  (\ref{uform}) is equivalent to (\ref{phiODE}).   Similarly, the unity condition (\ref{unitycondt}) becomes
\be\label{uzero1}
1=\hat{u}^0=\frac{\partial \hat{t}}{\partial \bar{t}}\bar{u}^0+\frac{\partial \hat{t}}{\partial\bar{r}}\bar{u}^0
=\left(-\xi\psi'+\psi\right)\bar{u}^0+\phi\bar{u}^0,
\ee
which by (\ref{uform}) is equivalent to (\ref{psiODE}).  Since the Jacobian derivatives (\ref{hattbar}), (\ref{hatrbar}) are all functions of $\zeta,$ it follows that the ($\hat{t},\hat{r}$) coordinate representation of metric (\ref{generalss}) must be of the form (\ref{zetass}).  $\Box$
\vspace{.2cm}

The next theorem states that the self-similar solutions defined by equations (\ref{ODE1})-(\ref{ODE3}) are distinct from the $k\neq0$ FRW spacetimes.

\begin{Theorem}  Spacetime metrics defined by solutions of equations (\ref{ODE1})-(\ref{ODE3}) are distinct from the $k\neq0$ Friedmann-Robertson-Walker spacetimes. 
\end{Theorem}
\vspace{.2cm}

\noindent{\bf Proof:}   The $k\neq0$ FRW spacetimes in co-moving coordinates are given by 
\be\label{FRWk}
ds^2=-dt^2+\frac{R(t)^2}{1-kr^2}dr^2+R(t)^2r^2d\Omega^2,
\ee
\cite{wein}.
Since by Theorem \ref{co-moving} we know each spacetime metric defined by a solution of equations (\ref{ODE1})-(\ref{ODE3}) can be mapped to a co-moving coordinate system ($\hat{t},\hat{r}$) in which the metric takes the form (\ref{zetass}),  in order to prove the theorem it suffices to prove that there is no coordinate mapping that takes (\ref{FRWk}) to a metric of form (\ref{zetass}) such that it preserves the co-moving condition.  So assume such a mapping  ($\hat{t},\hat{r})\rightarrow(t,r)$ does exist.  Now co-moving means fluid trajectories move along $r=const,$ so we must have
\be
r=g(\hat{r})\label{rhatmap},
\ee
for some function $g(\hat{r}).$   Since both (\ref{FRWk}) and (\ref{zetass}) are normalized so the coefficient  of $-dt^2$ is unity, it also follows that
\be
t=\hat{t}+h(\hat{r}),\label{thatmap}
\ee
for some function $h(\hat{r})$.  Thus,
\be
dr^2&=&g'(\hat{r})^2d\hat{r}^2,\nonumber\\
dt^2&=&d\hat{t}^2+2h'(\hat{r})d\hat{t}d\hat{r}+h'(\hat{r})^2d\hat{r}^2,\nonumber
\ee
implying that (\ref{FRWk}) transforms to the metric
\be\label{comovingmetric}
ds^2=-d\hat{t}^2+\left\{\frac{R^2g'}{1-kr^2}+(h')^2\right\}d\hat{r}^2+2h'\,d\hat{t}d\hat{r}+\bar{r}^2d\Omega^2,
\ee
and our assumption is that we must have
\be\label{Rcondt}
\left\{\frac{R^2g'}{1-kr^2}+(h')^2\right\}=\hat{R}(\zeta)^2,
\ee
and 
\be\label{Qcondt}
2h'(\hat{r})=\hat{Q}(\zeta)^2,
\ee
for some functions $\hat{R}(\zeta)$ and $\hat{Q}(\zeta)$, functions only of $\zeta=\frac{\hat{r}}{\hat{t}}.$
Condition (\ref{Qcondt}) immediately implies 
$$
h'(\hat{r})=\alpha=const.,\nonumber
$$
so (\ref{thatmap}) implies
\be
t=\hat{t}+\alpha\hat{r},\label{thatmap1}
\ee
for some constant $\alpha.$    Now denote the left hand side of  (\ref{Rcondt}) by $f,$  and set $S(t)\equiv R(t)^2$.  Then for a contradiction it suffices to show that the left hand side of (\ref{Rcondt}) cannot be a function of $\zeta$ when $k\neq0.$  But $f$ is a function of $\zeta$ if and only if  $$\frac{f_{\hat{t}}}{f_{\hat{r}}}=-\zeta,$$  so assuming it can we get
$$
f_{\hat{t}}=\frac{(1-kg(\hat{r})^2)S'g'}{(1-kg^2)^2},
$$
$$
f_{\hat{r}}=\frac{(1-kg(\hat{r})^2)S'g'}{(1-kg^2)^2},
$$
and we must have
$$
-\zeta=\frac{f_{\hat{t}}}{f_{\hat{r}}}=\frac{(1-kg^2)S'g'}{(1-kg^2)[\alpha S'g'+Sg'']+2kSg(g')^2},
$$
which after separating leads to 
\be\label{separate}
-\frac{2kgg'}{1-kg^2}-\frac{g''}{g'}=\left(\alpha+\frac{1}{\zeta}\right)\frac{S'}{S}.
\ee
The left hand side of (\ref{separate}) is a function of $\hat{r}$ alone, so differentiating with respect to $\hat{t}$ and noting $\frac{S'}{S}=2\frac{\dot{R}}{R}=2H$, ($H=H(t)$ the Hubble constant), we must have
\be
0=H+\left(\alpha\hat{r}+\hat{t}\right)H',\nonumber
\ee
or upon using $\alpha\hat{r}+\hat{t}=t,$
$$
\frac{H'}{H}=-\frac{1}{t},
$$
implying
\be\label{finalH}
H(t)=\frac{C_0}{t},
\ee
for some constant $C_0.$  Now (\ref{FRWk}) must satisfy the FRW-equations
\be\label{FRW-1}
H^2=\frac{\kappa}{3}\rho-k,
\ee
and 
\be\label{FRW-2}
\dot{\rho}=-3(\rho+p)H=-4\rho H.
\ee
Equations (\ref{finalH}) and (\ref{FRW-1}) imply
\be\label{frw1}
\rho=\frac{3}{\kappa}\left(\frac{C_0^2}{t^2}+k\right),
\ee
and (\ref{finalH}) together with (\ref{FRW-1}) imply
\be\label{frw2}
\rho=C_1t^{-4C_0},
\ee
for some constants $C_0$ and $C_1.$  But (\ref{frw1}) and (\ref{frw2}) are inconsistent unless $k=0$, (the case in which $\rho(t)$ is inverse square.)    It follows, then, that $f$ on the left hand side of (\ref{Rcondt}) cannot be written as a function of $\zeta$ when $k\neq0,$ and so there is no mapping of $k\neq0$ FRW to co-moving form (\ref{zetass}), and the theorem is proved.   $\Box$ 
\vspace{.2cm}

As a corollary of the proof we have:
\begin{Corollary}
In the case of the $k=0$ FRW metric (\ref{FRWk}), the mappings (\ref{rhatmap}), (\ref{thatmap}) can be taken as
\be
r&=&\ln(\hat{r}),\nonumber\\
t&=&\hat{t}+\alpha\hat{r},
\ee
for any constant $\alpha,$ and the co-moving form (\ref{comovingmetric}) of FRW in self-similarity variable $\zeta$ is 
\be
ds^2=-d\hat{t}^2+\frac{1}{\zeta}d\hat{r}^2+\alpha\; d\hat{t}\,d\hat{r}+\bar{r}^2d\Omega^2.
\ee
\end{Corollary}

\section{Leading Order Corrections to the Standard Model Induced by the Expanding Waves}\label{sect5}
\setcounter{equation}{0}

We extend and make rigorous the asymptotic estimates in \cite{smoltePNAS}.  To clarify the issues, note first that system (\ref{ODE1})-(\ref{ODE3}) takes the form
\be
\label{A1}
\xi\left(\begin{array}[pos]{c}
 A(\xi)\\G(\xi)\\v(\xi)
 \end{array}\right)_{\xi}=F\left(\begin{array}[pos]{c}
 A(\xi)\\G(\xi)\\v(\xi)
 \end{array}\right),
\ee
where $\xi\geq 0,$ and whose solutions we are concerned with in a neighborhood of $\xi=0.$  Thus consider a general $n\times n$ system of ODE's of the form
\be
\xi V_{\xi}=F(V).\label{system}
\ee
Define the change of independent variable
\be
\xi=e^s,\nonumber
\ee
and observe that (\ref{linearize3}) goes over to the autonomous system of ODE's
\be
U_s\equiv\frac{dU}{ds}=F(U),\label{s}
\ee
where
\be
U(s)=V(e^s),\label{U}
\ee
and note that for $V_0\in{\bf R}^n$,  $\lim_{s\rightarrow-\infty}U(s)=V_0$ iff $\lim_{\xi\rightarrow0}V(\xi)=V_0$, in which case $F(V_0)=0.$ 
Then we have the following elementary lemma:
\begin{Lemma} Consider a general $n\times n$ system of form (\ref{system}) in which $F$ is a smooth function of $V\in\bf{R}^n$, and assume that $V(\xi)$ is a smooth solution of (\ref{system}) in $\xi>0$ satisfying $V(\xi)\rightarrow V_0$ as $\xi\rightarrow0$, for some $V_0\in\bf{R}^n.$   Then $F(V_0)=0.$
\end{Lemma}

\noindent{\bf Proof:}  If $\xi\rightarrow0,$ then $s\rightarrow-\infty,$ so if $F(V_0)\neq0$, then the flow of the autonomous ODE (\ref{s}) takes small neighborhoods of $V_0$ to neighborhoods disjoint from $V_0,$ contradicting  $V(\xi)\rightarrow V_0.$ $\Box$

Now assume $V(\xi)\rightarrow V_0\in\bf{R}^n$ as $\xi\rightarrow0,$  so that by the lemma $F(V_0)=0.$ 
Expanding $F(V)$ in Taylor series about $V_0$ gives
\be
F(V)=dF(V_0)(V-V_0)+O(1)|V-V_0|^2,
\ee
so near $V_0,$ $V(\xi)$ satisfies
\be
\xi (V(\xi)-V_0)'=dF(V_0)(V(\xi)-V_0),\label{linearize1}
\ee
to leading order in $|V-V_0|.$  Setting $W\equiv W(\xi)=V(\xi)-V_0$ in (\ref{linearize1}) gives the leading order system
\be
\xi W_\xi=A\,W,\label{linearize2}\label{linearize3}
\ee
where $A$ is a constant $n\times n$ matrix with real entries.  We restrict to the case that $A$ has $n$ real and distinct eigenvalues.  Then let $B$ denote the $n\times n$ matrix that diagonalizes $A,$ so that
\be
B^{-1}AB={\rm diag}\left\{\lambda_1,...,\lambda_n\right\},\nonumber
\ee
where
\be
\lambda_1<\lambda_2<\cdots<\lambda_n,\nonumber
\ee
are the eigenvalues of $A.$ Setting $Z=B^{-1}W\in{\bf R}^n$ then reduces  (\ref{linearize3}) to the diagonal system,
\be
\xi Z'(\xi)=B^{-1}ABW, \label{linearize4}
\ee
which has the eigensolutions
\be
Z_i(\xi)&=&\xi^{\lambda_i}R_i,\label{linearsoln1}
\ee
where $(\lambda_i,{\bf R}_i)$ are the eigenpairs of $A$. 
From this we see that the only solutions $V(\xi)$ of (\ref{system}) satisfying $V(\xi)\rightarrow V_0$ as $\xi\rightarrow0$ correspond to the solutions $Z(\xi)$ that lie in the span of the eigensolutions $Z_i(\xi)$ corresponding to $\lambda_i>0.$

Note in particular that the solution $V(\xi)$ will only have the smoothness at $\xi=0$ allowed by the powers of the eigenvalues in (\ref{linearsoln1}).  However, in the special case when the positive eigenvalues $\lambda_i$ are all positive integers, (the case we find below), the span of the corresponding linearized solutions (\ref{linearsoln1}) will all be infinitely differentiable at $\xi=0.$  One can show that solutions of the nonlinear equations can be obtained by expanding solutions in powers of $\xi^{\lambda_i}$, so in the case of positive integer eigenvalues below, arbitrary smoothness can be assumed at $\xi=0,$ (c.f. Lemma \ref{lemma7} below). 

So assume from here on that we have a solution $V(\xi)$ of a nonlinear system (\ref{system}) such that $V(\xi)\rightarrow V_0\in{\bf R}^n$ as $\xi\rightarrow0,$ so that $F(V_0)=0$, and $A=dF(V_0)$ has $n$ real and distinct eigenvalues.   From (\ref{s}), (\ref{U}) 
note that for $V_0\in{\bf R}^n$,  $\lim_{s\rightarrow-\infty}U(s)=V_0$ iff $\lim_{\xi\rightarrow0}V(\xi)=V_0$, in which case $F(V_0)=0.$  Then the solutions $U(s)$ of (\ref{s}) corresponding to the positive eigensolutions (\ref{linearsoln1}) of the linearized system (\ref{linearize4}) are just solutions in the unstable manifold of the linearization of (\ref{s}) at $V_0,$ namely,
\be
\left\{U(s)=V_0+\sum_i\alpha_ie^{\lambda_is}R_i:\alpha_i\in{\bf R}\right\},
\ee 
where the sum on $i$ is over all non-negative eigenvalues $\lambda_i$ of $dF(V_0).$
 We thus have the following theorem:

\begin{Theorem}   The solutions $V(\xi)$ satisfying (\ref{linearize3}) together with the condition $V(\xi)\rightarrow V_0\in{\bf R}^n$ as $\xi\rightarrow0$ are {\em exactly} the solutions $U(s)=V(e^s)$ of (\ref{s}) in the unstable manifold of the hyperbolic rest point $V_0$, $F(V_0)=0.$   
\end{Theorem}

Consider now (\ref{ODE1})-(\ref{ODE3}), a system of form (\ref{linearize3}).   We know by Theorem \ref{thmSM} that the Standard Model  satisfies (\ref{ODE1})-(\ref{ODE3}) with,
\be
A=1-v^2,\ \ G=\psi_0\xi,\ \ \xi=\frac{2v}{\psi_0(1+v^2)},\label{SM}
\ee
where $\psi_0$ is the constant that corresponds to a change of time-scale, c.f. Lemma \ref{thmSM}.
Solving the last equation for $v$, (assuming the case $v\rightarrow0+$ as $\xi\rightarrow0+$), and from here on letting the subscript ``1" refer to the Standard Model, gives
\be
v_1(\xi)\equiv \frac{1-\sqrt{1-\psi_0^2\xi^2}}{\psi_0\xi}.\label{voneofxi}\label{SMv}
\ee
Thus (\ref{SM}) and (\ref{SMv}) define the functions $A_1(\xi),G_1(\xi),v_1(\xi)$ of the Standard Model, and so
as $\xi\rightarrow0$  in the Standard Model we have 
\be
\label{SMG}
 \left(\begin{array}[pos]{c}
 A_1(\xi)\\G_1(\xi)\\v_1(\xi)
 \end{array}\right)\equiv
\left(\begin{array}[pos]{c}
1-v_1(\xi)^2\\\psi_0\xi\\v_1(\xi)
 \end{array}\right)\rightarrow
 \left(\begin{array}[pos]{c}
1\\0\\0
 \end{array}\right).
\ee
We thus look for all solutions $(A(\xi),G(\xi),v(\xi))$ of (\ref{ODE1})-(\ref{ODE3}) that tend to the rest point $(1,0,0)$ as $\xi\rightarrow0,$
and these correspond to all solutions $U(s)=V(e^s)$ in the unstable manifold of rest point $(1,0,0)$ of (\ref{s}).   But note that (\ref{ODE1})-(\ref{ODE3}) are undetermined at $(A,G,v)=(1,0,0)$ without knowing the limit of the ratio $G/v$ in the limit $\xi\rightarrow0.$  

To remedy this, define the variable
\be
H=G/v,\label{H}
\ee
and consider (\ref{ODE1})-(\ref{ODE3}) in the nonsingular variables $(v,A,H),$ (re-ordered so as to diagonalize the linearized operator at $V_0$, c.f. (\ref{dF}) below.)  Thus without confusion, from here on let $V$ refer to the $3$-vector
\be
\label{newV}
V\equiv\left(\begin{array}[pos]{c}
 v\\A\\H
 \end{array}\right),
\ee
so that we recover $G$ from the identity 
$$
G=Hv.
$$
Again let $\xi=e^s$ and $U(s)=V(e^s)$ so that $\xi V_{\xi}=U(s)$.  Then equation (\ref{ODE3}), in terms of variables $V\equiv(v,A,H),$  becomes
\begin{eqnarray}
v_s\equiv \xi v_{\xi}&=&-\left(\frac{1-v^2}{2\left\{\cdot\right\}_D}\right)\left\{\left[(3+v^2)H-4\right]v+\frac{4\left(\frac{1-A}{A}\right)\left\{\cdot\right\}_N^*v}{(3+v^2)H-4}\right\},\nonumber\\
\label{ODE1H}
\ee
where
\begin{eqnarray}
\left\{\cdot\right\}_N\equiv\left\{\cdot\right\}_N^*v^2,\label{ODE1H1}
\end{eqnarray}
\begin{eqnarray}
\left\{\cdot\right\}_N^*=\left\{-2v^2+2(3-v^2)H-(3-v^4)H^2\right\},\label{ODE1H2}
\end{eqnarray}
\begin{eqnarray}
\left\{\cdot\right\}_D=\left\{(3v^2-1)-4Hv^2+(3-v^2)H^2v^2\right\};\label{ODE1H3}
\end{eqnarray}
Equation (\ref{ODE1}) in terms of variables $V\equiv(v,A,H)$  goes over to
\be
A_s\equiv \xi A_{\xi}&=&-\left[\frac{4(1-A)}{(3+v^2)H-4}\right];\label{ODE2H}
\end{eqnarray}
and equation (\ref{ODE2}) goes over to
\begin{eqnarray}
H_s\equiv \xi H_{\xi}=\frac{G_s}{v}-\frac{G}{v^2}v_s,
\label{ODE3H}
\end{eqnarray}
where
\begin{eqnarray}
\frac{G_s}{v}=-H\left\{\left(\frac{1-A}{A}\right)\frac{2(1+v^2)H-4}{(3+v^2)H-4}-1\right\},
\label{ODE3H1}
\end{eqnarray}
and
\begin{eqnarray}
\frac{G}{v^2}v_s=-H\left(\frac{1-v^2}{2\left\{\cdot\right\}_D}\right)\left\{(3+v^2)H-4+\frac{4\left(\frac{1-A}{A}\right)\left\{\cdot\right\}_N^*v}{(3+v^2)H-4}\right\}.
\label{ODE3H2}
\end{eqnarray}
The constraint (\ref{constraint2}) then becomes
\be
\kappa w=\frac{(1-A)H}{\left[(3+v^2)H-4\right]v}\label{constraintH}.
\ee
To summarize, system (\ref{ODE1H})-(\ref{ODE3H2}) shall be denoted 
\be
U_s\equiv \xi V_{\xi}=\label{Heqn}
\xi\left(\begin{array}[pos]{c}
 v'(\xi)\\A'(\xi)\\H'(\xi)
 \end{array}\right)=F(U),
\ee
where the change of variables 
\be
\xi=e^s,\ \ \ U(s)=V(e^s),\label{xitos}
\ee
transforms $\xi\frac{d}{d\xi}=\frac{d}{ds}$, so that in terms of $U(s)$, (\ref{Heqn}) is an autonomous system of three ODE's equivalent to (\ref{ODE1H}), ((\ref{ODE2H}) and (\ref{ODE3H}), namely
\be
U_s=F(U).\label{Usystem}
\ee
We now have that as $\xi\rightarrow0$ or $s\rightarrow-\infty$, the solution 
$$V_1(\xi)=(v_1(\xi),A_1(\xi),H_1(\xi)),$$ 
corresponding to the Standard Model, has the regular limit, (c.f. (\ref{linearsoln1}), (\ref{SMG})),
\be
\label{SMH}
U_1(s)=V_1(\xi)\equiv
 \left(\begin{array}[pos]{c}
 v_1(\xi)\\A_1(\xi)\\H_1(\xi)
 \end{array}\right)=
\left(\begin{array}[pos]{c}
v_1(\xi)\\1-v_1(\xi)^2\\\frac{2}{1+v_1(\xi)^2}
 \end{array}\right)\rightarrow
 \left(\begin{array}[pos]{c}
0\\1\\2
 \end{array}\right)\equiv V_0.
\ee
We now linearize  (\ref{Usystem}) about the nonsingular rest point $V_0=(0,1,2).$  

For this, write (\ref{Usystem}) as
\be
\left(\begin{array}[pos]{c}
 v\\A\\H
 \end{array}\right)_s=F(U)\equiv \left(\begin{array}[pos]{c}
f(U)\\g(U)\\h(U)
 \end{array}\right),
\ee
where $f,g,h$ are defined by the RHS of system (\ref{ODE1H}), (\ref{ODE2H}) and (\ref{ODE3H}), respectively.  
Then we obtain directly from the second equation (\ref{ODE2H}) that
\be
dg(V_0)\equiv \left(\frac{\partial g}{\partial v},\frac{\partial g}{\partial A},\frac{\partial g}{\partial H}\right)_{V_0}=(0,2,0).\label{dg}
\ee
Neglecting terms second order in $v$ and terms that vanish at $(0,1,2)$ on the RHS of the first equation (\ref{ODE2H}) gives 
\be
df(V_0)=-d\left\{H\left(\frac{1}{2(-1)}\right)\left(3Hv-4v\right)\right\}=(1,0,0);\label{df}
\ee
and similarly from the third equation (\ref{ODE3H}) we obtain
\be
dh(V_0)&=&-d\left\{\left(\left[\left(\frac{1-A}{A}\right)\frac{2H-4}{3H-4}-1\right]-\left[\frac{1}{2(-1)}(3H-4)\right]\right)\right\}_{V_0}\nonumber\\
&=&\left(0,-\frac{H(-A)}{A^2}\left(\frac{2H-4}{3H-4}\right),-\frac{d}{dH}\left(H\left\{-1+\frac{1}{2}(3H-4)\right\}\right)\right)_{V_0}\nonumber\\
&=&\left(0,0,-\left(-1+3H-2\right)\right)_{V_0}\nonumber\\
&=&(0,0,-3).\label{dh}
\ee 
Putting (\ref{dg})-(\ref{dh}) together yields the diagonal matrix
\be
dF(V_0)= \left(\begin{array}[pos]{c}
 df(V_0)\\dg(V_0)\\dh(V_0)
 \end{array}\right)=\left(\begin{array}[pos]{ccc}
1&0&0\\0&2&0\\0&0&-3
 \end{array}\right)\label{dF},
\ee
implying that $V_0$ is a hyperbolic rest point of system (\ref{Usystem}) with positive eigenvalues $\lambda_1=1,\lambda_2=2$ and negative eigenvalue $\lambda_3=-3.$  

We conclude that solutions $U(s)=V_0+W(s)$ of the linearized equations
\be
W_s=dF(V_0)\cdot W,\label{lineqn}
\ee
at the rest point $V_0$ of system (\ref{Heqn}), such that $U(s)$ tends to $V_0$ as $\xi\rightarrow0$, $s\rightarrow-\infty$, are precisely solutions in the $2$-dimensional unstable manifold ${\mathcal M}_0$ of $V_0$ 
\be
{\mathcal M}_{0}=\left(\begin{array}[pos]{c}
0\\1\\2
 \end{array}\right)+{\rm Span}\left\{\left(\begin{array}[pos]{c}
1\\0\\0
 \end{array}\right)e^s,\left(\begin{array}[pos]{c}
0\\1\\0
 \end{array}\right)e^{2s}\right\},\label{linearsolns}
\ee
so that 
\be
U(s)=\left(\begin{array}[pos]{c}
\alpha e^s\\ 1+\beta e^{2s}\\2
 \end{array}\right),
\ee
and
\be
V(\xi)=\left(\begin{array}[pos]{c}
\alpha \xi\\ 1+\beta \xi^2\\2
 \end{array}\right).
\ee
 It thus follows from the Stable Manifold Theorem and the Hartman-Grobman Theorem for hyperbolic rest points, \cite{coddle,hart},  that solutions $U(s)$ of the nonlinear system (\ref{Usystem}) that tend to the rest point $V_0$ as $s\rightarrow-\infty$ (like the Standard Model), must be precisely the solutions that lie in the $2$-dimensional unstable manifold $\mathcal{M}$ of rest point $V_0$ for the nonlinear system (\ref{Usystem}), where $\mathcal{M}$ is tangent to ${\mathcal M}_0$  at $V_0,$ and such that, in a neighborhood of $V_0$, nonlinear solutions in $\mathcal{M}$ are in $1-1$ correspondence with the linear solutions (\ref{linearsolns}) in ${\mathcal M}_0$, \cite{hart}.  In particular, it follows from (\ref{SMH}) that the Standard Model $V_1(\xi)=U_1(s)$ is the particular solution in ${\mathcal M}$ that corresponds to
 \be
 \alpha=\frac{\psi_0}{2},\ \ \ \beta=-\frac{\psi^2}{4}.\nonumber
 \ee 
 Thus in general we define the two constants $\psi_0$ and $a^2$ for the linearized solutions as 
  \be
 \alpha=\frac{\psi_0}{2},\ \ \ \beta=-\frac{a^2\psi^2}{4},\nonumber
 \ee       
so that the general solution of the linearized equations (\ref{ODE1})-(\ref{ODE3}) at rest point $V_0$ is given by
\be
v(\xi)&=&\frac{\psi_0}{2}\xi\nonumber\\
A(\xi)&=&1-\frac{a^2\psi_0^2\xi^2}{4}\label{linearizedwitha}\\
H(\xi)&=&2.
\ee
The condition $\psi_0>0$ guarantees an expanding solution ($v>0$) near $\xi=0,$  and the condition and $a^2>0$ guarantees a spacetime ``outside the black hole'' in the sense that  $0<A<1,$ near $\xi=0.$    In this paper we are interested in the case $\psi_0>0$, $a^2>0$ but when referring to a general solution in ${\mathcal M}_0$ we formally allow $\psi_0,a^2$ to be constants in ${\bf R}.$\footnotemark[15]\footnotetext[15]{It is interesting that equations (\ref{ODE1})-(\ref{ODE3}) admit solutions ``inside the black hole'' in a neighborhood of $\xi=0$.}

We now derive refined estimates for solutions $V(\xi)$ of the fully nonlinear system (\ref{Heqn}) that lie in the invariant manifold ${\mathcal M}$.  These estimates give the order at which a general solution $V(\xi)$ in ${\mathcal M}$ diverges from the Standard Model $V_1(\xi)$  near the rest point $V_0.$  For this, define ${\mathcal M}^*$ to be the subset of ${\mathcal M}$ consisting of all orbits except the unique orbit manifold corresponding to the (weaker) eigenvalue $\lambda=2.$  Thus the Standard Model lies in ${\mathcal M}^*$ because $\alpha\neq0,$ and all orbits in ${\mathcal M}^*$ enter the rest point tangent to the standard model as $s\rightarrow-\infty$.  

The main purpose of this section is to prove the following theorem:

\begin{Theorem}\label{theoremME}\label{thmME}
Let $V(\xi)=(v(\xi),A(\xi),H(\xi))$ be any solution of system (\ref{Heqn}) that lies in ${\mathcal M}^*$, so that $\lim_{\xi\rightarrow0} V(\xi)=V_0=(0,1,2),$
and let $V_1(\xi)\equiv(v_1(\xi),A_1(\xi),G_1(\xi)),$ with $H_1(\xi)\equiv G(\xi)v(\xi)$ be the solution for the Standard Model given in (\ref{SMG}), (c.f. (\ref{SM}), (\ref{voneofxi})).
Then there exist real constants $\psi_0,\ a^2$ such that the following estimates hold:
\be
v(\xi)&=&v_1(\xi)+\frac{(1-a^2)}{8}\psi_0^3\xi^3+O(1)|a-1|\xi^5,\label{vest}\\
A(\xi)&=&1-a^2v_1^2(\xi)+\frac{3a^2(1-a^2)}{16}\psi_0^4\xi^4+O(1)|a-1||\xi^6,\label{Aest}\\
H(\xi)&=&H_1(\xi)-\frac{1-a^2}{2}\psi_0^2\xi^2+O(1)|a-1|\xi^4\label{Hest}\\
&=&\frac{\psi_0\xi}{v(\xi)}+O(1)|a-1||\xi^4,\label{Hest1}
\ee
and
\be
G(\xi)&=&\psi_0\xi+\frac{a(1-a^2)}{32}\psi_0^5\xi^5+O(1)|a-1||\xi^7,\label{Gest}\\
A(\xi)B(\xi)\psi_{0}^{2}&=&1-\frac{a(1-a)}{16}\psi_0^4\xi^4+O(1)|a-1||\xi^6,\label{ABest}
\ee
where all of the right hand sides are functions of $\psi_0\xi$, and for convenience we let $O(1)$\footnotemark[16]\footnotetext[16]{In this paper $O(1)$ denotes a function bounded as $\xi\rightarrow0,$ and in the next section we will, without loss of generality, set $\psi_0=1.$} incorporate the constant $\psi_0$ in the error terms that vanish on the Standard Model $a=1.$
Moreover, for each choice of $\psi_0>0$, $a^2>0,$ there exists a solution 
\be\label{vasoln}
V(\xi)\equiv({\mathcal V}_a(\psi_0\xi),{\mathcal H}_a(\psi\xi),{\mathcal A}_a(\psi\xi))
\ee 
in ${\mathcal M}^*$ such that (\ref{vest})-(\ref{ABest}) hold, and this corresponds to the two parameter family of exact solutions of the Einstein equations
\be\label{realmetric}
ds^2=-\psi_{0}^{-2}{\mathcal B}_a(\psi_0\xi)d\bar{t}^2+\frac{1}{{\mathcal A}_a(\psi_0\xi)}d\bar{r}^2+\bar{r}^2d\Omega^2,
\ee
where
$$
\psi_{0}^{-2}{\mathcal B}_a(\psi_0\xi)\equiv B(\xi),
$$
and $B(\xi)$ is defined in (\ref{ABest}), so that ${\mathcal A}_a(\cdot)$ and ${\mathcal B}_a(\cdot)$ are independent of $\psi_0$.
\end{Theorem}
As a first comment, note that (\ref{ABest}) implies that
\be
\sqrt{AB}\equiv\sqrt{A(\xi)B(\xi)}=\frac{1}{\psi_0}\left\{1-\frac{a(1-a)}{32}\psi_0^4\xi^4\right\}+O(1)|a-1|\xi^6.\label{sqrtABest}
\ee
Now $\sqrt{AB}$ is the factor that converts invariant velocity $v$, $0\leq v<1,$ over to coordinate velocity $\bar{u}^1/\bar{u}^0$, where $\bar{u}^0$ and $\bar{u}^1$ are the time and radial components of the $4$-velocity of a particle moving at speed $v$ in SSC coordinates, respectively.  That is,  
$$
\frac{d\bar{r}}{d\bar{t}}=\frac{\bar{u}^1}{\bar{u}^0}=\sqrt{AB}\,v.
$$
Thus (\ref{sqrtABest}) tells us that near the center $\xi=0,$  time dilation when $a\neq1$ is remarkably close to time dilation in the Standard Model.

Note also that  (\ref{vest}), (\ref{Hest}) imply $$G\equiv G(\xi)=\psi_0\xi+O(1)|a-1|\xi^5$$ and $$AB=\frac{1}{\psi_0^2}+O(1)|a-1|\xi^4,$$  but they are not sufficient to determine the fourth and fifth order terms given in (\ref{Gest}) and (\ref{ABest}), respectively.  For example,  (\ref{vest}) and (\ref{Hest}) with $G\equiv Hv$ give
\vspace{.0001cm}
\be
G\equiv Hv&=&H_1v_1\left(1-\frac{1-a^2}{2H_1}\psi_0^2\xi^2+O(1)|a-1|\xi^4\right)\nonumber\\
&&\ \ \ \ \ \ \ \ \ \ \ \ \ \ \ \ \ \ \ \ \ \ \ \ \times\left(1+\frac{1-a^2}{8v_1}\psi_0^3\xi^3+O(1)|a-1|\xi^4\right)\nonumber\\
&=&G_1\left(1+O(1)|a-1|\xi^4\right),\label{Honevone}
\ee
where we used $H_1\equiv H_1(\xi)=2+O(1)\xi^2$ and $v_1\equiv v_1(\xi)=\frac{\psi\xi}{2}+O(1)\xi^3$ to see the canceling of the second order terms in the parentheses.  But (\ref{Honevone}) does not give the precise form of the fifth order terms in (\ref{Gest}).  Note also that (\ref{ABest}) implies the useful relation
\be
B=\frac{1}{\psi_0^2 A}\left\{1-\frac{a(1-a)}{16}\psi_0^4\xi^4\right\}+O(1)|a-1|\xi^6.\label{BBest}\\
\ee
In a different direction, note that (\ref{vest}) together with 
\be
v_1(\xi)=\frac{\psi_0\xi}{2}(1+v_1^2)=\frac{1}{2}\psi_0\xi+\frac{1}{8}\psi_0^3\xi^3+O(1)\xi^5\label{vone}
\ee
(c.f. (\ref{SM})), gives the first two non-trivial terms in $v(\xi)$ and $A(\xi)$ as
\be
v(\xi)=\frac{1}{2}\psi_0\xi+\frac{(2-a^2)}{8}\psi_0^3\xi^3+O(1)\xi^5,\label{vbyxi}
\ee
\be
A(\xi)=1-\frac{a^2\psi_0^2\xi^2}{4}+\frac{a^2(1-3a^2)}{16}\psi_0^4\xi^4+O(1)\xi^6.\label{Abyxi}\\
\ee
Also using $G_1=\psi_0\xi$ together with (\ref{vone}) gives
\be
H_1(\xi)\equiv\frac{G_1}{v_1}=2-\frac{1}{2}\psi_0^2\xi^2+O(1)\xi^4,\label{Hone}
\ee
and using this together with (\ref{Gest}) in (\ref{Hest}) gives
\be
H(\xi)=2-\frac{(2-a^2)}{2}\psi_0^2\xi^2+O(1)\xi^4,\label{Hbyxi}
\ee
but the $O(1)$'s in (\ref{vbyxi})-(\ref{Hbyxi}) do not vanish when $a=1.$  

Note finally that re-scaling  $\psi_0$ corresponds to re-scaling time in the nonlinear system (\ref{ODE1})-(\ref{ODE3}), and this corresponds to translation in the parameter $s$ in (\ref{Usystem}), an invariance of solutions of autonomous ODE's.  Changes in the parameter $a,$ however, correspond to real changes in the underlying spacetime.  

To complete the picture, the following corollary of Theorem \ref{theoremME} gives asymptotic formulas for the density, (c.f. (\ref{constraint1}):

\begin{Corollary}
When $a=1$, the density $\rho\equiv\rho_1$ of the Standard Model is given exactly by
\be
\kappa\rho_1=\frac{\phi_1(\xi)}{\bar{r}^2},
\ee
\be
\phi_1(\xi)=\frac{3}{2}(1-v_1^2)v_1^2,\label{density1}
\ee
where $v_1\equiv v_1(\xi)$ is the Standard Model velocity, c.f. (\ref{SM}).  When $a\neq1$, the density satisfies the asymptotic expression

\be
\kappa \rho=\frac{\phi(\xi)}{\bar{r}^2},\label{density0}
\ee
\be
\phi(\xi)=\phi_1(\xi)-\frac{3}{2}(1-a^2)v_1^2(\xi)+\frac{3}{2}(4a^2-1)(1-a^2)v_1^4(\xi)+O(1)|a-1|\xi^6.\nonumber\\\label{density}
\ee
\end{Corollary}

\noindent{\bf Proof:}  For (\ref{density1}), note that from (\ref{constraint1})
\be
\phi_1(\xi)=\frac{3(1+v_1^2)(1-A_1)H_1}{2H_1+(1+v_1^2)H_1-4}=\frac{3}{2}(1+v_1^2)(1-A_1),\nonumber
\ee
where we use the Standard Model identity
$$
(1+v_1^2)H_1-4=0.
$$
To derive (\ref{density}), start from (\ref{constraint1}) to get
\be
\phi(\xi)=\frac{3(1+v^2)(1-A)}{2\left(1+\frac{(1+v^2)H-4}{2H}\right)},\label{density01}
\ee
write
{\small\be
&\frac{3}{2}(1+v^2)(1-A)=\phi_1(\xi)+\frac{3}{2}\left\{-(1+v_1^2)(A-A_1)+(v^2-v_1^2)(1-A)\right\},&\nonumber\\\label{density2}
\ee}
and use  (\ref{Hest}) to estimate
\be
\frac{(1+v^2)H-4}{2H}=-\frac{1-a^2}{2}v_1^2+O(1)|a-1|\xi^4.
\label{density3}
\ee
 Putting (\ref{density2}) and (\ref{density3}) into (\ref{density01}) and using (\ref{density})
 and 
 $$(v^2-v_1^2)(1-A)=O(1)|a-1|\xi^6$$
 yields
 \be
&\phi(\xi)=\left\{\phi_1(\xi)-\frac{3}{2}(1+v_1^2)(A-A_1)\right\}\left(1+\frac{1-a^2}{2}v_1^2\right)+O(1)|a-1|\xi^6&\nonumber\\
&=\phi_1(\xi)-\frac{3}{2}(1+v_1^2)(A-A_1)+\phi_1(\xi)\frac{1-a^2}{2}v_1^2\ \ \label{density4}&\\
&\ \ \ \ \ \ \ \ \ \ \ -\frac{3}{2}(1+v_1^2)(A-A_1)\frac{(1-a^2)}{2}v_1^2+O(1)|a-1|\xi^6,&\nonumber
\ee
where by (\ref{Aest}) the three terms in (\ref{density4}) can be estimated by
 \be
&-(1+v_1^2)(A-A_1)=A-A_1+v_1^2(A-A_1)\ \ \ \ \ \ \ \ \ \ \ \ \ \ \ \ \ \ \ \ \ \ \ \ \ \ \ \ \ \ \ \ \ \ \ \ \ \ \ &\nonumber\\
&\ \ \ \ \ \ \ =(a^2-1)v_1^2+3a^2(1-a^2)v_1^4+(a^2-1)v_1^4+O(1)|a-1|\xi^6&\nonumber\\
&\ \ =-(1-a^2)v_1^2+(1-a^2)(3a^2-1)v_1^4+O(1)|a-1|\xi^6,&\nonumber\\\nonumber\\
&\phi_1(\xi)\frac{1-a^2}{2}v_1^2=\frac{3}{4}(1-a^2)v_1^4+O(1)|a-1|\xi^6,\ \ \ \ \ \ \ \ \ \ \ \ \ \ \ \ \ \ \ \ \ \ \ \ \ \ \ \ \ \ \ \ \ \ \ \ \ &\nonumber\\\nonumber\\
&-\frac{3}{2}(1+v_1^2)(A-A_1)\frac{(1-a^2)}{2}v_1^2=-\frac{3}{4}(1-a^2)^2v_1^4+O(1)|a-1|\xi^6.\ \ \ \ \ \ \ \ \ \ \ \ \ \ \ &\nonumber
 \ee
 Substituting these into (\ref{density4}) and collecting powers of $v_1$ yeilds (\ref{density}). $\Box$
 \vspace{.2cm}

Our strategy in the proof of Theorem  \ref{theoremME} is to first prove the theorem in the simpler case when we assume the $O(1)|a-1|$ terms in (\ref{vest})-(\ref{ABest}) are only $O(1)$ as $\xi\rightarrow0$.  This is simpler because we do not need to assume continuity or smoothness of the function $O(1)$ as $\xi\rightarrow0.$   This is the setting for Lemmas \ref{lemma1}- \ref{lemma6} below.   In the final lemma, Lemma \ref{lemma7} below, we will argue for the smoothness of these $O(1)$ terms in the limit $\xi\rightarrow0$, sufficient to bootstrap from $O(1)$ to $O(1)|a-1|.$ 

To start, recall that the invariant manifold ${\mathcal M}$ for the nonlinear system (\ref{Heqn}) is tangent to the invariant manifold ${\mathcal M}_0$ of the linearized system (\ref{lineqn}), at the rest point $V_0,$ so we can conclude that ${\mathcal M}$ is normal to the vector $(0,0,1)$ in $(v,A,H)$ space at $V_0,$ c.f. (\ref{linearsolns}).  Moreover, since solutions in ${\mathcal M}$ are associated with eigenvalues $\lambda=1,2,$ it follows that except for the unique orbit associated with the unstable manifold for eigenvalue $\lambda=2,$ (orbit tangent to $(0,1,0)$ at $V_0$), all solutions in ${\mathcal M}$ come into the rest point $V_0$, in backward time, tangent to the linearized solutions of the strongest eigenvalue, (smallest in {\it backward} time), $\lambda=1.$  Thus ${\mathcal M}^*$ is the set of solution trajectories of $U_s=F(U)$ that come into $V_0$ tangent to $(1,0,0).$    Our starting point in the proof of Theorem \ref{theoremME} is thus the following corollary of the Invariant Manifold Theorem:

\begin{Lemma}\label{invman}\label{lemma1}
Let $V(\xi)=(v(\xi),A(\xi),H(\xi))$ denote any solution of (\ref{Heqn}) in ${\mathcal M}^*$, so that $\lim_{\xi\rightarrow0}V(\xi)=V_0.$  Then $V(\xi)$ is a smooth function away from $\xi=0,$ and
\be
H(\xi)=2+O(1)\xi^2,\label{lemma1-1}\label{l1-1}
\ee
and there exists $\psi_0\neq0$ such that
\be
\left\|V(\xi)-\left[\left(\begin{array}[pos]{c}
0\\1\\2
 \end{array}\right)+\frac{\psi_0\xi}{2}\left(\begin{array}[pos]{c}
1\\0\\0
 \end{array}\right)\right]\right\|=w(\xi)\xi,\label{w}\label{lemma1-2}\label{l1-2}
\ee
where $O(1)$ denotes a function smooth for $\xi>0$ and bounded as $\xi\rightarrow0$, and $w(\xi)$ satisfies
\be
lim_{\xi\rightarrow0}w(\xi)=0.\label{lemma1-3}
\ee
\end{Lemma}  
Note that the Standard Model (\ref{SMH}) satisfies (\ref{l1-1}), (\ref{l1-2}) with the much stronger estimate $w(\xi)=O(1)\xi^2.$

We now prove a number of lemmas that improve on Lemma \ref{invman}.  The main technical result we use is the following:

\begin{Lemma}\label{lemmaprelim}
Let $u(\xi)$ be a solution of the scalar ODE
\be
\xi u_{\xi}=f(u,\xi),\label{basicODE}
\ee
such that $u(\xi)$ is smooth for $\xi>0,$ and bounded as $\xi\rightarrow0.$  Then:
\vspace{.2cm}

\noindent {\bf (a)}   If $f$ is a smooth function such that $f(u,0)$ has a finite number of non-degenerate rest points, then $u(\xi)$ must tend to a rest point of $f(u,0).$  That is, there exists a finite $u_0\in{\bf R}$ such that $f(0,u_0)=0,$ and $\lim_{\xi\rightarrow0}u(\xi)=u_0.$ 
\vspace{.2cm} 

\noindent {\bf (b)}   If equation (\ref{basicODE}) is of the form
\be
\xi u'(\xi)=\alpha u(\xi)(1+g(\xi)\xi^n),\label{ueqn}
\ee
where $n>0$, $\alpha>0$, and $g$ is a function continuous for $\xi>0$ and bounded in the limit $\xi\rightarrow0,$  then
\be
u(\xi)=u_0\xi^{\alpha}(1+O(1)\xi^n)\label{eqnb1}
\ee
as $\xi\rightarrow0,$ and the $sign\left\{u(\xi)\right\}$ is constant.
\vspace{.2cm}

\noindent {\bf (c)}  If equation (\ref{basicODE}) is of the form
\be
\xi u'(\xi)=-\alpha u(\xi)+\beta+g(\xi)\xi^n,\label{ueqnc1}
\ee
where $n>0$, $\alpha>0$, $\beta$ is real, and $g$ is a function continuous for $\xi>0$ and bounded in the limit $\xi\rightarrow0,$  then
\be
u(\xi)=\frac{\beta}{\alpha}+O(1)\xi^n\label{ueqnc}
\ee
as $\xi\rightarrow0.$
\end{Lemma}

\noindent {\bf Proof of (a):}  Note first that {\bf (a)} holds with $f(u,\xi)$ replaced by $f(u,0)$ in (\ref{basicODE}), because the substitution $\xi=e^s$ would transform the problem into
\be
\frac{d}{ds}u=f(u,0),\nonumber
\ee
a standard scalar ODE for which it is well known that solutions tend to $\infty$ or to rest points as $s\rightarrow-\infty$, $\xi\rightarrow0$.  Since $f(u,\xi)$ is continuous in $\xi,$ it follows that for $\epsilon$ sufficiently small, the function $f(u,\epsilon)$ has a finite number of non-degenerate rest points that are small perturbations of the rest points of $f(u,0).$  From this it follows that solutions of 
\be
\frac{d}{ds}u=f(u,\xi),\nonumber
\ee
must tend to $\infty$ or a rest point in the limit $s\rightarrow-\infty,$ $\xi\rightarrow0$ as well. 
\vspace{.2cm}

\noindent{\bf Proof of (b):}  Since $u(\xi)$ is smooth away from $\xi=0,$ and bounded as $\xi\rightarrow0,$ it follows by {\bf (a)} that $u(\xi)$ must tend to a rest point of the right hand side of (\ref{ueqn}) as $\xi\rightarrow0$, so 
\be
\lim_{\xi\rightarrow0}u(\xi)=0.\label{wofzerob}
\ee
Define $y(\xi)$ for $\xi>0$ by
\be
u(\xi)=y(\xi)\xi^{\alpha},\label{wofzero2b}
\ee
so that (\ref{wofzerob}) becomes
\be
\xi u'(\xi)\equiv \xi y'(\xi)\xi^{\alpha}+\alpha y(\xi)\xi^{\alpha}=\alpha y(\xi)\xi^{\alpha}(1+O(1)\xi^n),\label{wofzero3b}
\ee
which reduces to
\be
\frac{y'}{y}=O(1)\xi^{n-1}.\label{wofzero4b}
\ee
Integratinging (\ref{wofzero4b}) from $\xi$ to $\bar{\xi}$, $0<\xi<\bar{\xi}$, gives
\be
y(\xi)=y(\bar{\xi})~e^{O(1)(\bar{\xi}^n-\xi^n)},\label{aofxib}
\ee 
from which we conclude that $y(\xi)\rightarrow y_0\equiv u_0$ is bounded and finite as $\xi\rightarrow0.$  Thus taking $\xi\rightarrow0$ in (\ref{aofxib}) and replacing $\bar{\xi}$ by $\xi$ gives
\be
y(\xi)=u_0~e^{O(1)\xi^n}=u_0\left(1+O(1)\xi^n\right).\label{aofxib2}
\ee 
Putting (\ref{aofxib2}) into (\ref{wofzero2b}) gives (\ref{ueqnc}) as claimed.   From (\ref{aofxib}), the sign of $y(\xi)$ is either everywhere non-zero or identically zero, and so $sign\left\{u(\xi)\right\}$ is constant.

\vspace{.2cm}
\noindent {\bf Proof of (c):}    Since $u(\xi)$ is smooth away from $\xi=0,$ and bounded as $\xi\rightarrow0,$ it follows again by {\bf (a)} that $u(\xi)$ must tend to a rest point of the right hand side of (\ref{ueqnc}) as $\xi\rightarrow0$, so that
\be
\lim_{\xi\rightarrow0}u(\xi)=\frac{\beta}{\alpha}\equiv u_0.\label{wofzero}
\ee
Using (\ref{wofzero}) in (\ref{ueqnc}) then gives
\be
\xi(u-u_0)_{\xi}=-\alpha(u-u_0)+O(1)\xi^n.\label{wofzero1}
\ee
But the only way (\ref{wofzero1}) can be satisfied by a function $u(\xi)$ smooth for $\xi>0$, and {\em bounded} as $\xi\rightarrow0$, is if  
\be
w(\xi)=w_0+O(1)\xi^n.\label{wofzero2}
\ee
To see this, define $y(\xi)$ for $\xi>0$ by
\be
u-u_0=y(\xi)\xi^{-\alpha},\label{wofzero2a}
\ee
so that (\ref{wofzero2}) becomes
\be
\xi(u-u_0)_{\xi}\equiv \xi y'(\xi)\xi^{-\alpha}-\alpha y(\xi)\xi^{-\alpha}=-\alpha y(\xi)\xi^{-\alpha}+O(1)\xi^n\label{wofzero3}
\ee
which reduces to
\be
y'(\xi)=O(1)\xi^{n+\alpha-1}.\label{wofzero4}
\ee
Integrating (\ref{wofzero4}) from $0$ to $\xi$ then gives
$$
y(\xi)=y(0)+O(1)\xi^{n+\alpha},
$$
which in (\ref{wofzero2a}) gives
\be
u(\xi)=u_0+y(0)\xi^{-\alpha}+O(1)\xi^n.\label{wofzero2bsecond}
\ee
Since $u$ is bounded as $\xi\rightarrow0$, it follows that $y(0)=0$ in (\ref{wofzero2bsecond}), in which case (\ref{wofzero2bsecond}) gives (\ref{ueqnc}) .  $\Box$
\vspace{.2cm}

Now define
\be
A_a(\xi)\equiv 1-a^2v_1(\xi)\label{Afirst},
\ee 
so that 
\be
A_a(\xi)\equiv 1-\frac{a^2\psi_0^2\xi^2}{4}+O(1)\xi^4. \label{Asecond},
\ee 
and let $V_a(\xi)$ denote the approximate solution
\be
V_a(\xi)\equiv\left(\begin{array}[pos]{c}
v_a(\xi)\\A_a(\xi)\\H_a(\xi)
 \end{array}\right)=\left(\begin{array}[pos]{c}
v_1(\xi)\\1-\frac{a^2\psi_0^2v_1(\xi)}{4}\\\frac{2}{1+v_1(\xi)}
 \end{array}\right),
\ee
so that 
\be
v_a(\xi)&=&v_1(\xi)\label{va}\\
1-A_a(\xi)&=&a^2\left\{1-A_1(\xi)\right\},\\
H_a(\xi)&=&H_1(\xi),\label{Ha}
\ee 
and hence $V_a(\xi)$ differs from the exact Standard Model solution $V_1(\xi)$ only in the $A$-component.   Also note that as a function of $\xi,$
\be
V_a(\xi)\equiv\left(\begin{array}[pos]{c}
1\\2\\0
 \end{array}\right)+\left(\begin{array}[pos]{c}
\frac{\psi_0}{2}\\-\frac{a^2\psi_0^2\xi^2}{4}\\-\frac{\psi_0^2\xi^2}{2}
 \end{array}\right)+O(1)\xi^3,\label{vinxi}
\ee
but the $O(1)$ in (\ref{vinxi}) does not vanish when $a=1.$

\begin{Lemma}\label{lemma2}
Let $V(\xi)=(v(\xi),A(\xi),H(\xi))$ be any solution of (\ref{Heqn}) in ${\mathcal M}^*.$  Then there exists $a^2\in {\bf R}$ such that
\be
A(\xi)=A_a(\xi)+O(1)\xi^4.\label{lemma2-1}\label{l2-1}\label{l2-2}
\ee
(Again, we allow $\psi_0$ and $a^2$ to denote any real numbers when discussing all solutions in ${\mathcal M},$ but restrict to positive values when restricting to cosmological mODEls that perturb the Standard Model near $\xi=0$.)
\end{Lemma}

\noindent {\bf Proof:}    To verify (\ref{l2-2}), consider equation (\ref{ODE1}):
\be
\xi A_{\xi}=-\frac{(1-A)v}{(3+v^2)Hv-4v}=-\frac{4(1-A)}{(3+v^2)H-4}.\label{ODE1again}
\ee
Now since $V(\xi)$ exactly solves (\ref{ODE1again}), substituting (\ref{lemma1-1}) and (\ref{lemma1-2}) into (\ref{ODE1again}) gives
\be
\xi (1-A)_{\xi}=\frac{4(1-A)}{2(1+O(1)\xi^2)}=2(1-A)(1+O(1)\xi^2),\label{l2-3}
\ee
where again $O(1)$ is bounded as $\xi\rightarrow0.$  Equation (\ref{l2-3}) is an equation of form (\ref{eqnb1}) with $u=1-A,$ $\alpha=2,$ $n=2,$, so part {\bf (b)} of Lemma \ref{lemmaprelim} implies that

\be
1-A(\xi)=\frac{a^2\psi_0^2\xi^2}{4}+O(1)\xi^4,
\ee
where for convenience take,
\be
u_0=\frac{a^2\psi_0^2}{4},\nonumber
\ee
(no sign assumed on $a^2$).
Now substituting
$$
v_1(\xi)^2=\frac{\psi_0^2}{4}\xi^2+O(1)\xi^4,
$$
we obtain
$$
A(\xi)=1-a^2v_1(\xi)^2+O(1)\xi^4=A_a(\xi)+O(1)\xi^4,
$$
as claimed in (\ref{l2-2}).   $\Box$

The next lemma improves (\ref{l1-2}) for the function $v(\xi)$:

\begin{Lemma}\label{lemma3}
Let $V(\xi)=(v(\xi),A(\xi),H(\xi))$ be any solution of (\ref{Heqn}) in ${\mathcal M}^*$ such that (\ref{l2-1}) of Lemma \ref{lemma2} holds.  Then 
\be
v(\xi)=v_1(\xi)+O(1)\xi^3.\label{l3-1}\label{l3-2}
\ee
\end{Lemma}
Note that since $v_1(\xi)=\frac{\psi}{2}\xi+O(1)\xi^3,$ it follows from (\ref{l3-1}) that 
\be
v(\xi)=\frac{\psi}{2}\xi+O(1)\xi^3,\label{l3-2aa}
\ee
but in this case $O(1)$ does not vanish when $a=1.$

\noindent {\bf Proof:}  By Lemma \ref{lemma2} it suffices to prove that there exists $a^2\in{\bf R}$ such that $V(\xi)=(v(\xi),A(\xi),H(\xi))$ with $A(\xi)=A_a(\xi)+O(1)\xi^4$, such that (\ref{l3-2}) holds.
For this, consider equation (\ref{ODE1H}) in the form
\begin{eqnarray}
\xi v_{\xi}&=&-\left(\frac{1-v^2}{2\left\{\cdot\right\}_D}\right)\left(\left[(3+v^2)H-4\right]v+\frac{4\left(\frac{1-A}{A}\right)\left\{\cdot\right\}_N^*v}{(3+v^2)H-4}\right),\nonumber\\
\label{l3-3}
\ee
and observe that (\ref{l1-2}) implies that
$$
v(\xi)=O(1)\xi,
$$
as $\xi\rightarrow0.$  Using this together with (\ref{l1-1}) and (\ref{l2-1}), and applying them in (\ref{ODE1H1})-(\ref{ODE1H3}) gives
\be
H&=&2+O(1)\xi^2,\label{l3-4}\\
1-A&=&O(1)\xi^2,\label{l3-5}\\
\left\{\cdot\right\}_D&=&-1+O(1)v^2,\label{l3-6}\\
\left\{\cdot\right\}_N^*&=&O(1)v^2.\label{l3-7}
\ee
Using (\ref{l3-4})-(\ref{l3-7}) in (\ref{l3-3}) gives
\be
\xi v_{\xi}=-v+O(1)\xi^3.\label{l3-8}
\ee
To estimate (\ref{l3-8}), define $\alpha(\xi)$ by
\be
v_{\xi}=\alpha(\xi)\xi.\label{l3-9}
\ee
Putting (\ref{l3-9}) into (\ref{l3-8}) gives
\be
\alpha'(\xi)=O(1)\xi,\label{l3-10}
\ee
which integrates to
\be
\alpha({\xi})=\alpha(0)+O(1)\xi^2.\label{l3-11}
\ee
Substituting (\ref{l3-11}) into (\ref{l3-9}) and setting $\alpha(0)=\frac{\psi_0}{2}$ gives
$$
v(\xi)=\frac{\psi_0}{2}\xi+O(1)\xi^3,
$$
which, since $\frac{\psi_0}{2}\xi=v_1(\xi)+O(1)\xi^3$ as well, gives (\ref{l3-2}) as claimed.  $\Box$

We now use Lemmas \ref{lemma2} and \ref{lemma3} to improve the estimate (\ref{l1-1}) for $H(\xi):$

\begin{Lemma}\label{lemma4}
Let $V(\xi)=(v(\xi),A(\xi),H(\xi))$ be any solution of (\ref{Heqn}) in ${\mathcal M}^*$ satisfying (\ref{l2-1}) of Lemma \ref{lemma2} and (\ref{l3-1}) of Lemma \ref{lemma3}.  Then 
\be
G(\xi)&=&\psi_0\xi+O(1)\xi^5,\label{l4-1}\label{l4-5}\\
H(\xi)&=&\frac{\psi_0\xi}{v(\xi)}+O(1)\xi^4,\label{l4-2}\\
\sqrt{AB}&=&\frac{1}{\psi_0}+O(1)\xi^4,\label{l4-3}
\ee
where the constants $O(1)$ vanish on the Standard Model $a=1.$
\end{Lemma}

\noindent {\bf Proof:}  Note first that (\ref{l4-2}) and (\ref{l4-3}) follow from (\ref{l4-1}) via the identities $G=Hv$ and $\sqrt{AB}=G\xi,$ and thus since (\ref{l4-1}) is exact, ($G_1(\xi)=\psi_0\xi$), on the Standard Model, it follows that all constants $O(1)$ in (\ref{l4-1})-(\ref{l4-3}) are zero when $a=1.$ .

To verify (\ref{l4-5}), consider equations (\ref{ODE2}) and (\ref{ODE3H}) in the form 
\be
\xi G_{\xi}&=&-G\left\{\left(\frac{1-A}{A}\right)\frac{2(1+v^2)H-4}{(3+v^2)H-4}-1\right\}.\label{l4-6}\\
\ee
Note now that by (\ref{l3-1}),
\be
v(\xi)^2=v_1^2(\xi)+O(1)\xi^4,\label{l4-7}
\ee
and that on the Standard Model $V_1(\xi)$ we have the identity

\be
2(1+v_1^2)H_1-4=0,\label{l4-8}
\ee
with
\be
H_1=\frac{2}{1+v_1^2},\label{l4-9}
\ee
c.f. (\ref{SM}), (\ref{SMH}).  Also, by (\ref{l1-1}) we have
\be
H=H_1+O(1)\xi^2,\label{l4-10}
\ee
and by (\ref{l2-1})
\be
1-A=O(\xi^2).\label{l4-11}
\ee
Substituting (\ref{l4-7})-(\ref{l4-11}) into (\ref{l4-6}) gives
\be
\xi G_{\xi}=G\left\{1+O(\xi^4)\right\}.\label{l4-12}
\ee
Now putting the definition $G\equiv\frac{\xi}{\sqrt{AB}}$ into (\ref{l4-12}) gives
\be
\left(\sqrt{AB}\right)_{\xi}=O(1)\xi^3,\label{l4-13}
\ee
and integrating this from $0$ to $\xi$, (note that $\sqrt{AB}=\frac{\xi}{Hv}\rightarrow\frac{1}{\psi_0}$ as $\xi\rightarrow0$), gives

\be
\sqrt{AB}=\frac{1}{\psi_0}\left\{1+O(1)\xi^4)\right\},\label{l4-14}
\ee
which also gives
\be
AB=\frac{1}{\psi_0^2}\left\{1+O(1)\xi^4)\right\},\label{l4-15}
\ee
and
\be
\frac{1}{\sqrt{AB}}=\psi_0\left\{1+O(1)\xi^4)\right\}.\label{l4-16}
\ee
Now
$$
H=\frac{\xi}{v\sqrt{AB}},
$$
so using (\ref{l4-16}) in this gives
\be
H=\frac{\psi_0\xi}{v}\left\{1+O(1)\xi^4\right\}.\label{l4-17}
\ee
Thus since 
$$
G=Hv,
$$
and $v=\frac{\psi_0}{2}\xi+O(1)\xi^3$,
we conclude from (\ref{l4-17}) that
$$
G(\xi)=\psi_0\xi+O(1)\xi^5,
$$
as claimed in (\ref{l4-5}).  $\Box$

We now can give a proof of estimate (\ref{vest}), which gives the dependence of $v(\xi)$ at the third order in $\xi$, a refinement of (\ref{l3-2aa}):

\begin{Lemma}\label{lemma5}
Let $V(\xi)=(v(\xi),A(\xi),H(\xi))$ be any solution of (\ref{Heqn}) in ${\mathcal M}^*.$  Then there exists $a$ and $\psi_0$ such that
\be
v(\xi)=v_1(\xi)+\frac{1-a^2}{8}\psi_0^3\xi^3+O(1)\xi^5.\label{l5-1}
\ee
\end{Lemma}

\noindent {\bf Proof:}  Consider again equation (\ref{ODE1H}) in the form
\begin{eqnarray}
\xi v_{\xi}&=&-\left(\frac{1-v^2}{2\left\{\cdot\right\}_D}\right)\left(\left[(3+v^2)H-4\right]v+\frac{4\left(\frac{1-A}{A}\right)\left\{\cdot\right\}_N^*v}{(3+v^2)H-4}\right).\nonumber\\
\label{l5-2}
\ee
Since we now have 
\be
H&=&H_1+O(1)\xi^4\label{l5-3}\\
v&=&v_1+O(1)\xi^3\label{l5-4}\\
v^2&=&v_1^2+O(1)\xi^4,\label{l5-5}\\
\ee
we can  use these in (\ref{ODE1H1})-(\ref{ODE1H3}) to get

\be
\left\{\cdot\right\}_N^*&\equiv&\left\{-2+2(3-v^2)H-(3-v^4)H^2\right\}\nonumber\\
&=&\left\{-2+2(3-v_1^2)H_1-(3-v_1^4)H_1^2\right\}+O(1)\xi^4\nonumber\\
&=&\left\{1\right\}_N^*+O(1)\xi^4,\label{l5-6}\\
&=&-2+O(1)\xi^2,\label{l5-6b}\\
\left\{\cdot\right\}_D&\equiv&\left\{(3v^2-1)-4Hv^2+(3-v^2)H^2v^2\right\}\nonumber\\
&=&\left\{(3v_1^2-1)-4H_1v_1^2+(3-v_1^2H_1^2v_1^2\right\}+O(1)\xi^4\nonumber\\
&=&\left\{1\right\}_D+O(1)\xi^4,\label{l5-7}\\
&=&-1+O(1)\xi^2,\label{l5-7b}
\ee
where 
\begin{eqnarray}
\left\{1\right\}_N^*&\equiv&\left\{1\right\}_N^*(\xi)\nonumber\\
\left\{1\right\}_D&\equiv&\left\{1\right\}_D(\xi)\nonumber
\end{eqnarray}
denote $\left\{\cdot\right\}_N^*$, $\left\{\cdot\right\}_D$ with $(v_1(\xi),H_1(\xi))$ substituted for $(v,H)$, respectively.  Since $v$ is $O(1)\xi$, estimates (\ref{l5-3}), (\ref{l5-5}), (\ref{l5-6}) and (\ref{l5-7}) imply that we can replace $H$ by $H_1$, $v$ by $v_1$, $\left\{\cdot\right\}_N^*$ by  $\left\{1\right\}_N^*$ and $\left\{\cdot\right\}_D$ by $\left\{1\right\}_D$ in (\ref{l5-2}) and incur an error no greater than $O(1)\xi^5.$  That is, 

\begin{eqnarray}
\xi v_{\xi}&=&-\left(\frac{1-v_1^2}{2\left\{1\right\}_D}\right)\left[(3+v_1^2)H_1-4\right]v\label{l5-8}\\
&&\ \ \ \ \ \ \ \ \ \ +\left(\frac{1-v_1^2}{2\left\{1\right\}_D}\right)\frac{4\left(\frac{1-A}{A}\right)\left\{1\right\}_N^*v_1}{(3+v_1^2)H_1-4}+O(1)\xi^5.\nonumber
\ee
Note that the first two terms on the RHS of (\ref{l5-8}) are just $f(v_1,A_1,H_1)$ with $v$ in place of $v_1$ in the last place of the first term, and $A$ in place of $A_1$ in the second term.    Substituting
$$v=v_1+(v-v_1),$$
and
$$
\frac{1-A}{A}=\frac{1-A_1}{A_1}+\left[\frac{1-A}{A}-\frac{1-A_1}{A_1}\right]=\frac{1-A_1}{A_1}+\left[\frac{A_1-A}{A_1A}\right],
$$
into (\ref{l5-8}), gives
\begin{eqnarray}
\xi v_{\xi}&=&f(v_1,A_1,H_1)-\left\{\frac{1-v_1^2}{2\left\{1\right\}_D}\left[(3+v_1^2)H_1-4\right]\right\}(v-v_1)\nonumber\\
&&\ \ \ \ \ \ \ \ \ \ \ \ \ -\left\{\left(\frac{1-v_1^2}{2\left\{1\right\}_D}\right)\frac{4\left\{1\right\}_N^*v_1}{(3+v_1^2)H_1-4}\right\}\left[\frac{A_1-A}{A_1A}\right]\label{l5-9a}\\
&&\ \ \ \ \ \ \ \ \ \ \ \ \ \ \ \ \ \ \ \ \ \ \ \ \ \ \ \ \ \ \ \ \ \ \ \ \ \ \ \ \ \ \ \ \ \ \ \ \ \ \ \ \ \ \ \ \ \ \ \ \ \ \ \ \ \ \ +O(1)\xi^5,\nonumber
\ee
Now by (\ref{l2-1}),   
$$
1-A=a^2(1-A_1)+O(1)\xi^4,
$$
so
$$
A-A_1=\left(1-a^2(1-A_1)A_1\right)+O(1)\xi^4,\label{l2-13}
$$ 
and thus
\be
\frac{A_1-A}{A_1A}&=&\frac{(1-a^2)(1-A_1)}{\left[1-a^2(1-A_1)\right]A_1}+O(1)\xi^4\nonumber\\
&=&\frac{(1-a^2)(1-A_1)}{(1-a^2v_1^2)(1-v_1^2)}+O(1)\xi^4\nonumber\\
&=&(1-a^2)v_1^2+O(1)\xi^4.\label{l5-14}
\ee
Using this in (\ref{l5-9a}) gives

\begin{eqnarray}
\xi v_{\xi}&=&f(v_1,A_1,H_1)-\left\{\frac{1-v_1^2}{2\left\{1\right\}_D}\left[(3+v_1^2)H_1-4\right]\right\}_{I}(v-v_1)\nonumber\\
&&\ \ \ \ \ \ \ \ \ \ \ \ \ -(1-a^2)\left\{\left(\frac{1-v_1^2}{2\left\{1\right\}_D}\right)\frac{4\left\{1\right\}_N^*}{(3+v_1^2)H_1-4}\right\}_{II}v_1^3\label{l5-9}\\
&&\ \ \ \ \ \ \ \ \ \ \ \ \ \ \ \ \ \ \ \ \ \ \ \ \ \ \ \ \ \ \ \ \ \ \ \ \ \ \ \ \ \ \ \ \ \ \ \ \ \ \ \ \ \ \ \ \ \ \ \ \ \ \ \ \ \ \ +O(1)\xi^5.\nonumber
\ee
and use the notation that
\be
\left\{1\right\}_{I}&\equiv&\left\{1\right\}_{I}(\xi)\nonumber\\
\left\{1\right\}_{II}&\equiv&\left\{1\right\}_{II}(\xi)\nonumber
\ee
denote the two brackets $\left\{\cdot\right\}_{I}$ and $\left\{\cdot\right\}_{II}$ in (\ref{l5-9}), with the $1s$ indicating that both are evaluated on the Standard Model $V_1=V_1(\xi).$  Now since $v-v_1$ beside $\left\{1\right\}_{I}$ and $v_1^3$ beside $\left\{1\right\}_{II}$ are both $O(1)\xi^3,$ it follows that the $v_1^2$ terms in $\left\{1\right\}_{I}$ and $\left\{1\right\}_{II}$, being $O(1)\xi^2$, can be set equal to zero incurring an error no greater than $O(1)\xi^5.$  For the same reason,  $H_1=2+O(1)\xi^2$ can be replaced by $2$ in (\ref{l5-9}) to error $O(1)\xi^5.$  The result then is that
\begin{eqnarray}
\left\{1\right\}_{I}&=&\frac{1}{\left\{1\right\}_D}+O(1)\xi^2=-1+O(1)\xi^2\nonumber\\
\left\{1\right\}_{II}&=&\frac{\left\{1\right\}_N^*}{\left\{1\right\}_D}+O(1)\xi^2=2+O(1)\xi^2,\nonumber
\ee
where we have used (\ref{l5-6b}) and (\ref{l5-7b}).  Putting these in (\ref{l5-9}) gives

\begin{eqnarray}
\xi v_{\xi}&=&f(v_1,A_1,H_1)+(v-v_1)-2(1-a^2)v_1^3+O(1)\xi^5.\label{l5-9b}
\ee

To solve (\ref{l5-9b}) for $v-v_1$, set 
\be
v(\xi)=v_1(\xi)+w(\xi)\xi^3.\label{l5-10}
\ee
Substituting  (\ref{l5-10}) into (\ref{l5-9a}), and using that $v_1$ satisfies 
$$\xi (v_1)_{\xi}=f(v_1,A_1,H_1),$$ 
gives
\be
w'({\xi})\xi^4+3w(\xi)\xi^3=w(\xi)\xi^3+2(a^2-1)v_1^3+O(1)\xi^5,\nonumber
\ee
which reduces to, (using $v_1^3/\xi^3=\psi_0^3/8+O(1)\xi^2$),
\be
\xi w'({\xi})=-2w(\xi)+\frac{a^2-1}{4}+O(1)\xi^2.\label{weqn}
\ee
Equation (\ref{weqn}) is an equation of form (\ref{ueqnc}) with $u=w,$ $n=2$, $\alpha=2$,  and $\beta=\frac{a^2-1}{4}$, so Part ${\bf (c)}$ of Lemma \ref{lemmaprelim} implies
\be
w({\xi})=\frac{a^2-1}{8}+O(1)\xi^2.\label{wlast}
\ee
Putting (\ref{wofzero2}) together with (\ref{weqn}) and (\ref{wofzero}) gives

\be
v(\xi)=v_1(\xi)+\frac{a^2-1}{8}\xi^3+O(1)\xi^5,\nonumber
\ee
as claimed.   $\Box$
\vspace{.2cm}

With the third order dependence of $v$ established in Lemma \ref{lemma5}, we can now establish the dependence of $A$ and $H$ up to sixth and fourth order in $\xi$, as claimed in (\ref{Aest}) and (\ref{Hest}), respectively:

\begin{Lemma}\label{lemma6}
Let $V(\xi)=(v(\xi),A(\xi),H(\xi))$ be any solution of (\ref{Heqn}) in ${\mathcal M}^*.$  Then there exists $a$ and $\psi_0$ such that
\be
A(\xi)&=&A_a(\xi)+\frac{3a^2(1-a^2)}{16}\psi^4\xi^4+O(1)\xi^6,\label{Aest1}\\
H(\xi)&=&H_1(\xi)-\frac{1-a^2}{2}\psi^2\xi^2+O(1)\xi^4,\label{Hest2}
\ee
where $O(1)$ vanishes when $a=1,$ c.f. (\ref{Aest}), (\ref{Hest}) and (\ref{Afirst}).
\end{Lemma}

\vspace{.2cm}
\noindent{\bf Proof:}  We first establish (\ref{Hest2}).   For this, start with (\ref{l4-2}) and use (\ref{l5-1}) together with $H=2+O(1)\xi^2$ to estimate:
\be
H(\xi)&=&\frac{\psi_0\xi}{v}+O(1)\xi^4=\frac{\psi_0\xi}{v_1\left(1+\frac{1-a^2}{8v_1}\psi_0^3\xi^3+O(1)\xi^4\right)}+O(1)\xi^4\nonumber\\
&=&\frac{\psi_0\xi}{v_1}\left(1-\frac{1-a^2}{8v_1}\psi_0^3\xi^3+O(1)\xi^4\right)+O(1)\xi^4\nonumber\\
&=&H_1\left(1-\frac{1-a^2}{8\left(\frac{\psi_0\xi}{2}+O(1)\xi^3\right)}\psi_0^3\xi^3+O(1)\xi^4\right)+O(1)\xi^4\nonumber\\
&=&H_1-\frac{1-a^2}{2}\psi_0^2\xi^2+O(1)\xi^4,\nonumber
\ee
as claimed.  

To establish (\ref{Aest}), assume $A(\xi)$ is a solution (\ref{ODE2H}) written in the alternative form
\be
\xi (1-A)_{\xi}=\frac{4(1-A)}{(3+v^2)H-4}\label{oneminusAeqn}
\ee
satisfying (\ref{l2-2}), and define the function $\bar{A}(\xi)$ by
\be
A(\xi)=A_a(\xi)+\bar{A}\psi_0^4\xi^4.\label{Abar}
\ee
Then using (\ref{Hest}) together with the $v^2=v_1^2+O(1)\xi^4$, $1-A=\frac{a^2\psi_0^2\xi^2}{4}+O(\xi^4)$, and $H_1=2+O(\xi^2)$, gives
\be
\xi (1-A)_{\xi}&=&\frac{4(1-A)}{\left[(3+v_1^2)H_1-4\right](1-\frac{3}{4}(1-a^2)\psi_0^2\xi^2+O(\xi^3))}\nonumber\\\label{oneminusAeqn1}
&=&\frac{4(1-A)}{\left[(3+v_1^2)H_1-4\right]}(1+\frac{3}{4}(1-a^2)\psi_0^2\xi^2+O(\xi^3))\nonumber\\
&=&\frac{4(1-A)}{(3+v_1^2)H_1-4}+\frac{3a^2(1-a^2)}{8}\psi_0^4\xi^4+O(\xi^6)\nonumber\\
&=&\frac{4(1-A_a)}{(3+v_1^2)H_1-4}-2\bar{A}\psi_0^4\xi^4+\frac{3a^2(1-a^2)}{8}\psi_0^4\xi^4+O(\xi^6)\nonumber\\\label{oneminusAeqn2}
\ee
Putting (\ref{Abar}) into (\ref{oneminusAeqn2}) and using
$$
\xi(1-A_a)_{\xi}=\frac{4(1-A_a)}{(3+v_1^2)H_1-4}
$$
gives
\be
-\xi \bar{A}_{\xi}\xi^4+4\bar{A}\xi^4&=&-2\bar{A}\psi_0^4\xi^4+\frac{3a^2(1-a^2)}{8}\psi_0^4\xi^4+O(\xi^6),
\label{Abar1})
\ee
which simplifies to
\be
\xi\bar{A}_{\xi}=-2\bar{A}+\frac{3a^2(1-a^2)}{8}+O(\xi^2).\label{Abar2first}
\ee
Equation (\ref{Abar2first}) is an equation of form (\ref{ueqnc}) with $u=\bar{A},$ $n=2$, $\alpha=2$,  and $\beta=\frac{3a^2(1-a^2)}{8}$, so Part ${\bf (c)}$ of Lemma \ref{lemmaprelim} implies
\be
\bar{A}=\frac{3a^2(1-a^2)}{16}+O(\xi^2).\label{Abar2}
\ee
Putting (\ref{Abar2}) together with (\ref{Abar}) gives
\be
A=A_a+\frac{3a^2(1-a^2)}{16}\psi_0^4\xi^4+O(\xi^6),\nonumber
\ee
as claimed.   $\Box$
\vspace{.2cm}

We now get the dependence of $G$, $\sqrt{AB}$ at orders six and five in $\xi$, $\xi^6$, as claimed respectively in (\ref{Gest}), (\ref{ABest}), these being the first orders at which $G$ and $AB$ diverge from the Standard Model.

\begin{Lemma}\label{lemma6b}
Let $V(\xi)=(v(\xi),A(\xi),H(\xi))$ be any solution of (\ref{Heqn}) in ${\mathcal M}^*$. Then there exists $a$ and $\psi_0$ such that 

\be
G(\xi)&=&\psi_0\xi+\frac{a(1-a^2)}{32}\psi_0^5\xi^5+O(1)\xi^7,\label{l4-5b}\\
AB&=&\frac{1}{\psi_0^2}\left\{1-\frac{a(1-a)}{16}\psi_0^4\xi^4\right\}+O(1)\xi^6,\label{l4-2b}\\
\ee
where the constants $O(1)$ vanish on the Standard Model $a=1.$
\end{Lemma}

Again note that all constants $O(1)$ in (\ref{l4-5b}) and (\ref{l4-3}) are zero on the Standard Model $a=1.$ .

\noindent {\bf Proof:}  To verify (\ref{l4-5b}), use 
$$
v=v_1+\frac{1-a^2}{8}\psi_0^3\xi^3+O(1)\xi^5=v_1\left(1+\frac{1-a^2}{4}\psi_0^2\xi^3+O(1)\xi^4\right)
$$
in 
$$
H=\frac{\psi_0}{v}+O(1)\xi^4,
$$
to obtain
$$
H=\frac{\psi_0\xi}{v_1}\left(1-\frac{1-a^2}{4}\psi_0^2\xi^2+O(1)\xi^4\right),
$$
c.f.   (\ref{l5-1}), (\ref{l4-2}).  Using this together with
$$
v^2=v_1^2+O(1)\xi^4
$$
and 
$$
\frac{(1+v_1^2)\psi_0\xi}{v_2}=2
$$
in (\ref{l4-6}) gives, (c.f. (\ref{SM})), 
\be
\xi G_{\xi}=-G\left\{\left(\frac{1-A}{A}\right)\frac{(1+v_1^2)\psi_0\xi}{v_1}\frac{1-a^2}{4}\psi_0^2\xi^2+O(1)\xi^4+1\right\}\nonumber
\ee
Now using $v_1=\frac{\psi_0\xi}{2}+O(1)\xi^2$, $v_1^2=O(1)\xi^2$ and $1-A=\frac{a^2\xi^2}{4}+O(\xi^4)$ in this we obtain 
\be
\xi G_{\xi}&=&G\left\{\frac{a^2(1-a^2)}{8}\psi_0^4\xi^4+1+O(1)\xi^6.\right\}\label{goodG}
\ee
To integrate (\ref{goodG}) make the change of variables 
\be
G=\frac{\xi}{W},
\ee
so that $W\equiv\sqrt{AB}.$   Using this in (\ref{goodG}) gives
\be
\xi G_{\xi}&=&\xi\left(\frac{\xi}{W}\right)_{\xi}=\frac{\xi}{W}-\frac{\xi^2}{W^2}W_{\xi}\nonumber\\
&=&\frac{\xi}{W}\left\{\frac{a^2(1-a^2)}{8}\psi_0^4\xi^4+1+O(1)\xi^6\right\},\nonumber
\ee
which simplifies to
\be
\frac{W_{\xi}}{W}=-\left\{\frac{a^2(1-a^2)}{8}\psi_0^4\xi^3+O(1)\xi^5\right\}.\label{goodW}
\ee
Integrating (\ref{goodW}) from $0$ to $\xi$ then produces the formula

\be
W(\xi)=W(0)\left(1-\frac{a^2(1-a^2)}{32}\psi_0^4\xi^4+O(1)\xi^6\right).\label{goodW1}
\ee
Squaring (\ref{goodW1}) then gives (\ref{l4-2b}) in light of the fact that $W=\sqrt{AB}$, so $W(0)=\frac{1}{\psi_0}$.
Equation (\ref{l4-5b}) then follows from the identity $G=\frac{\xi}{\sqrt{AB}}.$     $\Box$

\begin{Lemma}\label{lemma7}
All of the constants $O(1)$ in (\ref{vest})-(\ref{ABest}) can be taken to be $O(1)|a-1|.$
\end{Lemma}

\noindent{\bf Proof:}  Since all of the equations (\ref{vest})-(\ref{ABest}) are exact when $a=1$ with $O(1)\equiv0,$ to verify the $O(1)$ is really $O(1)|a-1|,$ it suffices to prove that each error $O(1)$ represents a smooth function of $a$ and $\xi$ all the way into $\xi=0$.  This follows so long as we can show that any solution $V(\xi)=(v(\xi),A(\xi),H(\xi))$ of system (\ref{Heqn}) can be expanded uniquely in powers of $\xi$, with coefficients smooth functions of $a$, in a neighborhood of $\xi=0.$   This is true essentially because when $F$ is smooth, $H$ is a smooth function of $v$ and $A$ in ${\mathcal M}$, and solutions $v(\xi)$ and $A(\xi)$ in ${\mathcal M}$ can be expanded in powers of $\xi^{\lambda_1}$ and $\xi^{\lambda_2}$ where $\lambda_1=1$ and $\lambda_2=2$ are the eigenvalues of $dF(V_0)$, c.f. (\ref{lineqn}).   Since this is tedious to carry out in full detail, we demonstrate with the scalar example
\be
\xi u_{\xi}=f(u).\label{scalar}
\ee
Let $\lambda=f'(u_0)$ where $u_0$ is a rest point, $f(u_0)=0.$  Then write
\be
u=\sum_{n=0}^{\infty}b_n\xi^{n\lambda},\label{scalar1}
\ee
and assume $f$ has the Taylor expansion
\be
f(u)=f(u_0)+\sum_{n=0}^{\infty}\frac{f^{n}(u_0)}{n!}\left(u-u_0\right)^n.\label{scalar2}
\ee
Putting (\ref{scalar1}) and (\ref{scalar2}) into (\ref{scalar}) gives
\be
\lambda\sum_{n=1}^{\infty}nb_n\xi^{n\lambda}=f(u_0)+\sum_{n=0}^{\infty}\frac{f^{n}(u_0)}{n!}\left(\sum_{k=0}^{\infty}b_n\xi^{n\lambda}\right)^n.\label{scalar3}
\ee
Now since the Taylor series for $f$ converges, it follows that you can solve uniquely for $c_n$ such that
 \be
\sum_{n=0}^{\infty}\frac{f^{n}(u_0)}{n!}\left(\sum_{k=0}^{\infty}b_n\xi^{n\lambda}\right)^n=\sum_{n=0}^{\infty}c_n\xi^{n\lambda},\nonumber
\ee
where
\be
c_n=c_n(b_0,...,b_{n-1}),\nonumber
\ee
so that using  $f(u_0)=0$, (\ref{scalar3}) then becomes
 \be
\lambda\sum_{n=1}^{\infty}nb_n\xi^{n\lambda}=\sum_{n=0}^{\infty}c_n\xi^{n\lambda}.\nonumber
\ee
Equating powers of $\xi^{\lambda}$ we can solve inductively for the coefficients $b_n$ depending on the initial condition $b_0:$
\be
b_n=\frac{c_n(b_0,...,b_{n-1})}{n\lambda}.\label{ans}
\ee
It follows by standard theorems that  if $\lambda>0,$  (\ref{ans}) converges for each $b_0.$  The point now is that the solution will be continuous at $\xi=0$ only for positive $\lambda,$ and will have all classical derivatives if $\lambda$ is a positive integer;  and if $f$ depends on a parameter $a$, it is clear that each $b_n$ will be a smooth function of $a$ as well.  

From this is is not so difficult to see that since eigenvalues for solutions in ${\mathcal M}$ are both positive integers $\lambda_1=1$ and $\lambda_2=2,$ the larger eigenvalue being an integral multiple of the smaller one, the solution can similarly be expanded in powers of $\xi^{\lambda_1}=\xi$ in a neighborhood of $\xi=0,$ with convergence by analogous arguments.  It follows then that all coefficients in the expansion are continuous functions of $\xi$ and $a$, up to $\xi=0,$ from which it follows that each $O(1)$ is really $O(1)|a-1|$ given that $O(1)$ vanishes on the Standard Model $a=1.$ $\Box$

\vspace{.2cm}
\noindent {\bf Proof of Theorem \ref{theoremME}:}  Lemmas \ref{lemma5} through \ref{lemma7} together with (\ref{l4-2}) establish that every solution of (\ref{ODE1})-(\ref{ODE3}) in ${\mathcal M}^*$ enters the rest point $V_0$ according to the estimates (\ref{vest})-(\ref{Hest1}).   Conversely, the estimates (\ref{vest})-(\ref{ABest}) are strong enough to conclude that $V(\xi)$ comes into $V_0$, to leading order in $\xi,$ like solutions of the linearized equations.  Since, by the Hartman-Grobman Theorem, \cite{hart}, solutions of the nonlinear equations in ${\mathcal M}^*$ are in $1-1$ correspondence with solutions of the linearized system (\ref{lineqn}) in ${\mathcal M}_0$, it follows that for every choice of $\psi_ 0$ and $a^2$ there exists a solution of (\ref{ODE1})-(\ref{ODE3}) that comes into rest point $V_0$ like (\ref{vest})-(\ref{ABest}) in ${\mathcal M}^*$.  This completes the proof of Theorem \ref{theoremME}. 

Consider now the leading order corrections to the FRW metric implied by the two parameter family of solutions (\ref{vasoln}) of equations (\ref{one})-(\ref{three}), (\ref{div4}).  Expanding solutions in $\xi$ about $\xi=0$, we have shown that, modulo the scaling law, one eigen-solution tends to infinity as $\xi\rightarrow0$, and the other two satisfy $A(\xi)\rightarrow 1$, $B(\xi)\rightarrow 1,$ as $\xi\rightarrow0,$ for each value of the parameters $\psi_0$ and $a.$   Removing the singular solution, (corresponding to the eigenvalue $\lambda=3>0$ that blows up as $s\rightarrow-\infty$, $\xi\rightarrow0$),  (\ref{Aest}) and (\ref{ABest}) imply that what remains is a two parameter family of SSC spacetimes satisfying
\begin{eqnarray}
A(\xi)&=&\left(1-\frac{a^2v_1^2(\xi)}{4}\right)+O(1)|a-1|\xi^4,\label{firstcorrectionA}\\
B(\xi)&=&\frac{1}{\psi_0^2\left(1-\frac{a^2v_1^2(\xi)}{4}\right)}+O(1)|a-1|\xi^4,\label{firstcorrectionB}
\end{eqnarray}
that reduces exactly to the FRW Standard Model when $a=1$.  The parameter $\psi_0$ corresponds to the time rescaling symmetry of the SSC equations, and the parameter $a$ is a new parameter that changes the underlying spacetimes, and which we call the {\it acceleration paramter}. We thus have the following theorem:

\begin{Theorem}
The $2$-parameter family of bounded solutions (\ref{vasoln}) of  (\ref{one})-(\ref{three}), (\ref{div4}),  that extends FRW of the Standard Model in SSC coordinates to the spacetime metric (\ref{realmetric}), is given in terms of the two parameters $\psi_0$ and $a$, up to errors of order $\xi^4,$   by
\begin{eqnarray}
ds^2=-\frac{d\bar{t}^2}{\psi_0^2\left(1-\frac{a^2v_1^2(\xi)}{4}\right)}+\frac{d\bar{r}^2}{\left(1-\frac{a^2v_1^2(\xi)}{4}\right)}+\bar{r}^2d\Omega^2,\label{firstcorrection}\label{metrica}
\end{eqnarray}
where the velocity satisfies
\begin{eqnarray}
v(\xi)=v_1(\xi)+O(1)|a-1|\xi^3.\label{firstcorrectionv}
\end{eqnarray}
Here $\psi_0$ is the time-scaling parameter, $a=1$ corresponds to FRW, $v_1(\xi)$ denotes the SSC velocity of the Standard Model given in (\ref{FRWSSC1})-(\ref{FRWSSC4}), 
and $a\neq1$ introduces a new acceleration parameter which gives the leading order perturbation of FRW.  In particular (\ref{FRWSSC2})  gives 
$$
v_1(\xi)=\frac{\psi_0\xi}{2}+O(1)\xi^2,
$$
so (\ref{firstcorrectionv}) implies that the velocity $v$ is independent of the parameter $a$ up to second order in $\xi$.  
\end{Theorem}

In light of (\ref{FRWSSC1}), when $a=1$, (\ref{firstcorrection}) reduces exactly to the  FRW metric
\begin{eqnarray}
A(\xi)&=&\left(1-v_1^2\right),\\\label{firstcorrectionA1}
B(\xi)&=&\frac{1}{\psi_0^2\left(1-v_1^2\right)}.\label{firstcorrectionB1}
\end{eqnarray} 
Now the SSC coordinate representation of FRW depends only on $H$ and $\bar{r}$, both of which are invariant under the scaling $r\rightarrow\alpha r$, $R\rightarrow R/\alpha$  of the FRW metric (\ref{FRW}).  It follows that the SSC representation of FRW is independent of $\alpha,$ and therefore independent of our choice of scale for $R(t)$.  Thus without loss of generality we can assume throughout that the FRW metric is scaled exactly so that 
\begin{eqnarray}
R(t)=\sqrt{t},\label{Roft}
\end{eqnarray}
c.f. (\ref{FRW}) and  \cite{smolte}.  We can also remove the time rescaling freedom by setting $\psi_0=1.$\footnotemark[17]\footnotetext[17]{It is interesting that we needed to keep the time scale parameter $\psi_0$ in the analysis of the equations in order to cast $V_0$ as a rest point of a system of ODE's.  That is, the time scale invariance of the equations translates into translation in $s$ invariance in the autonomous system for $U(s).$  Once this is done, we are free to set $\psi_0=1.$} We conclude that to leading order, the $1$-parameter family of expanding wave perturbations of the FRW metric is given by
\begin{eqnarray}
ds^2=-\frac{d\bar{t}^2}{\left(1-\frac{a^2\xi^2}{4}\right)}+\frac{d\bar{r}^2}{\left(1-\frac{a^2\xi^2}{4}\right)}+\bar{r}^2d\Omega^2,\nonumber
\end{eqnarray}
with fourth order errors in $\xi$, and the velocity is given to leading order  by
\begin{eqnarray}
v=\frac{\xi}{2},\label{v}
\end{eqnarray}
{\em independent of $a$}, up to third order errors in $\xi$.

\section{A Foliation of the Expanding Wave Spacetimes into Flat Spacelike Hypersurfaces with Modified Scale Factor $R(t)=t^{a}$.}
\setcounter{equation}{0} 

To get insight into the geometry of the spacetime metric (\ref{metrica}) when $a\neq1,$
consider the extension of the FRW $(t,r)$ coordinate transformation (\ref{transt})-(\ref{transr}) to $a\neq1$ defined by
\begin{eqnarray}
\bar{t}&=&\left\{1+\frac{a^2\zeta^2}{4}\right\}t,\label{transthat}\\
\bar{r}&=&t^{a/2}r.\label{transrhat}
\end{eqnarray}
A straightforward caculation shows that the metric (\ref{metrica}) transforms to $(t,r)$-coordinates as
\begin{eqnarray}
ds^2=-dt^2+t^adr^2+\bar{r}^2d\Omega^2+a(1-a)\zeta dtd\bar{r}.\label{metricc}
\end{eqnarray}
Metric (\ref{metricc}) takes the form of a $k=0$ Friedmann-Robertson-Walker metric with a small correction to the scale factor, ($R_a(t)=t^{a/2}$ instead of $R(t)=t^{1/2}$), and a corrective mixed term.  In particular, the time slices $t=const.$ in (\ref{metricc}) are all flat space ${\mathcal R}^3,$ as in FRW, and the $\bar{r}=const$ slices agree with the FRW metric modified by scale factor $R_a(t).$  It follows that the $t=const.$ surfaces given by (\ref{transthat}), (\ref{transrhat}), define a foliation of spacetime into flat three dimensional spacelike slices.   Thus when $a\neq1$,  (\ref{metricc}) exhibits many of the flat space properties characteristic of the $a=1$ FRW spacetime. 

\section{Expanding Wave Corrections to the Standard Model in Approximate Comoving Coordinates}
\setcounter{equation}{0}

The metric (\ref{metricc}) is not co-moving with the velocity $v$ of (\ref{firstcorrectionv}), even at the leading order, when $a\neq1$.  To obtain (an approximate) co-moving frame, note that from (\ref{v}), $v$ is independent of $a$ up to order $\xi^3$, so it follows that even when $a\neq1,$ the inverse of the transformation (\ref{transt}), (\ref{transr}) gives, to leading order in $\xi$, a co-moving coordinate system for (\ref{metrica}) in which we can compare the Hubble constant and redshift vs luminosity relations for (\ref{metrica}) when $a\neq 1$ to the Hubble constant and redshift vs luminosity relations for $a=1$ FRW, as measured by (\ref{FRW}) and (\ref{hubble}).  Thus from here on, we take $(t,r)$-coordinates to be defined by the Standard Model coordinate map to FRW coordinates  (\ref{transt}), (\ref{transr}), which corresponds to taking $a=1$ in (\ref{transthat}), (\ref{transrhat}).  For this map we note that $\bar{r}=R(t)r$ gives $\bar{r}$ as a function of $(t,r)$, and by (\ref{FRWSSC3}), (\ref{FRWSSC4}) it follows that
\be
\zeta=\psi_0\xi+O(1)\xi^3\ as\ \xi\rightarrow0.\label{xizeta}
\ee

 \begin{Theorem}\label{FRWametric}
The inverse of the coordinate transformation (\ref{transt}), (\ref{transr}) maps (\ref{metrica}) over to $(t,r)$-coordinates as
\begin{eqnarray}
ds^2=F_a(\zeta)^2\left\{-dt^2+tdr^2\right\}+\bar{r}^2d\Omega^2,\label{metricb}
\end{eqnarray}
where
\begin{eqnarray}
F_a(\zeta)^2=\frac{1-\frac{\zeta^2}{4}}{1-\frac{a^2\zeta^2}{4}}=1+(a^2-1)\frac{\zeta^2}{4}+O(1)|a-1|\zeta^4,\label{errorF}
\end{eqnarray}
and the SSC velocity $v$ in (\ref{vest}) maps to the $(t,r)$-velocity
\begin{eqnarray}
w=-\frac{a^2-1}{8}\zeta^3+O(1)|a-1|\zeta^4.\label{vb}
\end{eqnarray}
\end{Theorem}

\noindent Note that by (\ref{vest}), (\ref{xizeta})
\be
w=v-v_1+O(1)|a-1|\zeta^4.\label{vminusvone}
\ee

\noindent{\bf Proof:}  Using  $v_1=\zeta/2$, the Jacobian of the transformation from SSC coordinates $(\bar{t},\bar{r})$ to $(t,r)$ coordinates is given in (\ref{J}),  namely,
$$
J\equiv\frac{\partial\bar{x}}{\partial x}=\left(\begin{array}[pos]{cc}
\psi_0&\psi_0\sqrt{t}\frac{\zeta}{2}\\
\frac{\zeta}{2}&\sqrt{t}
 \end{array}\right).
$$
Thus letting $\bar{g}$ denote the spacetime metric (\ref{metrica}), in $(t,r)$ coordinates $\bar{g}$ transforms to $g=J^{t}\bar{g} J$ where
 \be
g&=&\left(\begin{array}[pos]{cc}
\psi_0&\frac{\zeta}{2}\\
\psi_0\sqrt{t}\frac{\zeta}{2}&\sqrt{t}
 \end{array}\right)
 \left(\begin{array}[pos]{cc}
\frac{-1}{\psi_0^2(1-a^2v_1^2(\xi))}&0\\
0&\frac{1}{1-a^2v_1^2(\xi)}
 \end{array}\right)\left(\begin{array}[pos]{cc}
\psi_0&\psi_0\sqrt{t}\frac{\zeta}{2}\\
\frac{\zeta}{2}&\sqrt{t}
 \end{array}\right)\nonumber\\\nonumber\\
&=&\frac{1-\zeta^2/4}{1-a^2\zeta^2/4}\left(\begin{array}[pos]{cc}
-1&0\\
0&t
 \end{array}\right),
\ee
and so neglecting errors of order $|a-1|\zeta^4$, the transformed metric is
\be
ds^2=-\left(1-a^2\frac{\zeta^2}{4}\right)dt^2+t\left(1-a^2\frac{\zeta^2}{4}\right)dr^2+\bar{r}^2d\Omega^2.\label{metrictrans}
\ee
This confirms (\ref{metricb}) and (\ref{errorF}).

To establish (\ref{vb}),  let ${\bf \bar{u}}=(\bar{u}^0,\bar{u}^1)$ be the $4$-velocity of the particle in SSC coordinates, and ${\bf u}=(u^0,u^1)$ the $4$-velocity in $(t,r)$-coordinates, so that
\be
v=\frac{1}{\sqrt{AB}}\frac{d\bar{r}}{d\bar{t}}=\frac{1}{\sqrt{AB}}\,\frac{\bar{u}^1}{\bar{u}^0},\label{vAB}
\ee 
and similarly,
$$
w=\frac{dr}{dt}=\sqrt{t}\,\frac{u^1}{u^0},
$$ 
where we used (\ref{metricb}).  Now let $v=v_1+\hat{w}.$  We show $w=\hat{w}+O(1)|a-1|\zeta^4.$   For this note that the transformation law for vectors is
\be
\left(\begin{array}[pos]{c}
u^0\\
u^1
 \end{array}\right)=J^{-1}\left(\begin{array}[pos]{c}
\bar{u}^0\\
\bar{u}^1
 \end{array}\right).\label{Jinverse}
\ee
Let 
\be
R_0&=&\sqrt{t}\left(1,-\psi_0\frac{\zeta}{2}\right),\label{row0}\\
R_1&=&\left(-\frac{\zeta}{2},\psi_0\right),\label{row1}
\ee 
denote the top and bottom rows of $J^{-1}\left|J\right|$, c.f. (\ref{Jinv}).  Then  (\ref{ABest}) implies that in SSC coordinates, 
$$
\sqrt{AB}=\frac{1}{\psi_0}\left(1+O(1)|a-1|\xi^4\right),
$$
so since $\xi=O(\zeta)$,  (\ref{vAB}) gives

\be
\left(\begin{array}[pos]{c}
\bar{u}^0\\
\bar{u}^1
 \end{array}\right)=const.\left(\begin{array}[pos]{c}
\psi_0+e\\
v
 \end{array}\right),
\ee
where 
\be
e=O(1)|a-1|\zeta^4.
\ee
Using this in (\ref{Jinverse}) we have
\be
w&=&\sqrt{t}\frac{u^1}{u^0}=\sqrt{t}\frac{R_1\left(\begin{array}[pos]{c}
\psi_0+e\\
v
 \end{array}\right)}{R_0\left(\begin{array}[pos]{c}
\psi_0+e\\
v
 \end{array}\right)}
 =\sqrt{t}\frac{R_1\left(\begin{array}[pos]{c}
\psi_0\\
v_1
 \end{array}\right)+R_1\left(\begin{array}[pos]{c}
e\\
\hat{w}
 \end{array}\right)}{R_0\left(\begin{array}[pos]{c}
\psi_0+e\\
v
 \end{array}\right)} \nonumber\\
 &=&
 \sqrt{t}\frac{R_1\left(\begin{array}[pos]{c}
e\\
\hat{w}
 \end{array}\right)}{R_0\left(\begin{array}[pos]{c}
\psi_0+e\\
v
 \end{array}\right)}=\frac{\hat{w}}{1-\frac{\eta}{2}v}+O(1)|a-1|\zeta^4\nonumber\\
 \ee
where we used that 
\be
R_1\cdot(\psi_0,v_1)=0
\ee 
by (\ref{FRWSSC3}), true because $v_1$ is the SSC velocity of the Standard Model $a=1$, and the fluid is co-moving with respect to the FRW metric (\ref{FRW}) by Theorem \ref{thmSM}.   Thus, putting in the errors from (\ref{vest}) of Theorem \ref{thmME}, we have that
$$
w=\hat{w}+O(1)|a-1|\zeta^4=v-v_1+O(1)|a-1|\zeta^4=-\frac{a^2-1}{8}\zeta^3.\label{veltr}
$$ 
This completes the proof of Theorem \ref{FRWametric}.  $\Box$
\vspace{.2cm}

\noindent {\bf Remark:}  Theorem \ref{FRWametric} implies that, neglecting errors of order $O(1)|a-1|\xi^4$ in (\ref{Aest}) and (\ref{ABest}), the SSC spacetime $\bar{g}$ corresponding to parameter values $(\psi_0,a)$ takes the $(t,r)$ coordinate form (\ref{metricb}).  It follows, then, that any calculation based on undifferentiated metric coefficients from the approximate metric (\ref{metricb}) gives answers correct up to order $O(1)|a-1|\xi^4$ in the original metric (\ref{realmetric}); and any calculation based first derivatives of metric coefficients (\ref{metricb}) gives answers correct up to order $O(1)|a-1|\xi^3$ in the original metric (\ref{realmetric}).  Thus, since geodesics involve first derivatives of the metric, estimates based on geodesics of (\ref{metricb}) give answers correct up to order $O(1)|a-1|\xi^3$ in the original metric (\ref{realmetric}), with one important exception.  Since radial geodesics of (\ref{vb}) can be obtained directly from (\ref{metricb})  by setting $d\Omega=0$ and $ds=0$ without going to the geodesic equations, it follows that radial lightlike geodesics of (\ref{metricb}) agree with radial lightlike geodesics of the original metric (\ref{realmetric}) up to errors $O(1)|a-1|\zeta^4.$

The variable $\zeta=\bar{r}/t$ is a natural dimensionless perturbation parameter that has a physical interpretation in $(t,r)$-coordinates because, (assuming $c=1$ or $t\equiv ct$), $\zeta$ ranges from $0$ to $1$ as $\bar{r}$ ranges from zero to the horizon distance in FRW, (approximately the Hubble distance $c/H$), a measure of the furthest one can see from the center at time $t$ units after the Big Bang, \cite{wein};  that is,
\begin{eqnarray}\zeta&\approx&\frac{Dist}{Hubble\ Length}.\label{horizon}
\end{eqnarray}
Thus expanding in $\zeta$ gives an expansion in the fractional distance to the Hubble length, c.f. \cite{smolte}.
Note also that when $a=1$ we obtain the FRW metric (\ref{FRW}), where we have used $R(t)=\sqrt{t},$ c.f. (\ref{Roft}).  

Now for a first comparison of the relative expansion at $a\neq1$ to the expansion of FRW, define {\it the Hubble constant at parameter value a}, by
$$H_a(t,\zeta)=\frac{1}{R_a}\frac{\partial}{\partial t}R_a,$$
where
$$
R_a(t,\zeta)=F_a(\zeta)\sqrt{t},
$$
equals the square root of the coefficient of $dr^2$ in (\ref{metricb}).
Then one can easily show
$$H_a(t,\zeta)=
\frac{1}{2t}\left\{1-\frac{3}{8}(a^2-1)\zeta^2+O\left(|a^2-1|\zeta^4\right)\right\}.$$
We conclude that the fractional change in the Hubble constant due to the perturbation induced by expanding waves $a\neq1$ relative to the FRW of the Standard Model $a=1$, is given by
$$\frac{H_a-H}{H}=\frac{3}{8}(1-a^2)\zeta^2+O\left(|a^2-1|\zeta^4\right).$$

\section{Redshift vs Luminosity Relations and the Anomalous Acceleration}\label{sectionredlum}
\setcounter{equation}{0} 

In this section we obtain the $a\neq1$ corrections to the redshift vs luminosity relation of FRW up to order $\zeta^3$, as measured by an observer positioned at the center $\zeta=0$ of the expanding wave spacetimes described by the metric  (\ref{metrica}) when $a\neq1.$\footnotemark[18]\footnotetext[18]{This is of course a theoretical relation, as the pure radiation FRW spacetime is not transparent.}
   (Recall that $\zeta\equiv\bar{r}/t=r/\sqrt{t}$ measures the fractional distance to the horizon, c.f. (\ref{horizon}).)  Now redshift vs luminosity depends on motion of the observer, but is otherwise a coordinate independent relation. The physically natural coordinate system in which to do the comparison with FRW ($a=1$) would be co-moving with respect to the sources.  Thus we restrict to the coordinates $(t,r)$ defined by (\ref{transt}), (\ref{transr}), in which our one parameter family of expanding wave spacetimes are described, to leading order in $\zeta$, by the metric (\ref{metricb}).  In $(t,r)$ coordinates the spacetime is only co-moving up to order $\zeta^3,$ so we will need to incorporate the errors (from co-moving) below to get formulas up to order $\zeta^4.$   Note that the approximate metric  (\ref{metricb}) as well as the exact spacetime metrics (\ref{realmetric}) both reduce {\em exactly} to the FRW metric when $a=1,$ c.f. (\ref{FRW}),  (\ref{Roft}).  
   
   For our derivation of the redshift vs luminosity relation for (\ref{metricb}) we follow the development in Gron-Hervik \cite{gronhe}, page 289 ff.   The redshift vs luminosity relation calculation in \cite{gronhe} in the case of the Standard Model $a=1$ leads to 
\be\label{ell0}
d_{\ell}=2t_0z.
\ee
We now generalize the argument so that it applies to the spacetime metrics (\ref{metricb}) when $a\neq1$, assuming sources moving with arbitrary velocity $w.$  Thus assume radiation of frequency $\nu_e$ is emitted radially by a source moving at velocity $w$ relative to the co-moving observer at $(t_e,r_e)$, and observed at a later time $t=t_0$ at frequency $\nu_0$ at the center $r=0$ of the spacetime metric (\ref{metricb}).  Let $\bar{\nu}_e$ denote the (intermediate) frequency of the emitted radiation as measured by a co-moving observer fixed at position $r=r_e$ at time $t=t_e.$  In the Standard Model $a=1$ the formula (\ref{vb}) for $w$ reduces to  $w\equiv0$ because in this case the fluid is exactly co-moving in $(t,r)$-coordinates, implying that $\nu_e=\bar{\nu}_e$ when $a=1$.  But when $a\neq1,$ (\ref{vb}) implies $w\neq0$ and we must account for the case $\nu_e\neq\bar{\nu}_e$.

To start, let $\lambda_e$ denote the wavelength of the radiation emitted at ($t_e,r_e$)  and $\lambda_0$ the wavelength received at the center $\zeta\equiv\frac{\bar{r}}{t}=0$, (that is, at $\bar{r}=0$ at later time  $t_0$). Define
\begin{eqnarray}
L&\equiv&Absolute\ Luminosity=\frac{\rm Energy\ Emitted\ by\ Source}{\rm Time}\label{lum1}\\
\ell&\equiv& Apparent\ Luminosity=\frac{\rm Power\ Recieved}{\rm Area}\label{lum3}
\ee
and let
\be
d_{\ell}&\equiv& Luminosity\ Distance=\left(\frac{L}{4\pi\ell }\right)^{1/2}\label{lum2}\\
z&\equiv& Redshift\ Factor=\frac{\lambda_0}{\lambda_e}-1.\label{lum4}
\end{eqnarray}
Using two serendipitous properties of the metric (\ref{metricb}), namely, the metric is diagonal in co-moving coordinates, and there is no $a$-dependence on the sphere's of symmetry, it follows that the arguments in \cite{gronhe}, Section 11.8, 
 can be modified to give the following restatement of Theorem \ref{thmredshiftintro}, which extends the results of \cite{smoltePNAS}.

\begin{Theorem}\label{thmredshift}
The redshift vs luminosity relation, as measured by an observer positioned at the center $\zeta=0$ of the spacetime described by metric (\ref{metrica}), with velocity profile (\ref{firstcorrectionv}), is given up to fourth order in redshift factor $z$ by
\begin{eqnarray}
d_{\ell}=2ct_0\left\{z+\frac{a^2-1}{2}z^2+\frac{(a^2-1)(3a^2+5)}{6}z^3+O(1)|a-1|z^4\right\}\label{redvslum}\label{redvslumFRW}
\end{eqnarray}
where we use the fact, (c.f. (\ref{zetabyz}) below), that $z$ and $\zeta$ are of the same order as $\zeta\rightarrow0.$
\end{Theorem}
Again, note that when $a=1$, (\ref{redvslum}) reduces to (\ref{ell0}),
correct for the radiation phase of the Standard Model, \cite{gronhe}.
Thus (\ref{redvslum}) gives the leading order quadratic and cubic corrections to the redshift vs luminosity relation when $a\neq1$, thereby improving the quadratic estimate (6.5) of \cite{smoltePNAS}.  Since $(a^2-1)$ appears in front of the leading order correction in (\ref{redvslum}), it follows, (by continuous dependence of solutions on parameters), that the leading order part of any anomalous correction to the redshift vs luminosity relation of the Standard Model, observed at a time after the radiation phase, can be accounted for by suitable adjustment of the parameter $a$.   In particular, note that when $a>1$, the leading order corrections in (\ref{redvslum})  imply a blue-shifting of radiation relative to the Standard Model, as observed in the supernova data, \cite{cliffe}.

To establish (\ref{redvslumFRW}), let $P$ denote the energy per time (power) of radiation received at the mirror (of a reflecting telescope) of area ${\mathcal A}$, positioned at the coordinate center transverse to the radial direction, the radiation being emitted at a distant source moving at velocity $w$ at $(t_e,r_e)$, and received at $t=t_0,$ $r=0.$  We start with the following elementary relation, (c.f.  \cite{gronhe} page 289):
\be
P\equiv\frac{\Delta (energy)}{\Delta\tau_0}=L\cdot f_{\mathcal A}\cdot\frac{\nu_0}{\nu_e}\cdot\frac{\Delta \tau_e}{\Delta \tau_0}.\label{start1} 
\ee
Here
\be
L=\frac{\Delta (energy)}{\Delta \tau_e}\nonumber
\ee
is the absolute luminosity, the energy per time emitted by the source, c.f. (\ref{lum3}); the ratio of the frequencies, given by
\be
\frac{\nu_0}{\nu_e}=(1+z)^{-1},\nonumber
\ee
accounts for losses of energy due to redshifting at the source, (c.f. (\ref{null}) below); the ratio of proper times satisfies
\be
\frac{\Delta\tau_e}{\Delta\tau_0}=(1+z)^{-1},\nonumber
\ee
corrects proper time change at the receiver to proper time change at the source, (c.f. (\ref{nu}) and (\ref{null}) below); and finally $f_{\mathcal A}$ is defined to be {\it the fraction of the emitted radiation received at the mirror} ${\mathcal A}$.  In the case of the Standard Model $a=1,$ equation (11.116), page 289 of \cite{gronhe} gives
\be
f_{\mathcal A}=\frac{\mathcal A}{4\pi t_0r_e^2}.\label{fA}
\ee
We will need the following proposition, which establishes that when $a\neq1$, (\ref{fA}) holds subject to the correction factor $C_{a}$ given below in (\ref{CA}).

\begin{Proposition}\label{lemmatech}\label{areaprop}
When $a\neq1$, the value of $f_{\mathcal A}$ for the family of spacetime metrics (\ref{realmetric}) is given by
\be
f_{\mathcal A}=\frac{\mathcal A}{4\pi t_0r_e^2}C_{a},\label{fAreal}
\ee
where $C_{a}$ is given by
\be
C_{a}=1-\frac{a^2-1}{6}\zeta^2+O(1)|a-1|\zeta^3.\label{CA}
\ee.
\end{Proposition}
Proposition \ref{areaprop} solves what we call the {\it mirror problem}.  That is, it gives the ratio $C_{a}$ of an area ${\mathcal A}$ of light received from a distant source at a mirror positioned at the origin when $a\neq1$, to the corresponding area when $a=1,$ in the limit ${\mathcal A}\rightarrow0$   (the limit expressing that the mirror is small relative to the distance to the source.)  The result is important for the physical interpretation of the spacetimes when $a\neq1$ and the discussion and proof is the topic of Appendix \ref{mirrorappend}.

The proof of Theorem \ref{thmredshift} relies on the following lemma:
\begin{Lemma}\label{lemma11}  Assume that radiation of frequency $\nu_e$ is emitted by a source moving at velocity $w$ at $(t_e,r_e)$, and observed at a later time $t=t_0$ at frequency $\nu_0$ at the center $r=0$ of the spacetime metric (\ref{metricb}).  Then $r_e$ is related to $t_0$ by
\be
r_e=\frac{\zeta}{1+\frac{\zeta}{2}}\sqrt{t_0},\label{re}
\ee
where (for this section)
$$
\zeta=\frac{r_e}{\sqrt{t_e}},
$$
the luminosity distance $d_{\ell}$ is given by
\be
d_{\ell}&=&t_0(1+z)\frac{\zeta}{1+\frac{\zeta}{2}}\left(1+\frac{a^2-1}{12}\zeta^2\right)+O(1)|a-1|\zeta^4\nonumber\\
&=&t_0(1+z)\zeta\left(1-\frac{1}{2}\zeta+\frac{a^2+2}{12}\zeta^2\right)+O(1)|a-1|\zeta^4,\label{dell}
\ee
and the redshift $z$ observed at the origin is given by
\be
1+z=\left(1+\frac{\zeta}{2}\right)\frac{1}{F_a(\zeta)}\frac{1+w}{\sqrt{1-w^2}}.\label{mainlemmaredshift}
\ee

\end{Lemma}
\vspace{.2cm}
Postponing the proof of Lemma \ref{lemma11}, we now give the
\vspace{.2cm}

\noindent{\bf Proof of Theorem \ref{thmredshift}:} Since redshift vs luminosity as measured at a point in spacetime is a coordinate independent relation, we can obtain $d_{\ell}$ as a function of $z$ by substituting $w$ as given in (\ref{vb}) into (\ref{mainlemmaredshift}) to get $\zeta$ as a function of $z,$ and then substituting this expression for $\zeta$ into (\ref{dell}).  To accomplish this, start from (\ref{errorF}), 
\begin{eqnarray}
\frac{1}{F_a(\zeta)}=1-(a^2-1)\frac{\zeta^2}{8}+O(1)|a-1|\zeta^4,\label{errorF1}
\end{eqnarray}
so (\ref{mainlemmaredshift}) gives
\be
&1+z=\left(1+\frac{\zeta}{2}\right)\left(1-\frac{a^2-1}{8}\zeta^2\right)\left(1-\frac{a^2-1}{8}\zeta^3\right)+O(1)|a-1|\zeta^4\ \ \ \ \ \ \ \ \ \ \ \ \ &\nonumber\\
&=1+\frac{\zeta}{2}-\frac{a^2-1}{8}\zeta^2+\left(-\frac{a^2-1}{16}\zeta^3-\frac{a^2-1}{8}\zeta^3\right)+O(1)|a-1|\zeta^4&\nonumber
\ee
Thus
\be
1+z=1+\frac{\zeta}{2}-\frac{a^2-1}{8}\zeta^2-\frac{3}{16}(a^2-1)\zeta^3+O(1)|a-1|\zeta^4.\label{final1}
\ee
In particular (\ref{final1}) implies $\zeta=O(z)$, and using this we can solve (\ref{final1}) for $\zeta$ as a function of $z$. First
\be
z=\frac{\zeta}{2}+O(1)|a^2-1|\zeta^2.\label{zetabyz}
\ee
Putting this in (\ref{final1}) with $\zeta=O(z)$ gives
\be
z=\frac{\zeta}{2}-\frac{a^2-1}{2}z^2+O(1)|a-1|\zeta^3,\nonumber
\ee
so that 
\be
\zeta=2z+(a^2-1)z^2+O(1)|a-1|\zeta^3.\label{final2}
\ee
Using (\ref{final2}) into (\ref{final1}) gives
\be
z=\frac{\zeta}{2}-\frac{a^2-1}{2}z^2-\frac{a^2-1}{2}\left(a^2+2\right)z^3+O(1)|a-1|z^4,\nonumber
\ee
so that
\be
\zeta=2z+(a^2-1)z^2+(a^2-1)\left(a^2+2\right)z^3+O(1)|a-1|z^4.\label{zetabyz01}
\ee
Now by (\ref{dell}),
\be
d_{\ell}=t_0(1+z)\zeta\left(1-\frac{1}{2}\zeta+\frac{a^2+2}{12}\zeta^2\right)+O(1)|a-1|\zeta^4,\label{dell2}
\ee
and using (\ref{zetabyz01}) we can express the right hand side as
\be\label{zetabyz2}
(1+z)\zeta\left(1-\frac{1}{2}\zeta+\frac{a^2+2}{12}\zeta^2\right)&=&(1+z)z\left\{\cdot\right\}_I\left\{\cdot\right\}_{II}
\ee
where
\be\nonumber
\left\{\cdot\right\}_{I}=\left\{2+(a^2-1)z+(a^2-1)(a^2+2)z^2\right\},
\ee
\be\nonumber
\left\{\cdot\right\}_{II}=\left\{1-\left(z+\frac{(a^2-1)}{2}z^2\right)+\frac{a^2+2}{3}z^2\right\},
\ee
which upon collecting terms gives
\be
&(1+z)z\left\{\cdot\right\}_I\left\{\cdot\right\}_{II}=2z\left\{1+\frac{a^2-1}{2}z+\frac{(a^2-1)(3a^2+5)}{6}z^2\right\}&\label{zetabyz1}\\
&\ \ \ \ \ \ \ \ \ \ \ \ \ \ \ \ \ \ \ \ \ \ \ \ \ \ \ \ \ \ \ \ \ \ \ \ \ \ \ \ \ \ \ \ \ \ \ \ \ \ \ \ \ \ \ \ \ +O(1)|a-1|z^4&.\nonumber
\ee
Putting (\ref{zetabyz1}) into (\ref{dell2}) then gives (\ref{redvslum}), thereby completing the proof of Theorem \ref{thmredshift}. $\Box$.

It remains only to give the
\vspace{.2cm}

\noindent {\bf Proof of Lemma \ref{lemma11}:}  
We first establish equations (\ref{re})-(\ref{mainlemmaredshift}).  
To start, note that the coordinate $t$ measures geodesic time at fixed $r$ only when $a=1,$ so define geodesic time at fixed $r$ by
\be
d\tau=F_a dt,\label{dtau}
\ee
where by (\ref{errorF}),  $F_a\equiv F_a(\zeta)$ is given to leading orders by 
\begin{eqnarray}
F_a(\zeta)=1+(a^2-1)\frac{\zeta^2}{8}+|a-1|O(\zeta^4).\label{errorrootF}
\end{eqnarray}
Note that the radial lightlike geodesics for metric (\ref{metricb}) satisfy 
$$F_a^2(-dt^2+tdr^2)=0,$$
so the radial null geodesics in $(t,r)$-coordinates are given by
\be
\frac{dr}{dt}=\pm\frac{1}{\sqrt{t}},\label{nullgeo}
\ee
independent of $a.$ Integration gives
$$
\sqrt{t_e}=\sqrt{t_0}-r_e/2,
$$
and solving
\be
\zeta=\frac{r_e}{\sqrt{t_e}}=\frac{r_e}{\sqrt{t_0}-r_e/2}\nonumber
\ee
for $r_e$ gives
\be
r_e=\frac{\sqrt{t_0}\zeta}{1+\zeta/2},\label{null5}
\ee
establishing (\ref{re}). 

Consider next equation (\ref{dell}).  This is a direct consequence of Proposition \ref{areaprop} as follows.  Equations (\ref{start1})-(\ref{fA}) give
\be
P=\frac{L{\mathcal A} C_{a}}{(1+z)^2 4\pi r_e^2t_0},\label{Pnew}
\ee
and using (\ref{Pnew}) and (\ref{lum3}), the apparent luminosity $\ell$ is
\be
\ell\equiv\frac{P}{\mathcal A}=\frac{LC_{a}}{(1+z)^24\pi r_e^2t_0}.
\ee
Thus the luminosity distance $d_{\ell}$ satisfies
\be
d_{\ell}\equiv\left(\frac{L}{4\pi\ell}\right)^{1/2}=\frac{(1+z)r_e\sqrt{t_0}}{\sqrt{C_{a}}}=(1+z)\frac{t_0\zeta}{1+\frac{\zeta}{2}}\frac{1}{\sqrt{C_{a}}},\label{almostca}
\ee
where we have used the relation (\ref{null5}) between $\zeta$ and $r_e.$ Substituting  (\ref{CA}) into (\ref{almostca}) and collecting terms in $\zeta$ then establishes (\ref{dell}).

Finally, to establish (\ref{mainlemmaredshift}), consider  a null geodesic
$$
\frac{dr}{dt}=-\frac{1}{\sqrt{t}},
$$ 
emitted from $(t_e,r_e)$ and received at $r=0$ a later time $t_0$.  Then integrating gives
\be
r_e=-\int_{r_e}^{0}dr=\int_{t_e}^{t_0}\frac{dt}{\sqrt{t}}\label{null1}
\ee
Now  letting $\Delta t_0$ denote the time of one period for radiation measured at a given frequency at $t=t_0,$ $r=0,$ we have
$$
r_e=\int_{t_e}^{t_0}\frac{dt}{\sqrt{t}}=\int_{t_e+\overline{\Delta t}_e}^{t_0+\Delta t_0}\frac{dt}{\sqrt{t}},
$$
where $\overline{\Delta t}_e$ denotes the time of one period as measured by the {\it co-moving} observer positioned at the emitter at time $t=t_e$.  Thus
$$
\int_{t_e+\overline{\Delta t}_e}^{t_0+\Delta t_0}\frac{dt}{\sqrt{t}}-\int_{t_e}^{t_0}\frac{dt}{\sqrt{t}}=0,
$$
so
\be
\int_{t_0}^{t_0+\Delta t_0}\frac{dt}{\sqrt{t}}=\int_{t_e}^{t_e+\overline{\Delta t}_e}\frac{dt}{\sqrt{t}},\nonumber
\ee
and we therefore have
\be
\frac{\overline{\Delta t}_e}{\sqrt{t_e}}=\frac{\Delta t_0}{\sqrt{t_0}},\label{null2}
\ee
for sufficiently small $\Delta t_0.$  Now the frequency $\nu$ associated with the period $\overline{\Delta t}$ measured by a co-moving observer is
\be
\nu=\frac{1}{\overline{\Delta\tau}},\label{nu}
\ee
where $\overline{\Delta \tau}$ is the proper time interval associated with (\ref{metricb}),
\be
\overline{\Delta \tau}=F_a\overline{\Delta t}.\nonumber
\ee 
The ratio of the emitted to received frequency at given $a$ is then
\be
\frac{\overline{\nu}_e}{\nu_0}=\frac{F_{0}\sqrt{\Delta t_0}}{F_{e}\sqrt{\overline{\Delta t}_e}}=\frac{\sqrt{\Delta t_0}}{F_{a}(\zeta)\sqrt{\overline{\Delta t}_e}},\nonumber
\ee
where we use that $F_{0}\equiv F_a(0)=1$, $F_e\equiv F_a(\zeta_e)\equiv F_a(\zeta)$, and where from here on we use the notation 
\be
\zeta\equiv \zeta_e=\frac{r_e}{\sqrt{t_e}}.\label{newnotation}
\ee
Thus by (\ref{null2}) we conclude
\be
\frac{\overline{\nu}_e}{\nu_0}=\frac{\sqrt{t_0}}{F_{a}(\zeta)\sqrt{ t_e}}.\nonumber
\ee 
Now by (\ref{lum4}), 
\be
1+z=\frac{\nu_e}{\nu_0}=\frac{\overline{\nu}_e}{\nu_0}\frac{\nu_e}{\overline{\nu}_e}=\frac{\sqrt{t_0}}{F_a(\zeta)\sqrt{t_e}}\frac{\nu_e}{\overline{\nu}_e}\label{null}.\label{null3}
\ee
Thus by (\ref{re}) we have
\be
1+z=\frac{\sqrt{t_0}}{F_a(\zeta)\sqrt{t_e}}\frac{\nu_e}{\overline{\nu}_e}=\frac{\sqrt{t_0}\zeta}{r_eF_a(\zeta)}\frac{\nu_e}{\overline{\nu}_e},\nonumber
\ee
which upon using (\ref{null5}) gives
\be
1+z=\frac{1+\zeta/2}{F_a(\zeta)}\frac{\nu_e}{\overline{\nu}_e}.\label{null7}
\ee
We conclude that equation (\ref{mainlemmaredshift}) follows directly from (\ref{null7}) together with the following lemma:
\begin{Lemma}\label{lemmanu}
The following relation holds between the  frequency $\nu_e$ emitted by a source moving at velocity $w$ at $(t_e,r_e)$ and the frequency $\overline{\nu}_e$ as measured in the co-moving frame at $(t_e,r_e)$:
\be
\frac{\nu_e}{\overline{\nu}_e}=\frac{1+w}{\sqrt{1-w^2}}\label{nuratio}
\ee
\end{Lemma}
\vspace{.2cm}

\noindent {\bf Proof:}  Since the frequency defined in (\ref{nu}) is defined in terms of the invariant time interval $\Delta\tau$, it is defined independent of coordinates fixed with the same observer.  Thus to calculate $\nu_e/\overline{\nu}_e$ we can assume that $\overline{\nu}_e$ is the frequency measured in the (local) Minkowski frame fixed with an observer at $r=r_a,$ and $\nu$ is the frequency as measured by a second observer at the same point but moving with velocity $w$ with respect to the first observer.  Thus the result derives from the change of frequency formula for Lorentz transformations in special relativity.  To derive this formula, for this argument supress the angular variables held constant along radial motion, and let $(\bar{t},\bar{x})$ denote the (local) Minkowski frame fixed with a first observer co-moving with the metric at point $P=(t_e,r_e),$ and let $(t,x)$ be the (local) Minkowski frame fixed with the second observer moving with radial velocity $w$ with respect to the first observer.  Since Minkowski time changes agree with proper time changes we have
\be
&\overline{\nu}=\frac{1}{\overline{\Delta t}}&\nonumber\\
&\nu=\frac{1}{\Delta t}.&\nonumber
\ee
Now let $X$ denote the lightlike vector displacement of one period of the radial lightray at $(t_e,r_e)$.  Then by definition
\be
\Delta t=dt(X),\nonumber
\ee
and 
\be
\overline{\Delta t}=d\bar{t}(X).\nonumber
\ee
Thus
\be
\frac{\nu_e}{\overline{\nu}_e}=\frac{d\bar{t}(X)}{dt(X)}=\frac{\bar{\alpha}}{\alpha},\label{nu1}
\ee
where $\alpha$ and $\bar{\alpha}$ give the unbarred and barred components of $X$, respectively,
\be
X=\alpha\left\{\frac{\partial}{\partial t}+\frac{\partial}{\partial x}\right\}=\bar{\alpha}\left\{\frac{\partial}{\partial \bar{t}}+\frac{\partial}{\partial \bar{r}}\right\}.\nonumber
\ee
Now the Lorentz transformation that takes barred to unbarred coordinates  is
\be
\left(\begin{array}{c}
t\\x
\end{array}\right)=\left(\begin{array}{cc}
\frac{1}{\sqrt{1-w^2}}&\frac{w}{\sqrt{1-w^2}}\\ \frac{w}{\sqrt{1-w^2}}&\frac{1}{\sqrt{1-w^2}}\end{array}\right)\left(\begin{array}c
\bar{t}\\\bar{x}
\end{array}\right),\nonumber
\ee
so 
\be
\alpha=\left(\frac{1}{\sqrt{1-w^2}}+\frac{w}{\sqrt{1-w^2}}\right)\bar{\alpha}=\frac{1+w}{\sqrt{1-w^2}}\bar{\alpha},
\ee
and using this in  (\ref{nu1}) gives (\ref{nuratio}) as claimed, and thus the proof of (\ref{mainlemmaredshift}), and hence Lemma \ref{lemma11}, is complete.   $\Box$

\section{Appendix:  The Mirror Problem}\label{mirrorappend}
\setcounter{equation}{0} 

In this section we give the proof of Proposition \ref{areaprop}, Section  \ref{sectionredlum}, which gives the ratio $C_{a}$ of an area ${\mathcal A}$ of light received from a distance source at a mirror (telescope) positioned at the origin when $a\neq1$, to the corresponding area when $a=1,$ in the limit ${\mathcal A}\rightarrow0,$ (the limit expressing that the mirror is small relative to the distance to the source.)  To describe the mirror problem, consider light emitted from a distant source located at ($t_e,r_e$) and received at a mirror of area ${\mathcal A}$ positioned orthogonal to the line of sight at the center $r=\zeta=0$ of our spherically symmetric expanding spacetimes $a\neq1$, at a later time $t=t_0>t_e>0$.  The problem is to determine the fraction $f_{\mathcal A}$ of the area of the $2$-sphere emitting radiation at $r=r_e$, $t=t_e$ that reaches the mirror.  In the case of the Standard Model $a=1$, the center of the FRW ($t,r$)-coordinate system can be translated to any point.  Taking the center to be ($t_e,r_e$), light rays leaving the source at ($t_e,r_e$) will follow radial geodesics $d\Omega=0.$  It follows that the area of the (unit) $2$-sphere emitting radiation at $r=r_e$, $t=t_e$ that reaches the mirror ${\mathcal A}$ when $a=1$, is
$f_{\mathcal A}={\mathcal A}/4\pi t_0r_e^2,$ ($r_e^2=t_0r_e^2$ when $t_0=1$,  c.f. (\ref{fA})\footnotemark[19])\footnotetext[19]{To see this, consider a packet of lightlike radial geodesics covering angular area $\Omega$, emanating from an FRW coordinate center at $r=r_e$, and evolving up to an end at time $t=t_0.$  Such curves, being radial lightlike geodesics (\ref{nullgeo}), traverse the curves $\hat{r}=2\left(\sqrt{t}-\sqrt{t_e}\right)$, $\theta=\theta_0\in \Omega$, $t_e\leq t\leq t_0,$ where $\hat{r}$ is radial distance measured from the new center $r_e,$ and $\theta$ measures angles at center $r_e$.   Now setting $\xi=\sqrt{t}-\sqrt{t_0}$, these curves project into the curves at time $t\equiv t_0$ given by $\hat{r}=2\left(\xi\right)$, $\theta=\theta_0$, $0\leq \xi\leq r_e/2.$ Since in the Standard Model, $t=t_0$ is flat Euclidean space, the latter curves, being at fixed time $t=t_0,$ are just the straight lines in ${\mathcal R}^3$ emanating from center $\hat{r}=0$, sweeping out the angular region $\Omega$ at $\hat{r}=0$ and the area ${\mathcal A}$ at $\hat{r}=r_e,$ $r=0.$ It thus follows that the area at the end is ${\mathcal A}=\bar{r}_e^2\Omega,$ where $\bar{r_e}=R(t_0)r_e=\sqrt{t_0}r_e$ is spatial distance at received time $t=t_0.$  So the fractional area is $f_{\mathcal A}=\Omega/4\pi={\mathcal A}/4\pi t_0r_e^2,$ as claimed.}. 
When $a\neq1$, the $3$-spaces at fixed time are not {\it homogeneous} and {\it isotropic} about every point like the $a=1$ FRW, and the geodesics leaving the center of a coordinate system centered at ($t_e,r_e$) will not follow $d\Omega=0$ exactly. So there is a correction factor $C_{a}$ required in the formula (\ref{fA}) for $f_{\mathcal A}$ when $a\neq1.$  The goal of this section is to prove that (\ref{CA}) gives $C_{a}$ to order $O(1)|a-1|\zeta_e^3,$ where $\zeta_e=r_e/\sqrt{t_e}.$\footnotemark[20]\footnotetext[20]{In this section we change notation, and set $\zeta_e$ equal to the fixed value $\zeta_e=r_e/\sqrt{t_e}$, the position of the fixed source, and we let $\zeta=r/\sqrt{t}$ denote a variable that runs from the mirror at the origin to the source, $0\leq\zeta\leq\zeta_e$.}  We prove this for the approximate metric (\ref{metricb}), which agrees with the exact spacetime metric (\ref{realmetric}) up to order $O(1)|a-1|\zeta_e^4$, c.f. Theorem \ref{FRWametric}.  It suffices to prove the formula for $C_a$ for the approximate metric (\ref{metricb}) instead of the exact metric (\ref{realmetric}), because the argument we give below is based on the geodesic equations of motion which only involve first derivatives of metric components.   That is, the correction to the redshift vs luminosity relation of the standard model when $a\neq1$ is given in the bracket in formula (\ref{redvslumFRW}), and our argument will show that the first order correction to the geodesic equations from the standard model will yield the first order correction in the bracket, and the second order correction to the geodesics will yield the second order correction in the bracket, c.f. the discussion in the first paragraph of Section \ref{sectionredlum}).  Thus in this section we neglect the $\zeta^4$ errors in (\ref{metricb}). 

To start, note that by spherical symmetry with $d\Omega^2=d\theta^2+\sin^2{\theta}d\phi^2,$ it suffices to treat the single angle case of metric (\ref{metricb}) with $\theta\equiv\pi/2,$ $0\leq\phi\leq2\pi,$
\begin{eqnarray}
ds^2=F_a(\zeta)^2\left\{-dt^2+tdr^2\right\}+tr^2d\phi^2,\label{metricphi}
\end{eqnarray}
\begin{eqnarray}
F_a(\zeta)^2=1+\frac{a^2-1}{4}\zeta^2,\label{Fapprox}
\end{eqnarray}  
where $\zeta=r/\sqrt{t}$, and we neglect the $\zeta^4$ errors between (\ref{Fapprox}) and (\ref{metricb}).   That is, let $\hat{\phi}$ denote the angle that corresponds to the $\phi$ coordinate in spherical coordinates centered at the vertex $r=r_e$ where the radiation is emitted, (defined precisely below).  Now imagine a circular mirror of radius $r_0$ positioned at $r=0$ transverse to, and receiving light emitted from a star positioned at $r=r_e$.  Then the light that hits the circular boundary of the mirror follows a null geodesic that starts at $r=r_e,$ $t=t_e$ and ends at $r=0$, $t=t_0$.   Looking back toward the center $r=0$ from the new center at $r=r_e$, the null geodesics emanating from $r=r_e$ that hit the circular boundary of the mirror at $r=0$ are axially symmetric about the unique radial null geodesic ($\phi\equiv0$) that connects $t=t_e$, $r=r_e$ to the center of the mirror at $t=t_0$, $r=0.$  Thus by axial symmetry, the fractional change in $\hat{\phi}$ along the specific null-geodesic emanating from $r=r_e$, $t=t_e$ that hits the circular boundary of the mirror, specified by the condition $\hat{\theta}$ is fixed at $\pi/2$, will agree with the fractional change in angle along any of the other geodesics that leave $r=r_e$, $t=t_e$ and hit the circular boundary of the mirror, because they can be obtained one from another by rotation around the axis of symmetry joining the center of the mirror to the center of the star.  Now by (\ref{fAreal}), $C_a$ is the ratio of $f_{\mathcal A}$ when $a\neq1$ to the ratio of $f_{\mathcal A}$ when $a=1$, which is just the ratio $\frac{\Omega_a}{\Omega_1}$ of the angular area $\Omega_a$ of geodesics that hit the mirror when $a\neq1$, to the angular area $\Omega_1$ of geodesics that hit the mirror when $a=1$ in the limit of small angles. It follows from these considerations that
\be\label{Cphi}
C_{a}=c_{a}^2,
\ee
where 
\be\label{Cphi1}\label{cphi}
c_{a}=\lim_{\hat{\phi}_e\rightarrow0}\frac{\hat{\phi}_0}{\hat{\phi}_{e}},
\ee 
where $\hat{\phi}_0$ is the value of $\hat{\phi}$ at $t=t_0,$ $r=0$ for a lightlike geodesic of (\ref{metricphi}) emanating from $r=r_e,$ $t=t_e$ at angle $\hat{\phi}_e$, and received at the mirror at $t=t_0$, $r=0$.  (Note that $a=1$ is in the numerator of (\ref{cphi}) and in the denominator of  $\frac{\Omega_a}{\Omega_1}$ because the ratio of {\it angular} areas at $r=r_e$ of null geodesics that hit the same area at $r=0$, is the reciprocal of the ratio of {\it areas} that get hit at $r=0$ by a fixed  angular area at $r=r_e.$)   When $a=1$, $\hat{\phi}\equiv \hat{\phi}_e$ along all radial geodesics emanating from $r=r_e$ at angle $\hat{\phi}_e$ because the FRW Standard Model is homogeneous and isotropic about every point.   So $c_{a}^2$ does indeed give the ratio of the {\it area of light received from a distant source at a mirror positioned at the origin when $a\neq1$, to the corresponding area when $a=1,$ in the limit ${\mathcal A}\rightarrow0,$} the definition of $C_{a}.$  It remains only to estimate $c_{a}.$  

Proposition \ref{areaprop} is a direct consequence of the following theorem, which together with its refinement in Theorem \ref{lmain1} below, is the main result of this appendix. 
\begin{Theorem}\label{lmain}
The constant $c_a$ satisfies the asymptotic relation
\be
c_{a}=1-\frac{a^2-1}{12}\zeta_e^2+O(1)|a-1|\zeta_e^3\ \ \ as\ \ \  \hat{r}_e\rightarrow0,\label{cphi1}
\ee
where $\zeta_e=r_e/\sqrt{t_e}.$
\end{Theorem}
\vspace{.2cm}

\noindent To prove Theorem \ref{lmain}, we must show that $c_a$ as defined in (\ref{cphi}) satisfies the asymptotic relation (\ref{cphi1}).   For this, we construct the equations for the geodesic that starts out radially at $(t_e,r_e,\hat{\phi}_e)$, neglect terms of order $\hat{\phi}_0^2$, and expand $\hat{\phi}_0/\hat{\phi}_e$ in powers of $\zeta_e$ to obtain (\ref{cphi1}).  Since  we need to estimate the evolution of $\hat{\phi}(t)$ along such a geodesic,  it is convenient to transform the metric over to spherical coordinates ($\hat{t},\hat{r},\hat{\phi}$) centered at $\hat{r}=r_e-r=0$, assuming $\hat{t}=t.$  For this, we find,  (c.f. Figure 1),

\vspace{.2in}
\centerline {
\includegraphics[width=6in]{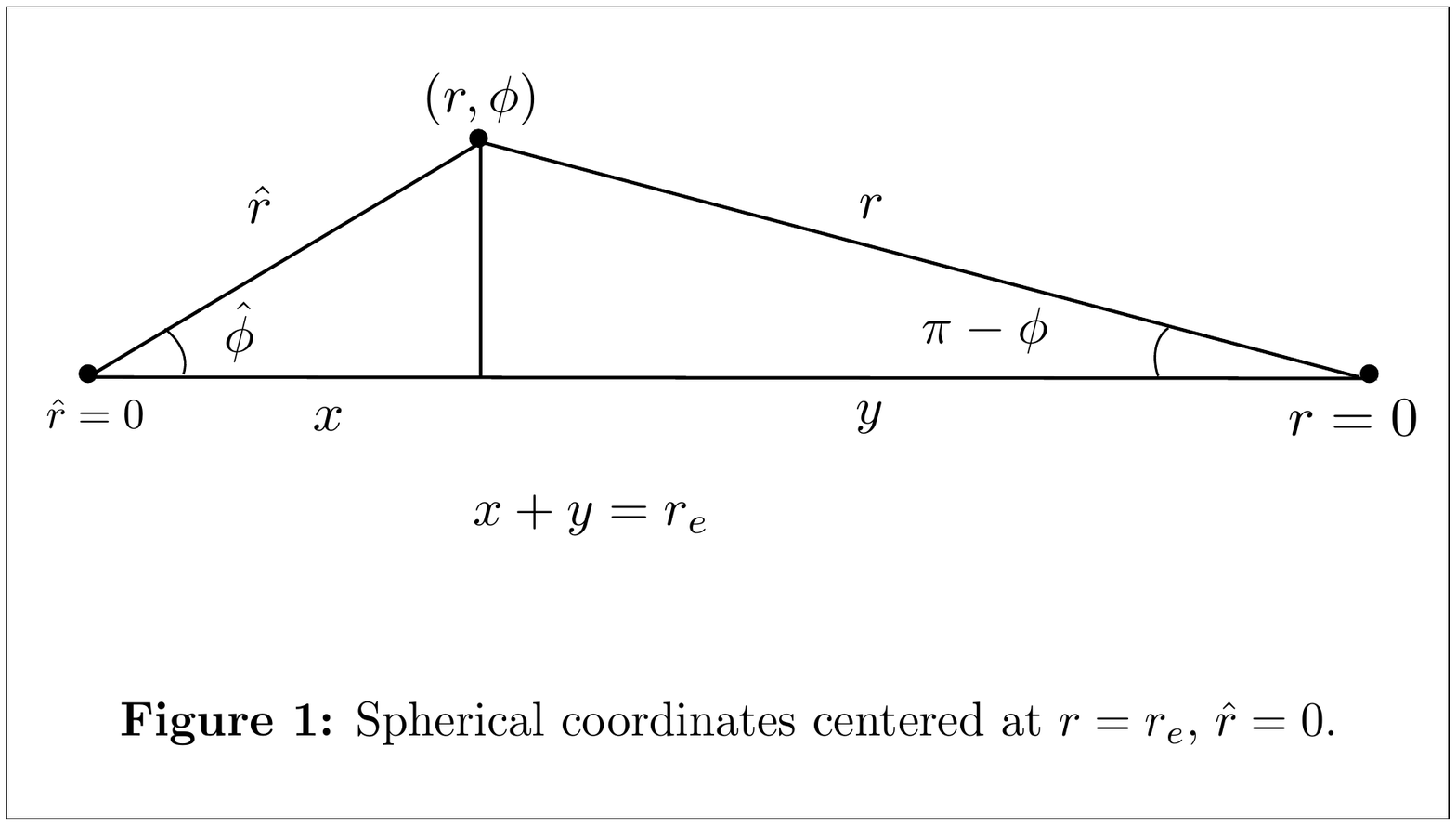}
}


\be
r\sin{\phi}&=&\hat{r}\sin{\hat{\phi}},\label{sphere1}\\
r\cos{\phi}&=&\hat{r}\cos{\hat{\phi}}-r_e,\label{sphere2}
\ee
so
\be
r^2&=&\hat{r}^2-2r_e\hat{r}\cos{\hat{\phi}}+r_e^2,\label{sphere3}\\
\tan{\phi}&=&\frac{\hat{r}\sin{\hat{\phi}}}{\hat{r}\cos{\hat{\phi}}-r_e}.\label{sphere4}
\ee
Thus the coordinate transformation from $(t,\hat{r},\hat{\phi})$ to $(t,r,\phi)$ is 
\be
t&=&t,\label{sphere}\\
r&=&\sqrt{\hat{r}^2-2r_e\hat{r}\cos{\hat{\phi}}+r_e^2},\label{sphere5}\\
\phi&=&\tan^{-1}{\left(\frac{\hat{r}\sin{\hat{\phi}}}{\hat{r}\cos{\hat{\phi}}-r_e}\right)}.\label{sphere6}
\ee
Using the notation $(x^0,x^1,x^2)=(t,r,\phi)$, $(\hat{x}^0,\hat{x}^1,\hat{x}^2)=(t,\hat{r},\hat{\phi})$, we compute
 \be
\frac{\partial x^1}{\partial \hat{x}^1}&=&\frac{\hat{r}-r_e\cos{\hat{\phi}}}{r}\label{sphere7}\\
\frac{\partial x^1}{\partial \hat{x}^2}&=&\frac{r_e\hat{r}\sin{\hat{\phi}}}{r}\label{sphere8}\\
\frac{\partial x^2}{\partial \hat{x}^1}&=&\frac{-r_e\sin{\hat{\phi}}}{(\hat{r}\cos{\hat{\phi}}-r_e)^2}~\frac{1}{\sec^2{\phi}}\label{sphere9}\\
\frac{\partial x^2}{\partial \hat{x}^2}&=&\frac{\hat{r}^2-r_e\hat{r}\cos{\hat{\phi}}}{(\hat{r}\cos{\hat{\phi}}-r_e)^2}~\frac{1}{\sec^2{\phi}},\label{sphere10}
\ee 
and
\be
\frac{\partial x^0}{\partial \hat{x}^j}=\delta^0_j,\ \ j=0,1,2.\nonumber
\ee 
Using the elementary relation,
$$
\sec{(\tan^{-1}{x})}=-\sqrt{1+x^2},
$$
 we get 
$$
\sec^2{\phi}=\frac{r^2}{(\hat{r}\cos{\hat{\phi}}-r_e)^2},
$$
by which we can replace (\ref{sphere9}), (\ref{sphere10}) with
 \be
\frac{\partial x^2}{\partial \hat{x}^1}&=&\frac{-r_e\sin{\hat{\phi}}}{r^2}\label{sphere11}\\
\frac{\partial x^2}{\partial \hat{x}^2}&=&\frac{\hat{r}^2-r_e\hat{r}\cos{\hat{\phi}}}{r^2}.\label{sphere12}
\ee 
\begin{Lemma}\label{lemmaid} The following identities hold:
\be\label{id1}
1=\left(\frac{\partial x^1}{\partial \hat{x}^1}\right)^2+r^2\left(\frac{\partial x^2}{\partial \hat{x}^1}\right)^2
\ee
\be\label{id2}
\hat{r}^2=\left(\frac{\partial x^1}{\partial \hat{x}^2}\right)^2+r^2\left(\frac{\partial x^2}{\partial \hat{x}^2}\right)^2
\ee
\be\label{id3}
0=\frac{\partial x^1}{\partial \hat{x}^1}\frac{\partial x^1}{\partial \hat{x}^2}+r^2\frac{\partial x^2}{\partial \hat{x}^1}\frac{\partial x^2}{\partial \hat{x}^2}.
\ee
\end{Lemma}
\vspace{.2cm}

\noindent{\bf Proof:}   Equations (\ref{id1})-(\ref{id3}) follow directly from (\ref{sphere7})-(\ref{sphere12}).$\Box$
\vspace{.2cm}

\noindent We omit the $t$-component, and write the resulting $2$-metric from (\ref{metricphi}) in $(r,\phi)$-coordinates as
$$
g_{ij}= diag\left\{F_a^2t,tr^2\right\},
$$
so that in $(\hat{r},\hat{\phi})$-coordinates
\be\label{ghat}
\hat{g}_{\alpha\beta}=\frac{\partial x^i}{\partial \hat{x}^{\alpha}}g_{ij}\frac{\partial x^j}{\partial \hat{x}^{\beta}}.
\ee
 \begin{Lemma}\label{lemmaghat} For $\alpha,\beta=1,2$,  $\hat{g}_{\alpha\beta}$ satisfies
\be\label{ghat1}
\hat{g}_{11}&=&t\left\{(F_a^2-1)\left(\frac{\hat{r}-r_e\cos{\hat{\phi}}}{r}\right)^2+1\right\}\\
\label{ghat2}
\hat{g}_{22}&=&t\left\{(F_a^2-1)\frac{r_e^2\hat{r}^2\sin^2{\hat{\phi}}}{r^2}+\hat{r}^2\right\}\\
\label{ghat3}
\hat{g}_{12}&=&t\left\{(F_a^2-1)\frac{(\hat{r}-r_e\cos{\hat{\phi}})(r_e\hat{r}\sin{\hat{\phi}})}{r^2}\right\}.
\ee
\end{Lemma}
\vspace{.2cm}

\noindent{\bf Proof:}  By (\ref{ghat}),
$$
\hat{g}_{11}=\hat{g}_{ii}\left(\frac{\partial x^i}{\partial \hat{x}^1}\right)^2=t\left\{(F_a^2-1)\left(\frac{\partial x^1}{\partial \hat{x}^1}\right)^2+\left(\frac{\partial x^1}{\partial \hat{x}^1}\right)^2+r^2\left(\frac{\partial x^2}{\partial \hat{x}^1}\right)^2\right\},
$$
and using identity (\ref{id1}) gives (\ref{ghat1}).  Similarly,
$$
\hat{g}_{22}=\hat{g}_{ii}\left(\frac{\partial x^i}{\partial \hat{x}^2}\right)^2=t\left\{(F_a^2-1)\left(\frac{\partial x^1}{\partial \hat{x}^2}\right)^2+\left(\frac{\partial x^1}{\partial \hat{x}^2}\right)^2+r^2\left(\frac{\partial x^2}{\partial \hat{x}^2}\right)^2\right\},
$$
and using identity (\ref{id2}) gives (\ref{ghat2}).  Finally,
$$
\hat{g}_{12}=\hat{g}_{ii}\frac{\partial x^i}{\partial \hat{x}^1}\frac{\partial x^i}{\partial \hat{x}^2}=t\left\{(F_a^2-1)\frac{\partial x^1}{\partial \hat{x}^1}\frac{\partial x^1}{\partial \hat{x}^2}+\frac{\partial x^1}{\partial \hat{x}^1}\frac{\partial x^1}{\partial \hat{x}^2}+r^2\frac{\partial x^2}{\partial \hat{x}^1}\frac{\partial x^2}{\partial \hat{x}^2}\right\},
$$
and using identity (\ref{id3}) gives (\ref{ghat3}). $\Box$ 
\vspace{.2cm}

\noindent It now follows that the metric (\ref{metricphi}) under transformation to $r_e$-centered coordinates $(t,\hat{r},\hat{\phi})$ goes over to the metric $\hat{g}$ given by
\begin{eqnarray}
&&ds^2=-F_a^2~dt^2+F^2_at~d\hat{r}^2+t\hat{r}^2~d\hat{\phi}^2\nonumber\\
&&\ \ \ \ \ \ \ \ \ \ \ \ \ \ \ \ \ +\frac{a^2-1}{4}\left\{-r_e^2\sin^2{\hat{\phi}}~d\hat{r}^2+r_e^2\hat{r}^2\sin^2{\hat{\phi}}~d\hat{\phi}^2\right.\label{metricphihat}\\
&&\ \ \ \ \ \ \ \ \ \ \ \ \ \ \ \ \ \ \ \ \ \ \ \ \ \ \ \ \ \ \ \ \ \ \ \ +\left.2(\hat{r}-r_e\cos{\hat{\phi}})r_e\hat{r}\sin{\hat{\phi}}~d\hat{r}d\hat{\phi}\right\},\nonumber
\end{eqnarray}
where we have simplified the $d\hat{r}^2$ component using
$$
(\hat{r}-r_e\cos{\hat{\phi}})^2=r^2-r_e^2\sin^2{\hat{\phi}}.
$$

To compute $c_a$, we need the equations for the null geodesic of (\ref{metricphihat}) that starts out in the radial direction from the point $(t_e,\hat{r}_e,\hat{\phi}_e)$ toward the original origin $r=0$ where the mirror ${\mathcal A}$ is positioned, the point labeled $(t_0,r_e,0)$ in $(t,\hat{r},\hat{\phi})$-coordinates.  To this end, using the notation  $(x^0,x^1,x^2)=(t,\hat{r},\hat{\phi})$, the three geodesic equations take the form
\be
\frac{d^2t}{d\lambda^2}&=&-\Gamma^0_{ij}\frac{d x^i}{d\lambda}\frac{d x^j}{d\lambda}\label{geo0}\\
\frac{d^2\hat{r}}{d\lambda^2}&=&-\Gamma^1_{ij}\frac{d x^i}{d\lambda}\frac{d x^j}{d\lambda}\label{geo1}\\
\frac{d^2\hat{\phi}}{d\lambda^2}&=&-\Gamma^2_{ij}\frac{d x^i}{d\lambda}\frac{d x^j}{d\lambda},\label{geo2}
\ee
where\footnotemark[21]\footnotetext[21]{We use the Einstein summation convention where by repeated up down indices are assumed summed, and indices are raised and lowered with the metric, \cite{wein}.}
\be
\Gamma^k_{ij}=\frac{1}{2}\hat{g}^{k\sigma}\left\{-\hat{g}_{ij,\sigma}+\hat{g}_{\sigma i,j}+\hat{g}_{j\sigma,i}\right\},\label{Gammas}\label{Gamma}
\ee
are the Christoffel symbols.   We restrict to null geodesics that satisfy the null condition $ds^2=0$, which holds identically on solutions of (\ref{geo1})-(\ref{geo2}) so long as it holds initially, \cite{wein}.   We call the {\it baseline null geodesic}  the radial null geodesic starting at $t=t_e$, $\hat{r}=0$, on which $\hat{\phi}\equiv0$, $\phi\equiv0$.  
The condition $d^2s=0$ when $\phi\equiv0$ imposes a condition independent of $a$,
\be
0=-dt^2+td\hat{r}^2,
\ee
which when integrated, assuming initial conditions $t=t_e$ when $\hat{r}=0$, gives the $a$-independent relation
\be
t=t_*(\hat{r})\equiv\left(\frac{1}{2}\hat{r}+\sqrt{t_e}\right)^2,\label{tsuba}
\ee
which holds on the baseline null geodesic.  Since $t=t_0$ when $\hat{r}=r_e,$ we also have
\be\label{ttzero}
t_0=\left(\frac{1}{2}r_e+\sqrt{t_e}\right)^2,
\ee 
for every $a.$   

Now although $t$ as a function of $\hat{r}$ along the baseline null geodesics is independent of $a$, complications arise because the relation between $\hat{r}$ and the natural geodesic parameter $\lambda$ depends on $a.$  To be precise, define the baseline null geodesic to be the solution 
\be\label{baselineformula}
(t,\hat{r},\hat{\phi})=(t_a(\lambda),\hat{r}_a(\lambda),0),
\ee
 of (\ref{geo0})-(\ref{geo2}),
\be\label{baselineformula2}
t_a(\lambda)=t_*(\hat{r}(\lambda)),
\ee
satisfying the initial conditions
\be
&t_a(0)=t_e,\ \  \frac{dt_a}{d\lambda}(0)=\frac{dt}{d\hat{r}}\frac{d\hat{r}}{d\lambda}|_{_{\lambda=0}}=\sqrt{t_e},&\\
&\hat{r}_a(0)=0,\ \ \frac{dr_a}{d\lambda}(0)=1,&\\
&\hat{\phi}(0)=0,\ \ \frac{d\hat{\phi}}{d\lambda}(0)=0.&
\ee

\begin{Lemma}\label{ageo}  The baseline null geodesic $\hat{r}_a$ satisfies the following identity used in the analysis of the $\hat{\phi}$-geodesic below:
\be\nonumber
\left(\frac{d\hat{r}_a}{d\lambda}\right)^{-1}\frac{d}{d\hat{r}}\frac{d\hat{r}_a}{d\lambda}&=&-\frac{1}{\sqrt{t}F_a^2}\left\{1-\frac{(a^2-1)(r_e-\hat{r})}{4\sqrt{t}}\right\}\\&=&-\frac{1}{\sqrt{t_e}}\left\{1+O(1)|a-1|\zeta_e\right\}\label{ageo3}.
\ee
More precisely, in the case $a=1$, the  baseline null geodesic is given by 
\be\label{tone}
t_1(\lambda)=t_e\left(1+\frac{3}{2\sqrt{t_e}}\lambda\right)^{2/3}=t_e\left(1+\frac{1}{\sqrt{t_e}}\lambda+O(1)\zeta_e^2\right),
\ee
\be\label{rone}
\hat{r}_1(\lambda)=2\sqrt{t_e}\left\{\left(1+\frac{3}{2\sqrt{t_e}}\lambda\right)^{1/3}-1\right\}=\lambda\left(1-\frac{1}{2\sqrt{t_e}}\lambda+O(1)\zeta_e^2\right),
\ee
and satisfies

\be\label{roneprime}
\frac{d\hat{r}_1}{d\lambda}=\left(1+\frac{3}{2\sqrt{t_e}}\lambda\right)^{-2/3}=1-\frac{1}{\sqrt{t_e}}\lambda+O(1)\zeta_e^2.
\ee
When $a\neq1$ we have
\be 
\frac{d\hat{r}_a}{d\lambda}(\lambda)=\frac{d\hat{r}_1}{d\lambda}+O(1)|a-1|\zeta_e^2,\label{ageo1}
\ee
and
\be\label{ageo2}
\hat{r}_a(\lambda)=\hat{r}_1(\lambda)+O(1)|a-1|\zeta_e^2.
\ee
\end{Lemma}
\vspace{.2cm}

\noindent {\bf Proof:}  The baseline null geodesic $r_a(\lambda)$ satisfies the geodesic equation
\be\label{icondtn}
\frac{d^2\hat{r}}{d\lambda^2}=-H^1_{ij}\frac{d\hat{x}^i}{d\lambda}\frac{d\hat{x}^j}{d\lambda},\ \ \ \hat{r}_a(0)=0,\ \ \ \frac{d\hat{r}_a}{d\lambda}=1,
\ee
where for this argument we let $H^1_{ij}$ denote the Christoffel symbols for the $2$-metric $h$ taken to be
\be\nonumber
ds^2=-F_a^2dt^2+F_a^2td\hat{r}^2=h_{ij}d\hat{x}^id\hat{x}^j,
\ee
where again $$F_a^2=1+\frac{a^2-1}{4}\frac{(r_e-\hat{r})^2}{t}=1+O(1)|a-1|\zeta_e^2,$$
$$\zeta^2=\frac{(r_e-\hat{r})^2}{t},\ \ \ \zeta_e^2=\frac{r_e^2}{t_e}.$$
Thus
\be
H^1_{00}&=&\frac{1}{2}h^{11}\left\{-h_{00,1}+h_{10,0}+h_{01,0}\right\}=\frac{1}{2}\frac{1}{F^2_at}\frac{\partial \left(F_a^2\right)}{\partial\hat{r}}=-\frac{(a^2-1)(r_e-\hat{r})}{8t^2F_a^2},\label{baseline1}\nonumber\\
H^1_{01}&=&\frac{1}{2}h^{11}\left\{-h_{01,1}+h_{10,1}+h_{11,0}\right\}=\frac{1}{2}\frac{1}{F^2_at}\frac{\partial \left(F_a^2t\right)}{\partial t}=\frac{1}{2tF_a^2},\label{baseline2}\nonumber\\
H^1_{11}&=&\frac{1}{2}h^{11}\left\{-h_{11,1}+h_{11,1}+h_{11,1}\right\}=\frac{1}{2}\frac{1}{F^2_at}\frac{\partial \left(F_a^2t\right)}{\partial\hat{r}}=-\frac{(a^2-1)(r_e-\hat{r})}{8tF_a^2}.\label{baseline3}\nonumber
\ee 
Putting these values in (\ref{icondtn}), multiplying through by $\left(\frac{d\hat{r}}{d\lambda}\right)^{-2}$ and using 
$$
\frac{dt}{d\hat{r}}=\sqrt{t},
$$
(c.f. (\ref{tsuba})), gives
\be\nonumber
\left(\frac{d\hat{r}_a}{d\lambda}\right)^{-1}\frac{d}{d\hat{r}}\frac{d\hat{r}_a}{d\lambda}&=&-H^1_{00}t-2H^1_{o1}\sqrt{t}-H^1_{11}\nonumber\\
&=&\frac{(a^2-1)(r_e-\hat{r})}{8F_a^2t}-\frac{1}{F_a^2\sqrt{t}}+\frac{(a^2-1)(r_e-\hat{r})}{8F_a^2t},\nonumber
\ee
which yields
\be\label{raeqnfinal}
\frac{d}{d\hat{r}}\frac{d\hat{r}_a}{d\lambda}=-\frac{1}{\sqrt{t}F_a^2}\left\{1-\frac{(a^2-1)(r_e-\hat{r})}{4\sqrt{t}}\right\}\left(\frac{d\hat{r}_a}{d\lambda}\right),
\ee
and this implies (\ref{ageo3}) by easy estimates.  Now (\ref{raeqnfinal}) is a first order ODE for $\frac{d\hat{r}_a}{d\lambda}$ with the initial condition $\ \frac{d\hat{r}_a}{d\lambda}(0)=1.$ The initial condition $\hat{r}_a(0)=0$ then determines $\hat{r}_a$ from  $\frac{d\hat{r}_a}{d\lambda}$. 

In the case $a=1$, (\ref{raeqnfinal}) reduces to
\be\label{raeqnfinal1}
\frac{d}{d\hat{r}}\frac{d\hat{r}_1}{d\lambda}=-\frac{1}{\sqrt{t}}\left(\frac{d\hat{r}_1}{d\lambda}\right),
\ee
which integrates exactly as follows.  First set $u=\frac{d\hat{r}_a}{d\lambda}$ so (\ref{raeqnfinal1}) becomes
$$
\frac{du}{d\hat{r}}=-\frac{1}{\sqrt{t}}u,
$$
then use $v=\sqrt{t}=\frac{1}{2}\hat{r}+\sqrt{t_e}$, $dv=\frac{1}{2}d\hat{r}$ to get
\be\nonumber
\int_{u_e}^{u}\frac{du}{u}=-\int_{\sqrt{t_e}}^{v}\frac{2dv}{v},
\ee
which, using $u_e=1,$  integrates to
\be\label{nexteqn}
\frac{d\hat{r}}{d\lambda}=u=\frac{t_e}{t}.
\ee
Solving (\ref{nexteqn}) gives $t=\left(\frac{1}{2}\hat{r}+\sqrt{t_e}\right)^2$, $dt=\sqrt{t}d\hat{r}$ giving
$$
t_{e}d\lambda=\sqrt{t}dt.
$$
Using $\lambda_e=0$, this integrates to
\be\label{lastlast}
\lambda=\frac{2}{3}\sqrt{t_e}\left\{\left(\frac{t}{t_e}\right)^{3/2}-1\right\},
\ee 
which upon solving for $t$ gives $t=t_1(\lambda)$ in agreement with (\ref{tone}).  Using $t=\left(\frac{1}{2}\hat{r}+\sqrt{t_e}\right)^2$ in (\ref{tone}) then gives (\ref{rone}), thereby also implying (\ref{roneprime}).  

To obtain (\ref{ageo1}), write (\ref{raeqnfinal}) in the form
\be\label{raeqnfinala}
\frac{d}{d\hat{r}}\frac{d\hat{r}_a}{d\lambda}=-\frac{1}{\sqrt{t}}\left\{1-O(1)|a-1|\zeta_e\right\}\left(\frac{d\hat{r}_a}{d\lambda}\right).
\ee
Following the argument above then leads to
\be\label{details0}
u=\frac{d\hat{r}_a}{d\lambda}=\frac{t_e}{t}+O(1)|a-1|\zeta_e^2,
\ee
which directly implies (\ref{ageo1}).  To see this, write (\ref{raeqnfinala}) as
\be\label{details1}
\frac{du}{d\hat{r}}=-\frac{u}{\sqrt{t}}\left\{1+O(1)\zeta_e\right\},
\ee
where we have incorporated $|a-1|$ into the $O(1)$.  Integrating (\ref{details1}) gives
$$
\int_{u_e}^u\frac{du}{u}=-\int_{\sqrt{t_e}}^v\frac{2dv}{v}(1+O(1)\zeta_e),
$$
which leads to
$$
u=\left(\frac{v}{\sqrt{t_e}}\right)^{-2}e^{\zeta_e\int_{\sqrt{t_e}}^v\frac{O(1)dv}{v}}.
$$
But we can estimate
$$
\left|\int_{\sqrt{t_e}}^v\frac{O(1)dv}{v}\right|\leq O(1)\int_{\sqrt{t_e}}^v\frac{dv}{v}\leq O(1)\ln{\frac{v}{\sqrt{t_e}}},
$$
so

$$
\left|e^{\zeta_e\int_{\sqrt{t_e}}^v\frac{O(1)dv}{v}}\right|\leq e^{O(1)\zeta_e\ln{\frac{v}{\sqrt{t_e}}}}=\left(\frac{v}{\sqrt{t_e}}\right)^{O(1)\zeta_e}=\left(1+\frac{\sqrt{t}-\sqrt{t_e}}{\sqrt{t_e}}\right)^{O(1)\zeta_e},
$$
and expanding leads to
\be\nonumber
\left|e^{\zeta_e\int_{\sqrt{t_e}}^v\frac{O(1)dv}{v}}\right|\leq 1+\frac{\sqrt{t}-\sqrt{t_e}}{\sqrt{t_e}}O(1)\zeta_e+O(1)\left(\frac{\sqrt{t}-\sqrt{t_e}}{\sqrt{t_e}}O(1)\zeta_e\right)\zeta_e,
\ee
which gives (\ref{details0}) because $\sqrt{t}-\sqrt{t_e}=\hat{r}/2=O(\hat{r_e}).$
 Integrating (\ref{ageo1}) gives (\ref{ageo2}) by easy estimates.    This completes the proof of Lemma \ref{ageo}.  $\Box$
\vspace{.2cm}

By a {\it small angle perturbation of the baseline null geodesic} we mean a null geodesic solution $(t,\hat{r},\hat{\phi})$ of (\ref{geo0})-(\ref{geo2}) defined for $0\leq\hat{r}\leq r_e$, satisfying the initial conditions at $\lambda=0,$ 
\be
(t,\hat{r},\hat{\phi})|_{_{\lambda=0}}=(t_e,0,\epsilon)\label{init1}
\ee
\be
\frac{d}{d\lambda}(t,\hat{r},\hat{\phi})|_{_{\lambda=0}}=\left(\left.\frac{dt_a}{d\lambda}\right|_{_{\lambda=0}}+O(\epsilon^2),~\left.\frac{d\hat{r}_a}{d\lambda}\right|_{_{\lambda=0}}+O(\epsilon^2),~0\right),\label{init2}
\ee
where we set the perturbation parameter $\epsilon$ equal to $\hat{\phi}_e.$  To estimate such geodesices at fixed $a$, we now define a consistent $O(\epsilon^2)$ asymptotic approximation for the small angle perturbation of the baseline null geodesic.  That is, we find an approximate solution involving a function $\bar{\phi}(\lambda)$ which we show satisfies the null geodesic equations to order $O(\epsilon^2)$ when $\epsilon=\hat{\phi}_e$, such that $\hat{\phi}(\lambda)=\epsilon\bar{\phi}(\lambda)$.  We accomplish this by defining an ansatz which satisfies (\ref{geo0}) and (\ref{geo1}) to $O(\epsilon^2)$, and such that setting the leading order $O(\epsilon)$ term in (\ref{geo2}) equal to zero, gives an equation for $\bar{\phi}(\lambda)$ which when satisfied, reduces (\ref{geo2}) to $O(\epsilon^2)$ as well.  

Our goal is now to prove the following theorem, which is a refinement of Theorem \ref{lmain}.

\begin{Theorem} \label{lfinal} \label{final}\label{lmain1} Fix $a$ and set $\epsilon\equiv\hat{\phi}_e$.
Then the following asymptotic approximation is valid for the geodesic equations (\ref{geo0})-(\ref{geo2}) as $\epsilon\rightarrow0$:
\be
\hat{r}(\lambda)&=&\hat{r}_a(\lambda)+\epsilon^2\bar{r}(\lambda)\label{asymr}\\
t(\hat{r})&=&t_a(\lambda)+\epsilon^2\bar{t}(\hat{r}),\label{asymt}\\
\hat{\phi}(\hat{r})&=&\epsilon\bar{\phi}(\hat{r}),\label{asymphi1}
\ee
where $\bar{r}(\lambda)$, $\bar{t}(\hat{r})$ and $\bar{\phi}(\hat{r})$ are $O(1)$ as $\epsilon\rightarrow0.$  More precisely, the approximate solution $\hat{r}=\hat{r}_a(\lambda)$, $t=t_a(\lambda)$, (defined in (\ref{baselineformula}), (\ref{baselineformula2})),  solve the $t$ and $\hat{r}$ geodesic equations (\ref{geo0}) and (\ref{geo1}) to order $O(\epsilon^2)$, and the $\hat{\phi}$ geodesic equation (\ref{geo2}) to order $\epsilon.$  Moreover, the $\epsilon$-order equation determined by (\ref{geo2}) then determines $\bar{\phi}$ to third order in $\zeta$ by
\be\label{barphianswer}
\bar{\phi}(r_e)&=&1-\frac{a^2-1}{12}\zeta_e^2+O(1)|a-1|\zeta_e^3\ \ \ as\ \ \ r_e\rightarrow0,\ \ \ 
\ee
$$\zeta_e\equiv\frac{r_e}{\sqrt{t_e}}.$$
Said differently, if $\bar{\phi}$ solves the $\hat{\phi}$-geodesic equation (\ref{geo2}) to order $\epsilon^2,$ then it must satisfy (\ref{barphianswer}) in the limit $\epsilon\rightarrow0,$ thereby verifying that (\ref{barphianswer}) is correct asymptotically in the limit $\epsilon\rightarrow0.$
\end{Theorem}
\vspace{.2cm}

\noindent Since $\epsilon=\hat{\phi}_e$ and $\bar{\phi}(r_e)=\hat{\phi}_0/\hat{\phi}_e,$ it follows from (\ref{cphi}) that 
$$c_a=\lim_{\epsilon\rightarrow0}\frac{\hat{\phi}_0}{\hat{\phi}_e}=\bar{\phi}(r_e),$$
and so (\ref{barphianswer}) verifies the asymptotic relation (\ref{cphi1}).  Thus Theorem \ref{lmain1} implies Theorem \ref{lmain}, and is more accurately a refined restatement of it.  
\vspace{.2cm}

\noindent In the proof of Theorem \ref{lfinal} we will need the following lemma.

\begin{Lemma}\label{lemmat} Assume $0\leq\hat{r}\leq r_e$, $t_e\leq t\leq t_0$, $\zeta_e\equiv\frac{r_e}{\sqrt{t_e}}$.  Then the ansatz (\ref{asymr})-(\ref{asymphi1}) implies the following asymptotic estimates:
\be\label{tzero}
t(\hat{r})=t_e\left\{1+\frac{\hat{r}}{2\sqrt{t_e}}\right\}^2+O(\epsilon^2)=t_e\left\{1+O(\zeta_e)\right\}\ \ \ as\ \ \ r_e,\epsilon\rightarrow0,
\ee
\be\label{dtdbarr}
\frac{dt}{d\hat{r}}=\sqrt{t}+O(\epsilon^2)\ \ \ as\ \ \ r_e,\epsilon\rightarrow0,
\ee
\end{Lemma}
\vspace{.2cm}

\noindent {\bf Proof:}  From (\ref{asymt}),
\be\label{tzero1}
t=\left(\frac{1}{2}r+\sqrt{t_e}\right)^2+O(\epsilon^2)=t_e\left(1+\frac{1}{2}\frac{\hat{r}}{\sqrt{t_e}}\right)^2+O(\epsilon^2),
\ee 
which implies (\ref{tzero}) for $t_e\leq t\leq t_0$.  Differentiating (\ref{tzero}) gives,
\be\label{tzero2}
\frac{dt}{d\hat{r}}&=&\frac{d}{d\hat{r}}\left(\frac{1}{2}\hat{r}+\sqrt{t_e}\right)^2+O(\epsilon^2)=\left(\frac{1}{2}\hat{r}+\sqrt{t_e}\right)+O(\epsilon^2)\nonumber\\
&=&\sqrt{t}+O(\epsilon^2)\ \ \ as\ \ \ r_e\epsilon\rightarrow0\nonumber
\ee 
as claimed in (\ref{dtdbarr}).
\vspace{.2cm}$\Box$
\vspace{.2cm}

\noindent It remains only to give the
\vspace{.2cm}

\noindent {\bf Proof of Theorem \ref{lfinal}:}    To analyze small angle perturbations of the baseline null  geodesics we need expressions for the Christoffel symbols (\ref{Gammas}).  Since we are only looking to solve the geodesic equations (\ref{geo0})-(\ref{geo2}) to order $\epsilon,$ we only need values for (\ref{Gammas}) up to order $\epsilon$,  but since the $\Gamma$'s involve derivatives with respect to $\hat{\phi}$, this requires that we use values for $\hat{g}_{ij}$ in (\ref{metricphihat}) up to order $\epsilon^2$.  On the other hand, the inverse matrix $\hat{g}^{-1}\equiv \hat{g}^{ij}$ enters (\ref{Gammas}) undifferentiated, so we need only compute $\hat{g}^{ij}$ to order $\epsilon$ .   Using that $\hat{\phi}=\bar{\phi}\epsilon=O(1)\epsilon,$ we can write (\ref{metricphihat}) in the matrix form,

\be\label{matrixhatg0}
\hat{g}_{ij}=\left(\begin{array}{ccc}
-F_a^2&0&0\\\\
0&F_a^2t-\frac{a^2-1}{4}r_e^2\hat{\phi}^2&\frac{a^2-1}{4}(\hat{r}-r_e)\hat{r}r_e\hat{\phi}\\\\ 
0&\frac{a^2-1}{4}(\hat{r}-r_e)\hat{r}r_e\hat{\phi}&t\hat{r}^2+O(\epsilon^2)\end{array}\right)+O(\epsilon^3),\label{ghatorder3}
\ee
which to order $\epsilon$ reduces to
\be
\hat{g}_{ij}=\left(\begin{array}{ccc}
-F_a^2&0&0\\\\
0&F_a^2t&\frac{a^2-1}{4}(\hat{r}-r_e)\hat{r}r_e\hat{\phi}\\\\ 
0&\frac{a^2-1}{4}(\hat{r}-r_e)\hat{r}r_e\hat{\phi}&t\hat{r}^2\end{array}\right)+O(\epsilon^2).\label{ghatorder2}
\ee
To compute the inverse $\hat{g}^{ij}$ to order $\epsilon,$ it suffices to take the inverse of (\ref{ghatorder2}), and a straightforward calculation gives
\be
\hat{g}^{ij}=\left(\begin{array}{ccc}
-\frac{1}{F_a^2}&0&0\\\\
0&\frac{1}{F_a^2t}&\frac{a^2-1}{4t^2F_a^2}\frac{(r_e-\hat{r})r_e}{\hat{r}}\hat{\phi}\\\\ 
0&\frac{a^2-1}{4t^2F_a^2}\frac{(r_e-\hat{r})r_e}{\hat{r}}\hat{\phi}&\frac{1}{t\hat{r}^2}\end{array}\right)+O(\epsilon^2).\label{matrixhatg}
\ee
\vspace{.2cm}

The next lemma records the relevant values of $\Gamma^{k}_{ij}$ to within errors necessary for evaluation of the $O(\epsilon)$ part of (\ref{geo2}) in the limit as $r_e,\epsilon\rightarrow0$.  

\begin{Lemma}\label{bracketlemma}\label{gammasall} As $r_e,\epsilon\rightarrow0$ we have
\be
\Gamma^2_{11}&=&\frac{a^2-1}{2}\,\frac{r_e}{t\hat{r}}\,\hat{\phi}-\frac{1}{2\hat{r}^2}\frac{\partial}{\partial \hat{\phi}}\left(F_a^2\right)t+E^2_{11},\label{G211}\\
\Gamma^2_{00}&=&\frac{1}{2\hat{r}^2t}\frac{\partial}{\partial \hat{\phi}}\left(F_a^2\right)+E^2_{00},\label{G200}\\
\Gamma^2_{01}&=&E^2_{01},\label{G201}\\
\Gamma^2_{02}&=&\frac{1}{2t}+O(\epsilon^2),\label{G022}\\
\Gamma^2_{12}&=&\frac{1}{\hat{r}}+O(\epsilon^2),\label{G122}
\ee
where the errors $E^2_{ij}$ are given by
\be\nonumber
E^2_{11}&=&O(1)\frac{|a-1|r_e^3}{t^2\hat{r}}\epsilon+O(\epsilon^2),\label{E211}\\\nonumber
E^2_{00}&=&O(1)\frac{|a-1|r_e^3}{t^3\hat{r}}\epsilon+O(\epsilon^2),\label{E200}\\\nonumber
E^2_{01}&=&O(1)\frac{|a-1|r_e^2}{t^2\hat{r}}\epsilon+O(\epsilon^2).\label{E201}\nonumber
\ee

\end{Lemma}
\vspace{.2cm}

\noindent {\bf Proof:}   \noindent Let the second term in ($\ref{Gamma}$) involving derivatives of the metric be denoted
\be\label{brackets}
\{ij,k\}\equiv \left\{-\hat{g}_{ij,k}+\hat{g}_{ki,j}+\hat{g}_{jk,i}\right\}.
\ee
To estimate these brackets we use the following easily verified identities:
\be\label{startidentities}
\hat{g}^{21}&=&\frac{a^2-1}{4F_a^2}\frac{(r_e-\hat{r})r_e}{t^2\hat{r}}\hat{\phi}+O(\epsilon^2),\nonumber\\
\hat{g}^{22}&=&\frac{1}{t\hat{r}^2}+O(\epsilon^2),\nonumber\\
\frac{\partial}{\partial\hat{r}}(F_a^2)&=&\left(\frac{a^2-1}{4}\right)\left(\frac{\hat{r}-r_e(1-\cos{\hat{\phi}})}{t}\right)+O(\epsilon^2),\nonumber\\
\frac{\partial}{\partial\hat{t}}(F_a^2)&=&\left(\frac{a^2-1}{4}\right)\left(\frac{\hat{r}-r_e(1-\cos{\hat{\phi}})}{t}\right)+O(\epsilon^2),\nonumber\\
\frac{\partial}{\partial\hat{\phi}}(F_a^2)&=&\left(\frac{a^2-1}{4}\right)\left(\frac{\hat{r}-r_e(1-\cos{\hat{\phi}})}{t}\right)+O(\epsilon^2),\nonumber
\ee
A straightforward calculation using these together with $\hat{g}_{ij}$ given up to order $\epsilon^2$ in (\ref{ghatorder3}), gives the following values for (\ref{brackets}):
\be
\begin{array}{ll}
\{11,2\}=(a^2-1)\hat{r}r_e\hat{\phi}-\frac{\partial}{\partial\hat{\phi}}F_a^2t\ \ \ \ &\{11,1\}=\frac{\partial}{\partial\hat{r}}(F_a^2)t\\
\{00,2\}=\frac{\partial}{\partial\hat{\phi}}(F_a^2)&\{00,1\}=\frac{\partial}{\partial\hat{r}}(F_a)^2\\
\{01,2\}=0&\{01,1\}=1\\
\{12,2\}=2t\hat{r}&\{12,1\}=O(\hat{\phi})\\
\{02,2\}=\hat{r}^2&\{02,1\}=0.
\end{array}\label{bracket21}
\ee
For example, the caclulation of $\{11,2\}$ above entails an interesting cancellation at the leading order as follows.  By (\ref{brackets}),
\be
\{11,2\}\equiv \left\{-\hat{g}_{11,2}+2\hat{g}_{21,1}\right\},
\ee
and by (\ref{ghatorder3}),
\be\label{g112}
-\hat{g}_{11,2}=2\frac{a^2-1}{4}r_e^2\hat{\phi}-\frac{\partial}{\partial \hat{\phi}}(F_a^2)t,
\ee
and
\be
2\hat{g}_{21,1}&=&2\frac{\partial}{\partial \hat{r}}\left\{\frac{a^2-1}{4}(\hat{r}-r_e)r_e\hat{r}\hat{\phi}\right\}\nonumber\\
&=&-2\frac{a^2-1}{4}r_e^2\hat{\phi}+(a^2-1)r_e\hat{r}\hat{\phi}.\label{2g211}
\ee
Adding (\ref{g112}) and (\ref{2g211}) verifies $\{11,2\}$ in (\ref{bracket21}). (Note the cancellation at leading order.)

Using the values in (\ref{bracket21}) we can compute:

\be\label{Gamma11}
\Gamma^2_{11}&=&\frac{1}{2}\hat{g}^{22}\{11,2\}+\frac{1}{2}\hat{g}^{21}\{11,1\}\nonumber\\
&=&\frac{a^2-1}{2}\frac{r_e}{t\hat{r}}\hat{\phi}-\frac{1}{2\hat{r}^2}\frac{d}{d\hat{\phi}}(F_a^2)+E^2_{11},
\ee
where
\be\nonumber
E^2_{11}=\frac{a^2-1}{8F_a^2}\frac{(r_e-\hat{r})r_e}{t\hat{r}}\frac{\partial}{\partial\hat{r}}(F_a^2)\hat{\phi}+O(\epsilon^2),
\ee
from which (\ref{G211}) and (\ref{E211}) follow.
\be\label{Gamma00}
\Gamma^2_{00}&=&\frac{1}{2}\hat{g}^{22}\{00,2\}+\frac{1}{2}\hat{g}^{21}\{00,1\}\nonumber\\
&=&\frac{1}{2\hat{r}^2t}\frac{d}{d\hat{\phi}}(F_a^2)+E^2_{00}
\ee
where
\be\nonumber
E^2_{00}=\frac{a^2-1}{8F_a^2}\frac{(r_e-\hat{r})r_e}{t^2\hat{r}}\frac{\partial}{\partial\hat{r}}(F_a^2)\hat{\phi}+O(\epsilon^2),
\ee
from which (\ref{G200}) and (\ref{E200}) follow.
\be\label{Gamma01}
\Gamma^2_{01}=\frac{1}{2}\hat{g}^{22}\{01,2\}+\frac{1}{2}\hat{g}^{21}\{01,1\}=E^2_{01}
\ee
where
\be\nonumber
E^2_{01}=\frac{a^2-1}{8F_a^2}\frac{(r_e-\hat{r})r_e}{t^2\hat{r}}\hat{\phi}+O(\epsilon^2),
\ee
from which (\ref{G201}) and (\ref{E201}) follow.
\be\label{Gamma12}
\Gamma^2_{12}&=&\frac{1}{2}\hat{g}^{22}\{12,2\}+\frac{1}{2}\hat{g}^{21}\{12,1\}\nonumber\\
&=&\frac{1}{\hat{r}}
+\frac{a^2-1}{8F_a^2}\frac{(r_e-\hat{r})r_e}{t^2\hat{r}}\hat{\phi}O(\hat{\phi})+O(\epsilon^2)\nonumber\\
&=&\frac{1}{\hat{r}}
+O(\epsilon^2)
\ee
giving (\ref{G122}).  And finally,
\be\label{Gamma02}
\Gamma^2_{02}&=&\frac{1}{2}\hat{g}^{22}\{02,2\}+\frac{1}{2}\hat{g}^{21}\{02,1\}\nonumber\\
&=&\frac{1}{2t}+O(\epsilon^2),
\ee
verifies (\ref{G022}) and the proof of the Lemma \ref{gammasall} is complete. $\Box$  
\vspace{.2cm}

\noindent To prove Theorem \ref{final} we must show that (\ref{asymr})-(\ref{barphianswer}) solve the geodesic equations (\ref{geo1})-(\ref{geo2}) to order $O(\epsilon^2)$.   To start we verify (\ref{geo2}).    Using that $\hat{\phi}=O(\epsilon)$, we can write (\ref{geo2}) as
\be
&-\frac{d^2\hat{\phi}}{d\lambda^2}=\Gamma^2_{00}\left(\frac{dt}{d\lambda}\right)^2+\Gamma^2_{11}\left(\frac{d\hat{r}}{d\lambda}\right)^2+2\Gamma^2_{01}\left(\frac{dt}{d\lambda}\right)\left(\frac{d\hat{r}}{d\lambda}\right)\ \ \ \ \ \ \ \ \ \ \ \ \ \ \ \ \ \ \ \ &\nonumber\\
&\ \ \ \ \ \ \ \ \ \ \ \ \ \ \ \ +2\Gamma^2_{02}\left(\frac{dt}{d\lambda}\right)\left(\frac{d\hat{\phi}}{d\lambda}\right)+2\Gamma^2_{12}\left(\frac{d\hat{r}}{d\lambda}\right)\left(\frac{d\hat{\phi}}{d\lambda}\right)+O(\epsilon^2),&\nonumber\\
\nonumber
\ee
the only derivative not appearing on the right hand side being $\left(\frac{d\hat{\phi}}{d\lambda}\right)^2$, which is $O(\epsilon^2).$  Multiplying through by $\left(\frac{d\hat{r}}{d\lambda}\right)^{-2}$ gives  
\be
&-\left(\frac{d\hat{r}}{d\lambda}\right)^{-1}\frac{d}{d\hat{r}}\frac{d\hat{\phi}}{d\lambda}=\Gamma^2_{00}\left(\frac{dt}{d\hat{r}}\right)^2+\Gamma^2_{11}+2\Gamma^2_{01}\left(\frac{dt}{d\hat{r}}\right)\ \ \ \ \ \ \ \ \ \ \ \ \ \ \ \ \ \ \ \ &\label{geostart}\\
&\ \ \ \ \ \ \ \ \ \ \ \ \ \ \ \ +2\Gamma^2_{02}\left(\frac{dt}{d\hat{r}}\right)\left(\frac{d\hat{\phi}}{d\hat{r}}\right)+2\Gamma^2_{12}\left(\frac{d\hat{\phi}}{d\bar{r}}\right)+O(\epsilon^2).&\nonumber\\
\nonumber
\ee
Now since by (\ref{dtdbarr}) we know $\frac{dt}{d\hat{r}}=\sqrt{t}+O(\epsilon^2)$, putting (\ref{G211})-(\ref{G122}) with (\ref{E211})-(\ref{E201}) in (\ref{geostart}), using (\ref{dtdbarr}) gives
\be
&-\left(\frac{d\hat{r}}{d\lambda}\right)^{-1}\frac{d}{d\hat{r}}\frac{d\hat{\phi}}{d\lambda}=\frac{1}{2\hat{r}^2}\frac{\partial}{\partial\hat{\phi}}(F_a^2)+E^2_{00}t+\frac{(a^2-1)r_e}{2t\hat{r}}\hat{\phi}-\frac{1}{2\hat{r}^2}\frac{\partial}{\partial\hat{\phi}}(F_a^2)\ \ \ \ \ \ \ \ \ \ \ \ \ \ \ \ \ \ \ \ &\nonumber\\
&\ \ \ \ \ \ \ \ \ \ \ \ \ \ \ \ +E^2_{11}+2E^2_{01}\sqrt{t}+2\left(\frac{1}{2\sqrt{t}}+\frac{1}{\bar{r}}\right)\frac{d\hat{\phi}}{d\hat{r}}+O(\epsilon^2).&\nonumber\\
\nonumber
\ee
(Note the cancellation between the $\Gamma^2_{11}$ and $\Gamma^2_{00}$ terms $\frac{1}{2\hat{r}^2}\frac{\partial}{\partial\hat{\phi}}(F_a^2)$.)

\noindent Neglecting $O(\epsilon^2)$ terms, then multiplying through by $\sqrt{t}\hat{r}\epsilon^{-1}$ and using (\ref{E211})-(\ref{E201}), gives the leading order equation for $\bar{\phi}$
\be
&-\sqrt{t}\hat{r}\left(\frac{d\hat{r}}{d\lambda}\right)^{-1}\frac{d}{d\hat{r}}\frac{d\bar{\phi}}{d\lambda}=\frac{(a^2-1)r_e}{2\sqrt{t}}\bar{\phi}+\sqrt{t}\left(\frac{\hat{r}}{\sqrt{t}}+2\right)\frac{d\bar{\phi}}{d\hat{r}}+O(1)|a-1|\left\{\cdot\right\}.&\nonumber\\
\nonumber
\ee
where
\be
\left\{\cdot\right\}=\left\{\frac{r_e^3}{t^{3/2}}+\frac{r_e^3}{t^{3/2}}+\frac{r_e^2}{t}\right\}=O(1)\zeta_e^2+O(\epsilon^2),
\ee
and we have used Lemma \ref{lemmat}.  Thus the leading order in $\epsilon$ equation for $\bar{\phi}$ is
\be
&-\hat{r}\sqrt{t}\left(\frac{d\hat{r}}{d\lambda}\right)^{-1}\frac{d}{d\hat{r}}\frac{d\bar{\phi}}{d\lambda}=\frac{(a^2-1)r_e}{2\sqrt{t}}\bar{\phi}+\sqrt{t}\left(\frac{\hat{r}}{\sqrt{t}}+2\right)\frac{d\bar{\phi}}{d\hat{r}}+O(1)|a-1|\zeta_e^2.\ \ \ \ \ &\nonumber\\\label{leadingorder0}
\ee
Now
\be
\left(\frac{d\hat{r}}{d\lambda}\right)^{-1}\frac{d}{d\hat{r}}\frac{d\bar{\phi}}{d\lambda}=\left(\frac{d\hat{r}}{d\lambda}\right)^{-1}\frac{d}{d\hat{r}}\left\{\frac{d\hat{r}}{d\lambda}\frac{d\bar{\phi}}{d\hat{r}}\right\}=\left(\frac{d\hat{r}}{d\lambda}\right)^{-1}\frac{d}{d\hat{r}}\left\{\frac{d\hat{r}}{d\lambda}\right\}\frac{d\bar{\phi}}{d\hat{r}}+\frac{d^2\bar{\phi}}{d\hat{r}^2},\nonumber
\ee
and by (\ref{asymr}), (\ref{ageo3}),
\be
\left(\frac{d\hat{r}}{d\lambda}\right)^{-1}\frac{d}{d\hat{r}}\left\{\frac{d\hat{r}}{d\lambda}\right\}=-\frac{1}{\sqrt{t}}\left\{1+O(1)|a-1|\zeta_e\right\}.
\nonumber
\ee
Putting this into (\ref{leadingorder0}), neglecting higher order terms, gives
\be\label{leadingorder}
&&\hat{r}\left\{1+O(1)|a-1|\zeta_e\right\}\frac{d\bar{\phi}}{d\hat{r}}-\hat{r}\sqrt{t}\frac{d^2\bar{\phi}}{d\hat{r}^2}\\
&&\ \ \ \ \ \ \ \ \ \ \ \ \ \ \ \ =\frac{(a^2-1)r_e}{2\sqrt{t}}\bar{\phi}+2\sqrt{t}\left(1+\frac{\hat{r}}{2\sqrt{t}}\right)\frac{d\bar{\phi}}{d\hat{r}}+O(1)|a-1|\zeta_e^2.\nonumber
\ee
Now using that $t$ as a function of $\hat{r}$ agrees with the baseline geodesic relation
$$
t=t_e\left(1+\frac{1}{2}\frac{\hat{r}}{\sqrt{t_e}}\right)^2,
$$
that $\hat{r}/\sqrt{t_e}$ is $O(1)\zeta_e$, and incorporating terms second order in $\zeta_e$ into $O(1)\zeta_e^2$, (\ref{leadingorder}) is equivalent to
\be\label{leadingorder1}
&-t_e\frac{\hat{r}}{t_e}\frac{d^2\bar{\phi}}{d\hat{r}^2}=\frac{(a^2-1)}{2}\zeta_e\bar{\phi}+2\sqrt{t_e}\left(1+\frac{1}{2}\frac{\hat{r}}{\sqrt{t_e}}\right)\frac{d\bar{\phi}}{d\hat{r}}+O(1)|a-1|\zeta_e^2.
\ee
Finally, we determine the constants $C_1$ and $C_2$ such that
\be
\bar{\phi}(\hat{r})&=&1+C_1\frac{\hat{r}}{\sqrt{t_e}}+C_2\left(\frac{\hat{r}}{\sqrt{t_e}}\right)^2+O(1)\left(\frac{\hat{r}}{\sqrt{t_e}}\right)^3.\label{phibar1}
\ee
(With this ansatz, $\bar{\phi}$ as well as equation (\ref{leadingorder1}) both reduce to functions of $\zeta_e$ at $\hat{r}=r_e$.)  Substituting this together with
\be
\frac{d\bar{\phi}}{d\hat{r}}(\hat{r})&=&\frac{1}{\sqrt{t_e}}\left\{C_1+2C_2\left(\frac{\hat{r}}{\sqrt{t_e}}\right)+O(1)\left(\frac{\hat{r}}{\sqrt{t_e}}\right)^2\right\},\nonumber\\
\frac{d^2\bar{\phi}}{d\hat{r}^2}(\hat{r})&=&\frac{1}{t_e}\left\{2C_2+O(1)\left(\frac{\hat{r}}{\sqrt{t_e}}\right)\right\},\nonumber
\ee
into (\ref{leadingorder1}), evaluating at $\hat{r}=\hat{r}_e$, and then collecting powers of $\zeta_e$, yields
\be
\left\{2C_1\right\}+\left\{6C_2+\frac{a^2-1}{2}+C_1\right\}\zeta_e+O(1)\zeta_e^2=0.\nonumber
\ee
It follows then that (\ref{phibar1}) meets the $\bar{\phi}$-geodesic equation (\ref{geo2}) to order $O(1)\zeta_e^3$ if we choose
\be
C_1=0,\ \ \ C_2=-\frac{a^2-1}{12}.\label{Cchoices}
\ee
We conclude that the function $\hat{\phi}(\bar{r})=\epsilon\bar{\phi}(\bar{r})$ that solves the $\hat{\phi}$-geodesic equation (\ref{geo2}) to order $O(\epsilon^2)$, satisfies the asymptotic relation in $\zeta_e$ given by
$$
\bar{\phi}=1-\frac{a^2-1}{12}\zeta_e^2+O(\zeta_e^2).
$$  
Since $\bar{\phi}=1$ solves (\ref{geo2}) identically in the FRW case $a=1,$ (in the Standard Model, the angle is constant along radial geodesics emanating from any center),  it follows by smoothness that when (\ref{Cchoices}) is assumed, the $O(1)$ in (\ref{phibar1}) is really $O(1)|a-1|$, and we have proven what was claimed in (\ref{barphianswer}).  

To complete the proof of Lemma \ref{final}, it remains only to show that the ansatz (\ref{asymr})-(\ref{asymphi1}) meets the $t$- and $\hat{r}$-geodesic equations (\ref{geo0}) and (\ref{geo1}) to order $\epsilon^2$ under the condition that $\bar{\phi}$ solves the $\epsilon$-order equation from the $\bar{\phi}$-geodesic equation (\ref{geo2}) that we just analyzed.   That is,  it suffices to prove the following lemma:

\begin{Lemma}\label{finalmirrorlemma}
Let $t_a(\lambda)$ and $\hat{r}_a(\lambda)$ be as defined in the baseline formula (c.f. (\ref{baselineformula}), (\ref{baselineformula2})), and assume that the approximate geodesic 
\be\label{theapproxsoln}
(t_a(\lambda),\hat{r}_a(\lambda),\epsilon \bar{\phi}_a(\lambda)),
\ee
solves the $\hat{\phi}$-geodesic equation to order $O(\epsilon^2).$  Then (\ref{theapproxsoln}) also solves the $t$- and $\hat{r}$-geodesic equations to order $O(\epsilon^2).$
\end{Lemma}
\vspace{.2cm}

\noindent {\bf Proof:}   \noindent   Define the metric $\tilde{g}$,
\be\label{radialgeo}
d\tilde{s}^2=-\tilde{F}_a^2dt^2+\tilde{F}_a^2td\hat{r}^2\equiv \tilde{g}_{ij}d\hat{x}^id\hat{x}^j,\label{approxhatt}
\ee 
so that $t_a(\lambda)$, $\hat{r}_a(\lambda)$ exactly solve the geodesic equations
\be\label{radialGt}
\frac{d^2t}{d\lambda^2}&=&-\tilde{\Gamma}^0_{ij}\frac{d\hat{x}^i}{d\lambda}\frac{d\hat{x}^j}{d\lambda},\\
\frac{d^2\hat{r}}{d\lambda^2}&=&-\tilde{\Gamma}^1_{ij}\frac{d\hat{x}^i}{d\lambda}\frac{d\hat{x}^j}{d\lambda},\label{radialGr}
\ee
 where
\be
\tilde{F}_a^2\equiv\tilde{F}_a^2(\tilde{\zeta})=1+\frac{a^2-1}{4}\tilde{\zeta}^2,\label{approxF}
\ee
\be\label{approxzeta}
\tilde{\zeta}^2\equiv\frac{\tilde{r}^2}{t},
\ee
\be\label{approxrtilde}
\tilde{r}=r_e-\hat{r}.
\ee
The point is that $\tilde{\zeta}=\tilde{r}/\sqrt{t}$ is obtained from $\zeta=r/\sqrt{t}$ by setting $\cos{\hat{\phi}}=0$ in the formula $r=\sqrt{r_e^2-2r_e\hat{r}\cos{\hat{\phi}}+\hat{r}^2}$, thereby incurring an error no greater than $O(\epsilon^2).$  A direct consequence of (\ref{radialGt})-(\ref{approxrtilde}) is that,
\be
\frac{\partial}{\partial\hat{r}}\,r^2&=&2\tilde{r}+O(\epsilon^2),\label{diffa}\\
\frac{\partial}{\partial\hat{\phi}}\,r^2&=&O(\epsilon),\label{diffb}\\
\frac{\partial}{\partial\hat{r}}\,\zeta^2&=&\frac{\partial}{\partial\hat{r}}\tilde{\zeta}^2+O(\epsilon^2).\label{diffc}
\ee
Now to prove that (\ref{theapproxsoln}) satisfies the equations (\ref{geo0}) and (\ref{geo1}) to $O(\epsilon^2)$,  it suffices only to show that
\be
\Gamma^{\alpha}_{ij}&=&\tilde{\Gamma}^{\alpha}_{ij}+O(\epsilon^2),\ \ i\neq2,\ j\neq2,\ \alpha=0,1\label{Gorder2}\\
\Gamma^{\alpha}_{2j}&=&\tilde{\Gamma}^{\alpha}_{2j}+O(\epsilon),\ \ j=0,1,2,\ \alpha=0,1.\label{Gorder1}
\ee
(We don't need $\alpha=2$ because that applies only to the $\hat{\phi}$-geodesic equation, and we only need an order $O(\epsilon)$ on any term that gets multiplied by $\frac{d\hat{\phi}}{d\lambda}$ because $\hat{\phi}$ and its derivatives are $O(\epsilon)$ asymptotically.)  

Using (\ref{diffa})-(\ref{diffc}) it is straightforward to obtain the following two tables giving the values of the brackets (\ref{brackets}) associated with the full metric $\hat{g}_{ij}$ taken to order $O(\epsilon^3)$ in (\ref{ghatorder3}). ($\left\{02,2\right\}$
\be
\begin{array}{ll}
\{00,0\}=-\frac{\partial F_a^2}{\partial t}\ \ \ \ &\{00,1\}=\frac{\partial F_a^2}{\partial \hat{r}}\\
\{11,0\}=-\frac{\partial}{\partial t}\left\{F_a^2t\right\}+O(\epsilon^2)&\{11,1\}=\frac{\partial }{\partial \hat{r}}(F_a^2t)+O(\epsilon^2)\\
\{01,0\}=-\frac{\partial F_a^2}{\partial \hat{r}}&\{01,1\}=\frac{\partial}{\partial t}(F_a^2t)+O(\epsilon^2)\\
\{02,0\}=-\frac{\partial F_a^2}{\partial \hat{\phi}}=O(\epsilon)&\{02,1\}=O(\epsilon)\\
\{12,0\}=O(\epsilon)&\{12,1\}=O(\epsilon);\\
\{22,0\}=O(1)&\{22,1\}=O(1)
\end{array}\label{bracketfinal}
\ee
and the nonzero brackets of form $\{ij,2\}$ that are lower order than $O(\epsilon^2)$ have the following orders as $\epsilon\rightarrow0:$
\be
\{00,2\}&=&O(\epsilon)\label{bracketfinala}\\
\{11,2\}&=&O(\epsilon)\label{bracketfinalb}\\
\{12,2\}&=&O(1).
\label{bracketfinal2}
\ee
Now it is also straightforward from (\ref{diffa})-(\ref{diffc}) that
 \be
 F_a^2&=&\tilde{F}_a^2+O(\epsilon^2),\nonumber\\
 \frac{\partial F_a^2}{\partial\hat{r}}&=& \frac{\partial \tilde{F}_a^2}{\partial\hat{r}}+O(\epsilon^2),\nonumber\\
  \frac{\partial F_a^2}{\partial t}&=& \frac{\partial \tilde{F}_a^2}{\partial t}+O(\epsilon^2),\nonumber\\
   \frac{\partial F_a^2}{\partial\hat{\phi}}&=&O(\epsilon),\nonumber
 \ee
 as $\epsilon\rightarrow0,$ so we can replace $F_a^2$ by $\tilde{F}_a^2$ in (\ref{bracketfinal}), incurring no change in the stated errors in $\epsilon$.

To verify (\ref{Gorder2}), (\ref{Gorder1}), consider first $\Gamma^0_{ij}.$  Then
$$
\Gamma^0_{ij}=\frac{1}{2}\hat{g}^{00}\{ij,0\}=-\frac{1}{2}F_a^2\{ij,0\}=-\frac{1}{2}\tilde{F}_a^2\{ij,0\}+O(\epsilon^2).
$$
Thus it suffices to verify  (\ref{Gorder2}), (\ref{Gorder1}) for $\{ij,0\}$ in place of $\Gamma^0_{ij}$.  But replacing $F_a^2$ by $\tilde{F}_a^2$ on the left hand side of (\ref{bracketfinal}) incurs no change in the $\epsilon$ order errors, and this verifies (\ref{Gorder2}), (\ref{Gorder1}) in this case.    

Consider next $\Gamma^1_{ij}.$  In this case 
$$
\Gamma^1_{ij}=\frac{1}{2}\hat{g}^{11}\{ij,1\}+\frac{1}{2}\hat{g}^{12}\{ij,2\},
$$
where by (\ref{matrixhatg}).
\be
\hat{g}^{11}&=&\frac{1}{tF_a^2}=O(1),\label{ghat11}\\
\hat{g}^{12}&=&-\frac{F_a^2-1}{t^2F_a^2}\frac{(\hat{r}-r_e)r_e}{\hat{r}}\epsilon\bar{\phi}=O\left(\epsilon\right).\label{ghat12}
\ee
Therefore the result (\ref{Gorder2}), (\ref{Gorder1}) applies after replacing $F_a^2$ by $\tilde{F}_a^2$ on the right hand side of (\ref{bracketfinal}), incurring no change in the stated errors in $\epsilon$. 

In short, the argument demonstrating that the $t$- and $\hat{r}$-geodesic equations (\ref{geo0}), (\ref{geo1}) agree with (\ref{radialGt}), (\ref{radialGr}) to order $\epsilon^2$ can be summarized as follows.  The main observation is that by (\ref{ghatorder2}), the metric $\hat{g}$ in (\ref{radialgeo}) agrees with $\tilde{g}$ and its inverse $\hat{g}^{-1}$ agrees with $\tilde{g}^{-1}$ to order $\epsilon^2$, except in the $\hat{g}_{12}$ and $\hat{g}^{12}$ components, respectively, which both contain an order $\epsilon$ error.  As a consequence we need only show that the derivatives $\hat{g}_{ij,k}$ that differ by order $\epsilon$ or by $O(1)$ from the corresponding derivatives in $\tilde{g}_{ij,k}$ end up multiplied by terms $O(\epsilon)$ and $O(\epsilon^2)$ in the $t$- and $\hat{r}$-geodesic equations (\ref{geo0}), (\ref{geo1}), respectively.  

Now the derivatives $\hat{g}_{ij,k}$ that differ by order $\epsilon$ from the corresponding derivatives in $\tilde{g}_{ij,k}$ are the $t$ and $\hat{r}$ derivatives of $\hat{g}_{12}$, together with the $\hat{\phi}$ derivative of a metric entry {\it other than} $\hat{g}_{12}$;  and the only derivative $\hat{g}_{ij,k}$ that differs by $O(1)$ from $\tilde{g}_{ij,k}$ is $\hat{g}_{12,2}$.  Now derivatives of metric entries with respect to $\hat{\phi}$ can only appear as terms in $\Gamma^0_{ij}$ or $\Gamma^1_{ij}$ with $i=2$ or $j=2$, or else as terms of the form $\hat{g}^{12}\{ij,2\}$ in $\Gamma^1_{ij}$, with $i,j$=$0$ or $1$.  But $\Gamma^k_{i2}$ is always multiplied by $\frac{d\hat{\phi}}{d\lambda}$ in the geodesic equations, so these epsilon order errors become  $O(\epsilon^2)$ in the equation, and can be neglected.  Thus it remains only to see that the $\epsilon$-order derivatives $\hat{g}_{12,0}$, $\hat{g}_{12,1}$ end up multiplied by order $\epsilon$ in the $t$- and $\hat{r}$-geodesic equations  (\ref{geo0}), (\ref{geo1}), and that the zero order term $\hat{g}_{12,2}$ ends up multiplied by order $\epsilon^2$.  But  the derivative $\hat{g}_{12,0}$ appears only in $\Gamma^0_{12}$ and $\Gamma^1_{02}$, and $\hat{g}_{12,1}$ appears only in $\Gamma^1_{12}$ and $\Gamma^1_{11}$, and in each instance these derivatives are multiplied by either the order epsilon term $\hat{g}^{12}$ or else by the order $O(\epsilon)$ term $\frac{d\hat{\phi}}{d\lambda}$ in the $t$- and $\hat{r}$-geodesic equations (\ref{geo0}), (\ref{geo1});  and the order one term $\hat{g}_{12,2}$ only appears in $\Gamma^1_{22}$, which ends up multiplied by the $O(\epsilon^2)$ term $\frac{d^2\hat{\phi}}{d\lambda^2}$, as claimed.

  This completes the proof of Lemma \ref{finalmirrorlemma}, and the proof of Lemma \ref{final} is complete.


\section{Concluding Remarks}
\setcounter{equation}{0} 

We have constructed a one parameter family of general relativistic expansion waves which at a single parameter value, reduces to what in this paper we call the FRW spacetime, the Standard Model of Cosmology during the radiation epoch.  The discovery of this family is made possible by a remarkable coordinate transformation that maps the FRW metric in standard co-moving coordinates, over to Standard Schwarzschild coordinates (SSC) in such a way that all quantities depend only on the  single self-similar variable $\xi=\bar{r}/\bar{t}$.  Note that it is not evident from the FRW metric in standard co-moving coordinates that self-similar variables even exist, and if they do exist, by what ansatz one should extend the metric in those variables to obtain nearby self-similar solutions that solve the Einstein equations exactly.   The main point is that our coordinate mapping to SSC form, explicitly identifies the self-similar variables as well as the metric ansatz that together accomplish such an extension of the metric. 

The self-similarity of the FRW metric in SSC suggested the existence of a reduction of the SSC Einstein equations to a new set of ODE's in  $\xi.$  Deriving this system from first principles then establishes that the FRW spacetime does indeed extend to a three parameter family of expanding wave solutions of the Einstein equations.  This three parameter family reduces to an (implicitly defined) one parameter family by removing a scaling invariance and imposing regularity at the center.  The remaining parameter $a$ changes the expansion rate of the spacetimes in the family, and thus we call it the {\it acceleration parameter}.   Transforming back to (approximate) co-moving coordinates, the resulting one parameter family of metrics is amenable to the calculation of a redshift vs luminosity relation, to third order in the redshift factor $z,$ leading to the relation (\ref{redvslumintro}).  It follows by continuity that the leading order part of an anomalous correction to the redshift vs luminosity relation of the Standard Model observed {\em after} the radiation phase, can be accounted for by suitable adjustment of parameter $a$.  

These results suggest an interpretation that we might call a {\it Conservation Law Scenario} of the Big Bang.  That is, it is well known that highly interactive oscillatory solutions of conservation laws decay in time to non-interacting waves, (shock waves and expansion waves), by the mechanisms of wave interactions and shock wave dissipation.  The subtle point is that even though dissipation terms are neglected in the formulation of the equations, there is a canonical dissipation and consequent loss of information due to the nonlinearities, and this can be modeled by shock wave interactions that drive solutions to non-interacting wave patterns.  (This viewpoint is well expressed in the celebrated works \cite{lax,glim,glimla}).  Since the one fact most certain about the Standard Model is that our universe arose from an earlier hot dense epoch in which all sources of energy were in the form of radiation, and since it is approximately uniform on the largest scale but highly oscillatory on smaller scales\footnotemark[22]\footnotetext[22]{In the Standard Model, the universe is approximated by uniform density on a scale of a billion light years or so, about a tenth of the radius of the visible universe, \cite{wein}.  The stars, galaxies and clusters of galaxies are then evidence of large oscillations on smaller scales.}, one might reasonably conjecture that decay to a non-interacting expanding wave occurred during the radiation phase of the Standard Model, via the highly nonlinear evolution driven by the large sound speed, and correspondingly large modulus of {\it Genuine Nonlinearity}\footnotemark[23]\footnotetext[23]{Again, {\it Genuine Nonlinearity} is in the sense of Lax, a measure of the magnitude of nonlinear compression that drives decay, c.f.,  \cite{lax}.}, present when $p=\rho c^2/3$, c.f. \cite{smolteeuler}.   Our analysis has shown that FRW is just one point in a family of non-interacting, self-similar expanding waves, and as a result we conclude that some further explanation is required as to why, on some length scale, decay during the radiation phase of the Standard Model would not proceed to a member of the family satisfying $a\neq1.$  If decay to $a\neq1$ did occur, then the galaxies that formed from matter at the end of the radiation phase, (some $379,000$ years after the Big Bang), would be displaced from their anticipated positions in the Standard Model at present time, and this displacement would lead to a modification of the observed redshift vs luminosity relation.  In short, the displacement of the fluid particles, (i.e., the displacement of the co-moving frames in the radiation field), by the wave during the radiaton epoch leads to a displacement of the galaxies at a later time.    In principle such a mechanism could account for the anomalous acceleration of the galaxies as observed in the supernova data.  Of course, if $a\neq1$, then the spacetime within the expanding wave has a center, and this would violate the so-called {\it Copernican Principle}, a simplifying assumption generally accepted in cosmology, at least on the scale of the wave (c.f. the discussions in \cite{temptalk} and \cite{copihudjgd}).  Moreover, if our Milky Way galaxy did not lie within some threshold of the center of expansion, the expanding wave theory would imply unobserved angular variations in the expansion rate.  In fact, all of these observational issues have already been discussed recently in  \cite{cliffela,copihudjgd,cliffe}, (and references therein), which explore the possibility that the anomalous acceleration of the galaxies might be due to a local {\it void} or under-density of galaxies in the vicinity of the Milky Way.\footnotemark[24]\footnotetext[24]{The size of the center, consistent with the angular dependence that has been observed in the actual supernova and microwave data, has been estimated to be about $15$ megaparsecs, approximately the distance between clusters of galaxies, roughly $1/200$ the distance across the visible universe, c.f. \cite{copihudjgd, cliffela,cliffe}.}   Our proposal then, is that the one parameter family of general relativistic self-similar expansion waves derived here are possible end-states that could result after dissipation by wave interactions during the radiation phase of the Standard Model is completed, and such waves could thereby account for the appearance of a local under-density of galaxies at a later time.\footnotemark[25]\footnotetext[25]{The following back of the envelope calculation from \cite{tempsmquest} provides a ballpark estimate for what we might expect the extent of the remnants of one of these expanding waves might be today if our thesis is correct that wave interactions and dissipation by strong nonlinearities during the radiation phase were the primary mechanisms involved in formation of the wave.  For this, note that matter becomes transparent with radiation at about 300,000 years after the Big Bang, so we might estimate that the wave should have emerged by about $t_{endrad}\approx 10^5$ years after the Big Bang.  At this time, the distance of light-travel since the Big Bang is about $10^5$ lightyears.  Since the sound speed $c/\sqrt{3}\approx .58 c$ during the radiation phase is comparable to the speed of light, we could estimate that dissipation that drives decay to the expanding wave might reasonably be operating over a scale of $10^5$ lightyears by the end of the radiation phase.   Now in the $p=0$ expansion that follows the radiation phase, the scale factor (that gives the expansion rate) evolves like
$$R(t)=t^{2/3},$$ 
\cite{wein}, so a distance of $10^5$ lightyears at $t=t_{endrad}$ years will expand to a length $L$ at present time $t_{present}\approx10^{10}$ years by a factor of 
$$
\frac{R(t_{present})}{R(t_{endrad})}\approx\frac{\left(10^{10}\right)^{2/3}}{\left(10^5\right)^{2/3}}=10^{4.7}\geq 5\times10^4.
$$
It follows then that we might expect the scale of the wave at present time to extend over a distance of about 
$$
L=5\times10^5\times 10^4=5\times10^9\ \ lightyears.
$$
This is a third to a fifth of the distance across the visible universe, and agrees well with the extent of the under-density void region quoted in the Clifton-Ferriera paper, with room to spare.}   

In any case, the expanding wave theory is testable.  For a first test, we propose next to evolve the quadratic and cubic corrections to redshift vs luminosity recorded here in relation (\ref{redvslumintro}), valid at the end of the radiation phase, up through the $p\approx 0$ stage to present time in the Standard Model,  to obtain the present time values of the quadratic and cubic corrections to redshift vs luminisity implied by the expanding waves, as a function of the acceleration paramter $a.$  Once accomplished, we can look for a best fit value of $a$ via comparison of the  quadratic correction at present time to the quadratic correction observed in the supernova data, leaving the third order correction at present time as a prediction of the theory.  That is,  in principle, the predicted third order correction term could be used to distinguish the expanding wave theory from other theories (such as dark energy) by the degree to which they match an accurate plot of redshift vs luminosity from the supernove data,   (a topic of the authors' current research).   The idea that the anomalous acceleration might be accounted for by a local under-density in a neighborhood of our galaxy was expounded in the recent papers \cite{cliffela,cliffe}.  
Our results here might then give an accounting for the source of such an under-density.  

In summary, the expanding wave theory could in principle give an explanation for the observed anomalous acceleration of the galaxies within classical general relativity, with classical sources.  In the expanding wave theory, the so-called anomalous acceleration is not an acceleration at all, but is a correction to the Standard Model due to the fact that we are looking outward into an expansion wave.  The one parameter family of non-interacting, self-similar, general relativistic expansion waves derived here, are all possible end-states that could result by wave interaction and dissipation due to nonlinearities back when the universe was filled with pure radiation sources.  And when $a\neq1$ they introduce an anomalous acceleration into the Standard Model of cosmology.   Unlike the theory of Dark Energy, this provides a possible explanation for the anomalous acceleration of the galaxies that is not {\it ad hoc} in the sense that it is derivable exactly from physical principles and a mathematically rigorous theory of general relativistic expansion waves.  In particular, this explanation does not require the {\it ad hoc} assumption of a universe filled with an as yet unobserved form of energy with anti-gravitational properties, (the standard physical interpretation of the cosmological constant),  in order to fit the data.  The idea that the anomalous acceleration might be accounted for by a local under-density in a neighborhood of our galaxy was expounded in the recent papers \cite{cliffela,cliffe}.   Our results here might then give an accounting for the source of such an under-density.  

In conclusion, these new general relativistic expanding waves provide a new paradigm to test against the Standard Model.  Even if they do not in the end explain the anomalous acceleration of the galaxies, one has to believe they are present and propagating on some scale, and their presence represents an instability in the Standard Model in the sense that an explanation is required as to why small scale oscillations have to settle down to large scale $a=1$ expansions instead of $a\neq1$ expansions, (either locally or globally), during the radiation phase of the Big Bang.

\end{document}